\pdfoutput=1 

\documentclass[preprint]{elsarticle}

\usepackage{verbatim}                         
\usepackage[margin=1in]{geometry}    

\usepackage{color, soul}                       

\usepackage{graphicx}    

\usepackage{todonotes}    

\usepackage{url}    

\biboptions{round, semicolon, authoryear}   

\usepackage{amsmath}
\usepackage{amsfonts}
\usepackage{amssymb}  

\usepackage{tabularx}    

\usepackage{array}
\newcolumntype{K}[1]{>{\raggedright\let\newline\\\arraybackslash\hspace{0pt}}m{#1}}    

\usepackage{booktabs}    
\usepackage[justification=centering]{caption}

\usepackage{pdflscape}    

\begin{document}

\begin{frontmatter}

\title{Machine Learning Meets Microeconomics:\\The Case of Decision Trees and Discrete Choice}

\author[tbrathwaite]{Timothy Brathwaite\corref{cor1}}
\ead{timothyb0912@gmail.com}

\author[avij]{Akshay Vij}
\ead{Akshay.Vij@unisa.edu.au}

\author[jwalker]{Joan L. Walker}
\ead{joanwalker@berkeley.edu}

\cortext[cor1]{Corresponding Author}

\address[tbrathwaite]{Department of Civil and Environmental Engineering, University of California at Berkeley\\ 116 McLaughlin Hall, University of California, Berkeley, CA, 94720-1720}
\address[avij]{Institute for Choice, University of South Australia\\ Level 13, 140 Arthur Street, North Sydney, NSW 2060}
\address[jwalker]{Department of Civil and Environmental Engineering, University of California at Berkeley\\ 111 McLaughlin Hall, University of California, Berkeley, CA, 94720-1720}

\begin{abstract}
In the 1960's, the logistic regression model from statistics and the binary probit model from psychology were linked with random utility theory, thereby connecting such methods with economic theory. Since then, the fields of statistics, computer science, and machine learning have created numerous methods for modeling discrete choices. However, these newer methods have not been derived from or linked with economic theories of human decision making. We believe this lack of economic interpretation is one reason discrete choice modelers have been slow to adopt these newer methods.

Our paper begins bridging this gap by providing a microeconomic framework for decision trees: a popular machine learning method. Specifically, we show how decision trees represent a non-compensatory decision protocol known as disjunctions-of-conjunctions and how this protocol generalizes many of the non-compensatory rules used in the discrete choice literature so far. Additionally, we show how existing decision tree variants address many economic concerns that choice modelers might have. Beyond theoretical interpretations, we contribute to the existing literature of two-stage, semi-compensatory modeling and to the existing decision tree literature. In particular, we formulate the first bayesian model tree, thereby allowing for uncertainty in the estimated non-compensatory rules as well as for context-dependent preference heterogeneity in one's second-stage choice model. Using an application of bicycle mode choice in the San Francisco Bay Area, we estimate our bayesian model tree, and we find that it is over 1,000 times more likely to be closer to the true data-generating process than a multinomial logit model (MNL). Qualitatively, our bayesian model tree automatically finds the effect of bicycle infrastructure investment to be moderated by travel distance, socio-demographics and topography, and our model identifies diminishing returns from bicycle lane investments. These qualitative differences lead the bayesian model trees to produce forecasts that directly align with the observed bicycle mode shares in regions with abundant bicycle infrastructure such as Davis, CA and the Netherlands. In comparison, the forecasts of the MNL model are overly optimistic.
\end{abstract}

\begin{keyword}
Decision Trees \sep Non-compensatory Decision Protocols \sep Discrete Choice \sep Two-stage Decision Making \sep Machine Learning \sep Semi-compensatory Models
\end{keyword}
\end{frontmatter}

\section{Introduction}
\label{sec:introduction}
During the 1960s and 1970s, Daniel McFadden spearheaded the use of discrete choice techniques within economics, and in 2000, he was awarded a Nobel Prize for this work \citep{university_of_california_2000, manski_2001_daniel}. By his own account \citep{mcfadden_2001_economic}, McFadden's major contribution was \textit{not} the creation of the conditional logit\footnote{Note that the conditional logit model is also commonly referred to as the multinomial logit (MNL) model.} model---a model that is still one of the most widely used discrete choice methods today. Indeed, the concept of a random utility maximization model was created earlier by Jacob Marschak (\citeyear{marschak_1960_binary}), and statistical models that are nearly equivalent to McFadden's conditional logit model had already been introduced by David Cox (\citeyear{cox_1966_some}). According to McFadden,
\begin{quotation}
``The reason my formulation of the MNL model has received more attention than others that were developed independently during the same decade seems to be the direct connection that I provided to consumer theory [...].'' \citep[p. 354]{mcfadden_2001_economic}.
\end{quotation}
Put simply, the great contribution of McFadden's work is that he connected an existing statistical model of discrete outcomes with economic theory \citep{manski_2001_daniel}.

In the more than fifty years since McFadden's pioneering efforts, the fields of machine learning and statistics have produced a vast array of methods that, like discrete choice models, predict the probability that a given discrete outcome will be realized out of a finite set of discrete alternatives. We now have decision trees, kernel machines, neural networks, and much more \citep{bishop_2006_pattern, friedman_2008_elements, murphy_2012_machine}. In general, these new techniques often display superior predictive ability compared to traditional discrete choice models \citep{fernandez_2014_do, wainer_2016_comparison}. However, despite this smorgasbord of accurate methods, discrete choice modelers have mostly restricted themselves to econometric techniques that are descended from McFadden's conditional logit model \citep{manski_2001_daniel}.

We hypothesize that one reason machine learning models have not made greater inroads amongst discrete choice modelers is because these models have not been linked to economic theories of human decision-making. Moshe Ben-Akiva (\citeyear{ben_1973_structure}), one of the earliest discrete choice researchers, once wrote that ``a model can duplicate the data perfectly, but may serve no useful purpose for prediction\footnote{Note that the sort of prediction being referred to is prediction in the face of a policy change. This type of prediction is characteristic of causal inference whereby one predicts the effects of external manipulation of environmental conditions.} if it represents erroneous behavioral assumptions.'' Though written in the 1970's, we believe that this sentiment still pervades the field of discrete choice modeling and econometrics more broadly \citep{einav_2014_data, bajari_2015_demand, bajari_2015_machine}. As a result, econometricians do not make frequent use of alternative techniques from machine learning and statistics. Such methods may be useful for prediction under stationary conditions, but they are considered black-boxes that lack a theoretical basis for interpreting and understanding human behavior.

In contrast to newer techniques from statistics and machine learning, almost all discrete choice models in the literature are rooted in the theory of utility maximization \citep{train_2009_discrete}, and even competing discrete choice models are based on alternative behavioral theories such as regret minimization \citep{chorus_2012_random}. Overall, theory-based econometric techniques appear to have become dominant within econometrics because behavioral theories provide a way to understand and interpret one's model outputs beyond in-sample and out-of-sample predictive accuracy. Machine learning methods have yet to provide this additional framework and linkage with economic theory.

In this paper, we aim to bridge this method-versus-theory gap by continuing to merge existing quantitative techniques with economic principles. Our contributions to the literature are as follows. First, we take a popular machine learning method---decision trees---and we connect it to economic theory. To do so, we provide a microeconomic framework for the interpretation of decision trees. In particular, we show that decision trees correspond to a non-compensatory, microeconomic decision protocol known as ``disjunctions-of-conjunctions'' \citep{hauser_2010_disjunctions}. Using this perspective, we explain how many of the varieties of decision trees address and can be motivated by microeconomic considerations such as analyst uncertainty or heterogeneity in one's non-compensatory behaviors. Additionally, our economic viewpoint suggests new additions to the existing body of decision tree techniques---additions that should lead to not only richer econometric models, but to more accurate statistical models overall.

Second, by combining decision trees with traditional discrete choice models, we advance the state of the art in the modeling of semi-compensatory decision making. We discuss how decision trees allow us to more flexibly represent non-compensatory behaviors than previously possible. Moreover, we show that our two-stage, semi-compensatory model jointly models how non-compensatory decision protocols influence both choice set formation and preference heterogeneity\footnote{We are aware that, in the discrete choice literature, the term preference heterogeneity has been used ambiguously. In some cases, preference heterogeneity refers to differences in the general preference for an alternative, irrespective of attributes of the alternative \citep{bhat_1998_accommodating}. In other cases, preference heterogeneity is taken to also include the coefficients that are multiplied by an alternative's attributes when using a linear-in-parameters choice model specification \citep{kamakura_1996_modeling}. In still other cases, preference heterogeneity is taken to also include choice set heterogeneity \citep{vij_2014_preference}. In this paper, we use preference heterogeneity to include all of the coefficients in one's linear-in-parameters choice model specification. If one is using a non-linear or non-parametric choice model specification, we are also using preference heterogeneity to include differences in the systematic utility functions for different individuals.}.

Finally, our third contribution is an empirical demonstration of the aforementioned techniques to the choice of travel mode in the San Francisco Bay Area. We show that the semi-compensatory models fit the data better than traditional models based solely on utility-maximization, and we show that the semi-compensatory models lead to a number of policy implications that are not readily uncovered by traditional discrete choice models. Through this application, we illustrate the quantitative and qualitative benefits that can come from combining economic theory with machine learning and modern statistical methods.

Structurally, the rest of our paper is organized as follows. In Section \ref{sec:decision-tree-explanation}, we provide an econometrically accessible introduction to decision trees. Here, we focus on decision trees as a statistical tool. Next, Section \ref{sec:decision-tree-economics} describes the microeconomic theories of non-compensatory decision making that are related to decision trees, and it shows how decision trees algorithmically represent these concepts. Here, we focus on the ways that decision trees are motivated by particular decision making principles. In Section \ref{sec:literature-review}, we review how the aforementioned microeconomic concepts have been operationalized in the discrete choice literature so far, and we make note of how decision trees address the theoretical and practical difficulties with these previous implementations. Section \ref{sec:decision-tree-variants} then details the various types of decision trees, including combined decision-tree/discrete-choice models. Specifically, we orient our discussion around the ways these decision tree variants address economic considerations that might prevent choice modelers from using decision trees in their work. In Section \ref{sec:empirical-application}, we formulate a new decision-tree/discrete-choice model, and we apply the model to the choice of travel mode in the San Francisco Bay Area. We describe the data used for this application, and we discuss the greater fit and unique insights provided by our semi-compensatory model in comparison to models based purely on utility-maximization. Finally, Section \ref{sec:conclusion} concludes.

\section{Decision trees explained}
\label{sec:decision-tree-explanation}
In this section, we provide a brief description of decision trees, targeting econometricians as our main audience. We will first provide an explanation of what a decision tree is, and we will use a highly simplified example to demonstrate how they can be used. After this, we will give a brief description of some of the (many) ways that decision trees are estimated from data. Here, again, we will focus on comparing and contrasting these estimation methods with techniques that econometricians are familiar with.

\subsection{What are decision trees and how do we use them?}
\label{sec:what-are-dtrees}
In simple terms, decision trees are a set of ``if-then'' statements that are used to predict a given quantity\footnote{We realize that our definition of decision trees is broad. Our definition includes models such as regression trees, classification trees, decision lists, and decision tables \citep{rivest_1987_learning, loh_2011_classification}. For this paper's purposes, these models are similar enough to merit a joint description.} \citep{loh_2011_classification}. Etymologically, decision trees get their name because they are often represented graphically as a tree: an acyclic set of nodes connected by directed edges, with each node connected to at most one preceding node, beginning with a single ``root'' node that has no edges pointing into it, and terminating with a set of ``output'' nodes \citep{meila_2000_learning, rokach_2005_top}. Each path from the root node to an output node represents one of the ``if-then'' statements that make up the tree. These if-then statements must partition the space of explanatory variables into a set of mutually exclusive regions (corresponding to the output nodes) that span the entire space of explanatory variables \citep{lemon_2003_classification}. Then, when making predictions about a decision maker, the ``if'' condition is used to determine the region/output-node the decision maker is in, and the corresponding ``then'' statement is used to provide the desired prediction. In a discrete choice context, such predicted quantities might be (1) the probability that a particular alternative is considered or (2) the probability with which an alternative is chosen.

\begin{figure}
\centering
\includegraphics[width=0.5\textwidth]{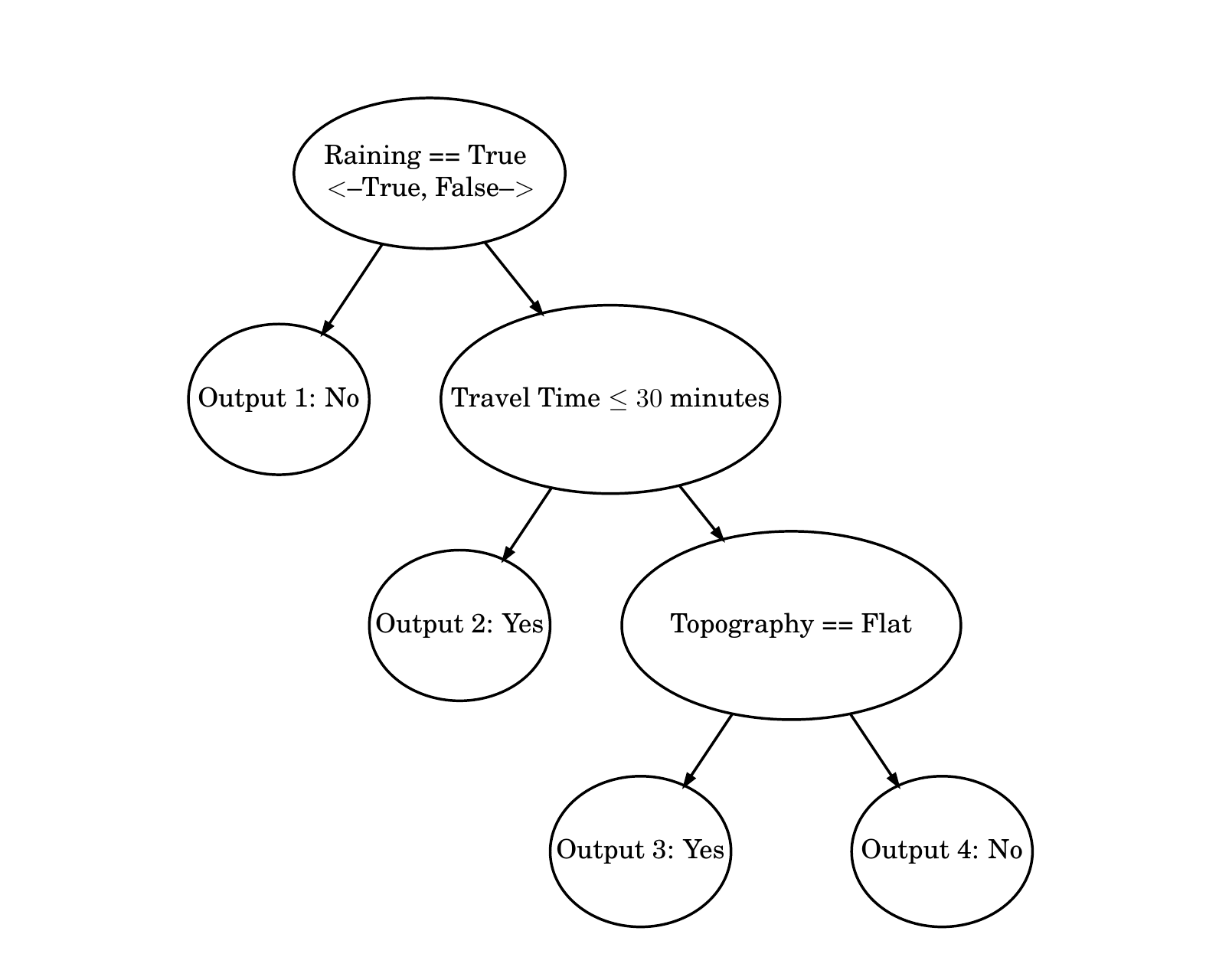}
\caption{Example decision tree for bicycle consideration}
\label{fig:tree-for-bicycle-consideration}
\end{figure}

To continue our explanation of what decision trees are and how they can be used, we will now provide a concrete, but highly stylized example of choice set generation, conditional on a given decision tree. In the discussion that follows, we realize that modelers may have many valid reservations about the realism of our example. It suffices to say that concerns about the deterministic nature the choice sets generated by our tree (shown in Figure \ref{fig:tree-for-bicycle-consideration}), concerns about the explicit discontinuities in the tree, and concerns about how such a tree could be estimated can all be addressed. Our example only features these qualities for simplicity of discussion. We note that in some contexts, deterministic choice sets are not uncommon: for example, when individuals are making residential location choices, some housing options may be deterministically excluded because the rents violate the individuals' income constraints \citep{kaplan_2012_development, zolfaghari_2013_simplified, bhat2015comprehensive}. Moreover, decision trees that probabilistically predict an individual's choice set can be estimated. These considerations will be discussed in Section \ref{sec:decision-tree-variants}. Concerns about the explicit discontinuities in our tree can be relaxed by considering individual heterogeneity in the split points of a tree or in the very structure of the tree being used. Like the issue of estimating trees that probabilistically predict an individual's choice set, concerns about individual heterogeneity are discussed in Section \ref{sec:decision-tree-variants}. Lastly, the estimation of decision trees will be discussed in Subsection \ref{sec:how-to-estimate-dtrees}.
 
Now, disclaimers aside, imagine that we are modeling the choice set formation behavior of travellers who are choosing the mode by which they will travel. Further, assume that our population of individuals has only two commuting alternatives: bicycle and public transit, and assume that public transit is always considered. Finally, Figure \ref{fig:tree-for-bicycle-consideration} shows the decision tree that represents the assumed choice set formation process in our hypothetical population. Here, \textit{Raining} is either \textit{True} or \textit{False}, \textit{Travel Time} is measured in minutes, \textit{Topography} is either \textit{Flat} or \textit{Hilly}, and the dependent variable (bicycle consideration) is either \textit{Yes} or \textit{No}. From the tree in Figure \ref{fig:tree-for-bicycle-consideration}, a number of useful observations can be made. First, there are four output nodes, two of which result in bicycle being considered and two that result in bicycle not being considered. Secondly, we see that bicycle consideration is a function of weather (raining or not), travel time, and topography. Now, to use the tree to make predictions for a given individual, one must traverse the tree from top to bottom, ending at one of the tree's output nodes. The rules for traversing the given\footnote{We note in passing that the traversal rules may change from tree to tree, based on author preference, but they should always be explicitly stated.} decision tree are that if the condition in a decision node (i.e. a non-output node) is True, then one goes to the left and if the condition is False, one goes to the right.

So, what can one use the tree in Figure \ref{fig:tree-for-bicycle-consideration} for? First, the tree and its predictions can be directly used to inform policies. For instance, a municipality trying to increase bicycle usage must first ensure that bicycle is considered as a mode of travel. Based on this example's tree, the municipality might subsidize the relocation costs for individuals that wish to move to a location that is 30 minutes away or closer to their workplace. Such subsidies would help push bicycle into the choice sets of individuals, thereby increasing the expected number of bicycle commuters. Secondly, the tree in Figure \ref{fig:tree-for-bicycle-consideration} might be used as part of a larger model building effort. For instance, one might use the tree in Figure \ref{fig:tree-for-bicycle-consideration} to inform a two-stage model of travel mode choice. At the first stage, an individual's choice set is modeled. By assuming that individuals must travel to work and that public transit is always considered, our example is left with two possible choice sets: \{Public Transit\} and \{Public Transit, Bicycle\}. The choice sets in this example are based on whether bicycle is considered or not, and the probabilities of these choice sets (i.e. the first stage in Manski's two-stage models) can be written as follows:
\begin{equation}
\begin{aligned}
P \left( C = \left\lbrace \textrm{Public Transit, Bicycle} \right\rbrace \mid x, \textrm{tree} \right) &= P \left( \textrm{Bicycle considered} \mid x, \textrm{tree} \right)\\
&= \sum _r P \left[ \textrm{Bicycle considered} \mid T \left( x \right) = r \right] P \left[ T \left( x \right) = r \right]\\
\textrm{where } C &= \textrm{An individual's choice set.}\\
r &\in \left\lbrace 1, 2, 3, 4 \right\rbrace\\
r &= \textrm{A specific region demarcated by the decision tree.}\\
T \left( x \right) &= \textrm{The region an individual belongs in based on $x$ and the tree.}\\
x &= \textrm{Explanatory variables for an individual.}
\end{aligned}
\end{equation}

\begin{figure}
\centering
\includegraphics[width=0.75\textwidth]{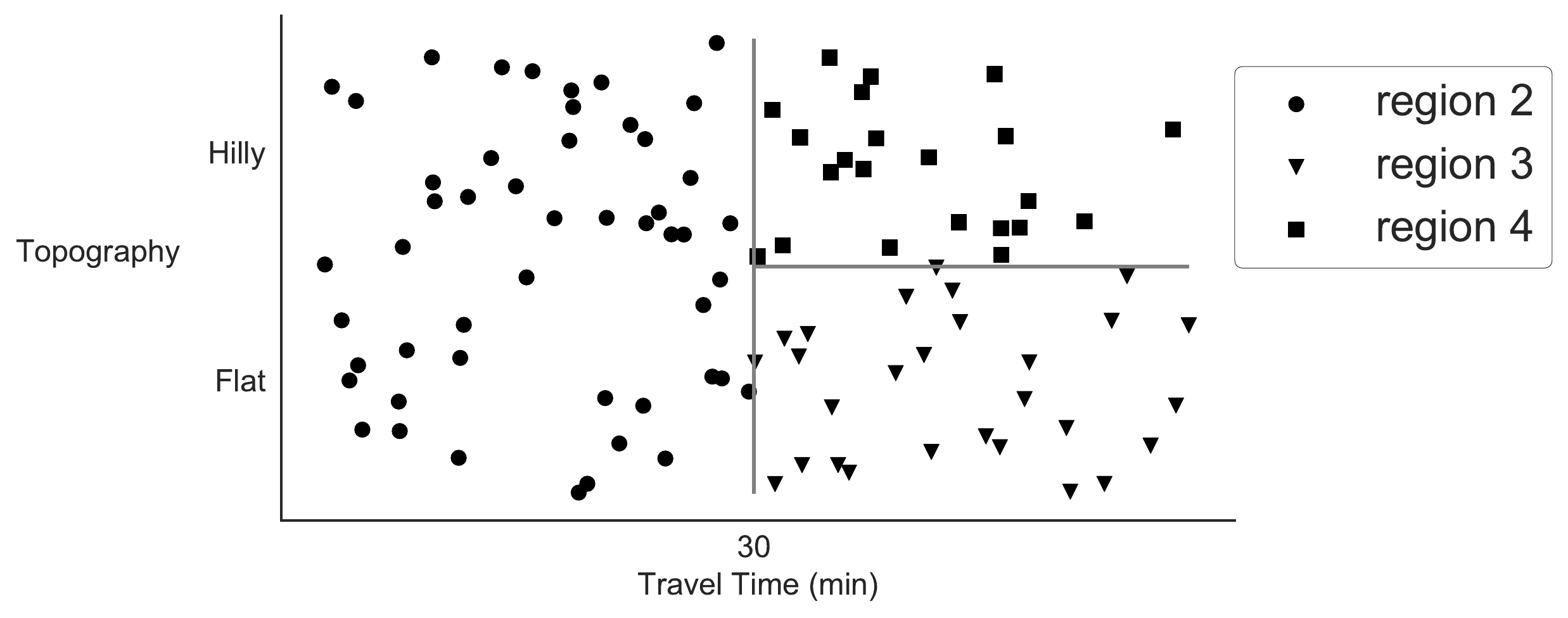}
\caption{Regions 2-4 of example decision tree for bicycle consideration}
\label{fig:regions-of-tree}
\end{figure}

For most decision trees, $T \left( x \right)$ is a deterministic function\footnote{The primary exception to this is a ``probabilistic'' decision tree, also known as a ``soft'' or ``fuzzy'' decision tree, where $T \left (x \right)$ is a probabilistic function. These decision tree variants will be discussed in Section \ref{sec:decision-tree-variants}. The other exception is where the case of measurement error where the value $x$ is unknown and modeled with a probability distribution of its own.} such that, given explanatory variables $x$, an observation is deterministically assigned to a given region/output-node $r$. For our example, regions 2-4 are graphically depicted in Figure \ref{fig:regions-of-tree}. Because our $T \left( x \right)$ is deterministic, $P \left[ T \left( x \right) = r \right]$ is either 1 or 0, and the same is true of the probability of bicycle consideration, conditional on being in a given region. In all cases, we can expand $P \left[ T \left( x \right) = r \right]$ to more explicitly show how each explanatory variable contributes to the likelihood of an observation being in a given region.

Specifically, we note that each ``if'' statement in the decision tree can be written as the union of elementary conditions, typically\footnote{Exceptions to this statement come from decision trees that are not ``axis-aligned,'' such as oblique decision trees that use inequalities with linear combinations of variables for their ``if'' conditions \citep{murthy_1994_system, ittner_1996_non}.} with one such elementary condition per explanatory variable. For instance, let $x_1$ denote whether it is raining, let $x_2$ denote the bicycle travel time between an individual's home and work, and let $x_3$ denote the topography between an individual's home and work. Additionally, let $S_{rk}$ denote the set that variable $x_k$ must be in for an individual to belong to region $r$. Using these variables, we can write the region corresponding to the first output node as $S_{11} = \left\lbrace \textrm{True} \right\rbrace$, $S_{12} = \left[ 0, \infty \right)$, and $S_{13} = \left\lbrace \textrm{Flat, Hilly} \right\rbrace$. These sets reflect the fact that output node 1 is the region of the variable space where \textit{Raining} is True and where any values of \textit{Travel Time} or \textit{Topography} are valid. With this notation, we can express the probability of bicycle consideration as follows:

\begin{equation}
\label{eq:consideration-likelihood}
\begin{aligned}
P \left( \textrm{Bicycle considered} \mid x, \textrm{tree} \right) &= \sum _r P \left[ \textrm{Bicycle considered} \mid T \left( x \right) = r \right] P \left[ T \left( x \right) = r \right]\\
&= \sum _r P \left[ \textrm{Bicycle considered} \mid T \left( x \right) = r \right] P \left[ \bigcap _k x_k \in S_{rk} \right]\\
&= \sum _r \left\lbrace P \left[ \textrm{Bicycle considered} \mid T \left( x \right) = r \right] \prod _k P \left[ x_k \in S_{rk} \right] \right\rbrace
\end{aligned}
\end{equation}

The equation above shows how, conditional on a given decision tree, one can form the sorts of probability statements that are common in the first stage of two-stage choice models with non-compensatory rules for choice set formation \citep{gilbride_2004_choice, cantillo2006discrete}. Moreover, if one's decision tree was being used to directly predict the probability of a given alternative, one's likelihood function would be formed analogously. Besides being transparent about how the structure of the tree translates to one's likelihood equations, Equation \ref{eq:consideration-likelihood} highlights the link to the non-compensatory decision protocol known as disjunctions-of-conjunctions \citep{hauser_2010_disjunctions}. Though we will delay a detailed discussion of this protocol to Section \ref{sec:decision-tree-economics}, we point out here that logical disjunctions are algebraically represented as summations and logical conjunctions are algebraically represented as products \citep{gilbride_2004_choice}. Equation \ref{eq:consideration-likelihood} shows that when modeling bicycle consideration with a decision tree, our probabilities of interest are explicitly given as a summations of products (i.e as disjunctions-of-conjunctions). Importantly, such a decision protocol generalizes the typical conjunctive or disjunctive rules that are used in choice models that represent non-compensatory processes. See Section \ref{sec:decision-tree-economics} for further discussion and explanation of this point.

\subsection{How do we estimate decision trees?}
\label{sec:how-to-estimate-dtrees}
In the previous subsection, we explained what decision trees are and (conditional on a specific decision tree) what one can do with them. In this subsection, we turn to the question of how such decision trees are estimated from data and how such estimation techniques differ from those commonly employed in the discrete choice literature.

To begin, discrete choice modelers are most likely to be familiar with estimation techniques such as maximum likelihood, method of moments, and bayesian Markov Chain Monte Carlo (MCMC) methods \citep{train_2009_discrete}. Of these techniques, only bayesian MCMC methods have been applied to the estimation of decision trees \citep{chipman_1998_bayesian, denison_1998_bayesian, letham_2015_interpretable, pratola_2016_efficient}. We believe that the main reason for this discrepancy in estimation methods is that decision trees are not continuous functions. Instead, they are explicitly discontinuous functions of the explanatory variables (e.g. at a particular node, should we split on \textit{Travel Time} or \textit{Travel Cost}?). Maximum likelihood, if it is to be performed at all can no longer rely on gradients and hessians, so enumeration and comparison of all decision trees is necessary. However, enumeration of all possible decision trees is NP-hard \citep{ruggieri_2017_enumerating}. Since it is computationally prohibitive to enumerate all possible decision trees and assess their log-likelihoods, maximum likelihood estimation of decision trees is typically viewed as infeasible. Similarly, since the method of moments and its generalizations require continuous moment functions \citep{hansen_1982_large}, these estimation techniques cannot be used to estimate decision trees.

As highlighted in the last paragraph, estimation of decision trees is severely hindered by the discontinuous nature of the trees and the fact that explicit enumeration of all possible trees is computationally prohibitive. Due to these challenges, most estimation techniques (both bayesian and frequentist) use approximations and heuristics. By far, the most common frequentist heuristic is to use a greedy algorithm to estimate the tree \citep{rokach_2005_top}. Here, one recursively performs a search over all variables and values of those variables to pick the variable and value combination that best meets some ``splitting criteria.'' After finding the best variable and value pair, the dataset is split according to the chosen pair. The process is then repeated for each subset of the data: those meeting the chosen condition and those not meeting the condition. The greedy estimation of the decision tree will terminate once some stopping criteria is met (e.g. no output node should contain less than 5 observations). After the initial estimation of the decision tree, some estimation methods ``prune'' the initial tree by removing nodes according to a ``pruning criterion'' \citep{mingers_1989_empirical, esposito_1997_comparative}. Differing methods and criteria for splitting, stopping, and pruning all lead to different types of decision trees \citep{loh_2014_fifty, rokach_2014_data}. Moreover, besides the greedy approach just described, there exist a number of other frequentist tree estimation techniques such as using genetic algorithms \citep{barros_2012_survey} or branch-and-bound algorithms \citep{angelino_2017_learning}. Though we cannot perform an exhaustive review of the various decision tree estimation techniques, good surveys of this material can be found in \citet{murthy_1998_automatic, rokach_2005_top, barros_2012_survey}, and \citet{lomax_2013_survey}.

For the bayesian estimation of decision trees, a prior is placed over the space of possible decision trees, the likelihood is formed using equations similar to Equation \ref{eq:consideration-likelihood}, and then an MCMC algorithm is used to sample from the posterior distribution of possible decision trees \citep{chipman_1998_bayesian, denison_1998_bayesian, letham_2015_interpretable, pratola_2016_efficient}. At first glance, this seems exactly the same as what is always done in a bayesian estimation. However, since the set of all possible decision trees is huge and discrete, the MCMC algorithms do not typically ``explore'' the entire posterior distribution of trees \citep{chipman_1998_bayesian}. The approximation is that the MCMC methods typically only explore part of the posterior since these algorithms are limited by however much time an analyst has to let the algorithm run. If the MCMC algorithm is run for long enough, the hope is that ``high accuracy'' sections of the posterior are explored, such that one samples from the trees that are most predictive of the choices in one's dataset. Note, unlike the frequentist estimation methods where trees are defined based on how they are estimated, differing priors or differing MCMC methods lead to differences in how the space of decision trees is explored, but it is uncommon to speak of ``different'' bayesian decision trees. Such differentiation is likely unnecessary because, given an impractically long time, all bayesian MCMC techniques will explore the entire posterior of trees.

Finally, we pause to make a few passing remarks about the properties of the various estimators for decision trees. In standard discrete choice modeling, much importance is placed on having consistent and efficient estimators. The greedy estimation techniques described above for decision trees have long been proven to be consistent, non-parametric estimators of underlying data-generating processes \citep{gordon_1980_consistent, gordon_1984_almost, toth_2011_building}. Bayesian techniques have also demonstrated their consistency in simulation \citep{letham_2015_interpretable}, though formal proofs are still missing. In terms of efficiency, however, it is not clear that this notion is meaningful for decision tree models. In particular, the notion of an ``efficient'' estimator being one that achieves the Cramer-Rao lower bound is no longer meaningful since the parameter space (the number of splits in the tree, variables and values being split on, and the tree structure) is discrete and increases with the size of one's data set (i.e. it is not fixed). If one views efficiency as being inversely related to the variance of one's estimator, then it is known that estimation techniques that generate a large number of candidate trees and then select the best one tend to be less variable than the greedy methods described above \citep{tibshirani_1999_model}. Nevertheless, whether or not other variations on the notion of efficiency can be shown to apply to decision trees is beyond the scope of this paper and will not be investigated.

\section{Decision trees: The link with microeconomics}
\label{sec:decision-tree-economics}
In Section \ref{sec:decision-tree-explanation}, we described what decision trees are, how a given decision tree can be used, and how decision trees might be estimated. Additionally, in both Sections \ref{sec:introduction} and \ref{sec:decision-tree-explanation}, we noted that decision trees correspond to a non-compensatory decision protocol known as disjunctions-of-conjunctions \citep{hauser_2010_disjunctions}. In this section we will review this microeconomic interpretation of decision trees in detail. Initially, we will briefly describe standard discrete choice models and their use of compensatory decision protocols. Then we will motivate the need for non-compensatory decision protocols, and in Subsection \ref{sec:noncompensatory-description}, we will proceed to describe a number of such behavioral strategies. We will begin with simple non-compensatory protocols and proceed to describe further generalizations of such strategies until we arrive at disjunctions-of-conjunctions: a focal point of our paper. Finally, in Subsection \ref{sec:dtrees-and-disjunctions-of-conjunctions}, we will mathematically show how decision trees represent disjunctions-of-conjunctions.

To start, we note that compensatory decision protocols are decision making strategies where, for a given alternative, ``high levels of satisfaction with one attribute compensate for low levels of satisfaction with [other]'' attributes \citep{foerster1979mode}. As readers are likely aware, almost all discrete choice models used in practice and research are based on compensatory decision processes, with utility-maximization being the most common example\footnote{We are aware of the increasing number of discrete choice models that are being estimated under the assumption of regret-minimizing behavior. However, such models are still compensatory in nature, and therefore retain many of the properties we describe in the context of utility-maximization.} \citep{swait_2001_non, truong_2015_modeling}. However, counter to prevailing practices, behavioral economists and psychologists have presented much evidence that individuals frequently depart from standard notions of utility maximization and rationality \citep{foerster1979mode, bronner_1982_decision, tversky_1986_rational, conlisk_1996_why}. Spurred by these observations, a steady but small stream of research has both called for and proposed new models of human decision making that explicitly incorporates the possibility of non-utility maximizing choice behavior \citep{simon_1955_behavioral, tversky_1972_elimination, gigerenzer_1996_reasoning, leong_2012_embedding}. Such alternative methods of decision making are typically referred to as non-compensatory decision rules or non-compensatory decision protocols. They are called non-compensatory because they do not always allow positive attributes of a given alternative to compensate for negative attributes of that same alternative. Additionally, since non-compensatory decision rules do not typically require the evaluation of all attributes of all alternatives, they better capture the limited cognitive resources of decision makers \citep{simon_1955_behavioral, young_1984_non, swait_2001_non} and are therefore thought to be more behaviorally realistic.

\subsection{Non-compensatory decision rules}
\label{sec:noncompensatory-description}
Thus far, some of the non-compensatory decision processes that have been detailed in the discrete choice literature include: dominance \citep{cascetta2009dominance}, lexicography \citep{kohli2007representation}, elimination-by-aspects \citep{tversky_1972_elimination}, satisficing \citep{stuttgen2012satisficing}, conjunctive rules, disjunctive rules, subset-conjunctive rules, and disjunctions-of-conjunctions. Of these, conjunctive and disjunctive rules are quite prevalent in the literature, and all of the last four non-compensatory rules are related to decision trees. We therefore describe the last four non-compensatory decision protocols below, and in Section \ref{sec:literature-review}, we review how these four protocols have been previously incorporated into discrete choice models.

\begin{description}
\item [Conjunctive Rules \citep{coombs_1951_mathematical, dawes_1964_social}] \hfill \newline
Using a conjunctive decision rule, an individual only considers alternatives that meet all of a given number of requirements. For instance, an individual making a residential location choice may only consider housing options that meet his or her requirements on the maximum amount of rent \textbf{and} the distance from the individual's workplace location. The ``and'' statement is what distinguishes this decision rule as conjunctive. As noted in Subsection \ref{sec:what-are-dtrees}, conjunctive statements are algebraically represented using products.

\item [Disjunctive Rules \citep{coombs_1951_mathematical, dawes_1964_social}] \hfill \newline
Using a disjunctive decision rule, individuals only consider alternatives that meet at least one of a given set of requirements. For instance, continuing with the residential choice example, an individual may only consider housing options that are within a given distance from their workplace location \textbf{or} that are within a given distance from major public parks. The ``or'' statement is what distinguishes this decision rule as disjunctive. As noted in Subsection \ref{sec:what-are-dtrees}, disjunctive statements are algebraically represented using sums.

\item [Subset-Conjunctive Rules \citep{jedidi_2005_probabilistic}] \hfill \newline
Subset-conjunctive rules are a generalization of both conjunctive rules and disjunctive rules. Using a subset-conjunctive decision rule, an individual only considers alternatives that meet a certain number of requirements. Using another residential location choice example, consider an individual who would like to live within one mile of a major public park, who would like to live within two miles of his or her workplace, who would like to pay less than \$1,000 per month in rent (but is flexible), and who would like to live within one mile of a subway station. Under a subset-conjunctive rule, this individual would consider any housing units that meet some number of these four requirements. For instance, this individual might consider any housing units that meet at least three of these four requirements. Note that if this individual only considered housing units that met all four requirements, then this would be equivalent to a conjunctive decision rule with four requirements. Likewise, if this individual only required housing units to meet one of the four requirements, then this would be equivalent to a disjunctive decision rule. Algebraically, subset-conjunctive rules are therefore sums of products, with the restriction that each product term have a given number elements (one for each requirement that should be met).

\item [Disjunctions-of-Conjunctions \citep{hauser_2010_disjunctions}] \hfill \newline
Disjunctions-of-conjunctions generalize the conjunctive, disjunctive, and subset-conjunctive decision rules. Under a disjunctions-of-conjunctions decision protocol, an individual will consider any alternative that meets at least one of a given set of conjunctive conditions. Each condition may differ in the number of requirements that compose the conjunction. Algebraically, then, disjunctions-of-conjunctions are expressed as sums of products with no constraints on the number of elements in each product.

Consider once more the residential choice example. If, for instance, our decision maker was more concerned about rent than the other requirements, he or she might consider any housing unit that required less than \$1,000 per month in rent and that met one of the remaining three requirements. Additionally, he or she might consider any housing unit that was simultaneously within one mile of a major park, within one mile of a subway station, and within two miles of his or her workplace. In this case, only one of the following four conjunctive conditions needs to be met in order for a housing unit to be considered:
\begin{itemize}
\item rent less than \$1,000 per month and housing unit within one mile of a major public park

\item rent less than \$1,000 per month and housing unit within two miles of the individual's workplace

\item rent less than \$1,000 per month and housing unit within one mile of a subway station

\item housing unit within one mile of a major public park and within one mile of a subway station and within two miles of the individual's workplace.
\end{itemize}

As can be seen from the example above, if the individual had only one condition for consideration, we would have a conjunctive rule. If the individual had only one requirement in each of the four conditions above, then we would have a disjunctive rule. Similarly, if we expanded the first three conditions above so that they each included a third requirement, we would once again have the subset-conjunctive rule whereby any housing unit with three of the four requirements would be considered.
\end{description}

Before moving on to Subsection \ref{sec:dtrees-and-disjunctions-of-conjunctions}, we pause to briefly summarize why we believe the link between disjunctions-of-conjunctions and decision trees is important. First, as noted above, conjunctive rules and disjunctive rules are seen as important information processing strategies, and they have been applied in many choice modeling efforts \citep{foerster1979mode, swait_2001_non, gilbride_2004_choice, elrod_2004_new, martinez_2009_constrained, hauser_2010_disjunctions, hess2012allowing, kaplan_2012_development, zolfaghari_2013_simplified, truong_2015_modeling}. Being a generalization of these two rules, disjunctions-of-conjunctions may also be an important decision making strategy, but it has seldom been tested in choice modeling contexts. We think a major reason for this lack of choice modeling application is because there have not been easy or straightforward ways to estimate such rules. Linking disjunctions-of-conjunctions to decision trees gives researchers a way to estimate disjunctions-of-conjunctions by drawing upon well established methods of estimating decision trees. Additionally, once disjunctions-of-conjunctions can be estimated by themselves, it is then possible to estimate such strategies in combination with the compensatory procedures used in standard discrete choice models. We pursue this strategy later, in Section \ref{sec:decision-tree-variants} and Section \ref{sec:empirical-application}.

\subsection{Linking decision trees with disjunctions-of-conjunctions}
\label{sec:dtrees-and-disjunctions-of-conjunctions}
As described in the previous subsection, disjunctions-of-conjunctions are highly flexible non-compensatory decision protocols. Here, we highlight how decision trees mathematically represent the relationships implied by disjunctions-of-conjunctions.

First, we define the necessary notation. Let $b$ represent a primitive boolean statement, i.e. a specific requirement. Such a statement is an equality or inequality that is not composed of any other equalities or inequalities. For instance, $x == 2$ and $x \leq 5$ are primitive boolean statements but ($\left( x_1 == 2 \right) * \left( x_2 \leq 5 \right)$) is not a primitive boolean statement because it is composed of two boolean statements. Additionally, if $b$ is True, then we say that $b = 1$, and if $b$ is False, then we say that $b = 0$.

With this notation, conjunctive rules can be expressed as:
\begin{equation}
\begin{aligned}
\textrm{if } \left( \prod _{i = 1} ^{B} b_i \right) &== 1 \Rightarrow y\\
\textrm{where } B &= \textrm{the total number of requirements in the rule.}\\
\textrm{``$\Rightarrow$''} &= \textrm{``then'' or ``implies''.}\\
y &= \textrm{some outcome.}
\end{aligned}
\end{equation}
In words, this is read as ``if all requirements, $b_i$, are met, then $y$''. This follows because each $b_i$ must be True (i.e. must be met) in order for that $b_i$ to equal 1, and we need all $b_i$ to equal 1 in order for $\prod _{i = 1} b_i$ to evaluate to 1.

Similarly, a disjunctive rule can be expressed as:
\begin{equation}
\textrm{if } \left( \sum _{i = 1} ^{B} b_i \right) \geq 1 \Rightarrow y
\end{equation}
In words, this is read as ``if at least one (i.e. if any) of the requirements $b_i$ are met, then $y$''. This follows because any requirement $b_i$ that is not met will cause that $b_i$ to evaluate to 0. If at least one requirement is met, then the corresponding $b_i$'s will evaluate to 1, and then $\sum _{i = 1} b_i$ will be greater than or equal to 1.

With these building blocks, we turn immediately to the case of disjunctions-of-conjunctions\footnote{Subset-conjunctive rules will be expressed as a special case of the formula for disjunctions-of-conjunctions.}. In words, the use of disjunctions-of-conjunctions requires statements such as ``if at least one of some set of conjunctive conditions is met, then $y$.'' To mathematically express such a statement, we will introduce additional symbols. The first symbol, $p$, will represent conjunctive conditions, i.e. products of primitive boolean statements. As noted in Subsection \ref{sec:noncompensatory-description}, in disjunctions-of-conjunctions, the various conjunctive conditions need not have the same number of requirements. To account for this, we will index the various conjunctive conditions by $i$, and we will use $\mid p_i \mid$ to denote the number of requirements that make up $p_i$. Finally, we will use the symbol, $b_j ^i$, to indicate the $j$'th primitive boolean statement (i.e. the $j$'th requirement) in conjunctive statement $p_i$. With this additional notation, our disjunctions-of-conjunctions statement can now be expressed as:
\begin{equation}
\begin{aligned}
\textrm{if } \left( \sum _{i = 1} ^{D} p_i \right) &\geq 1 \Rightarrow y\\
\textrm{if } \left( \sum _{i = 1} ^{D} \prod _{j = 1} ^{\mid p_i \mid} b_j ^i \right) &\geq 1 \Rightarrow y\\
\textrm{where } D &= \textrm{the total number of conjunctive conditions.}
\end{aligned}
\end{equation}
From the first line, we mathematically see the disjunction (i.e. the summation) of conjunctive conditions. The second line shows the conjunction (i.e. the product) of requirements. Now, for subset-conjunctive rules, we merely impost the constraint that $\mid p_i \mid$ be equal to some constant value for all $p_i$. This is equivalent to saying that each conjunctive condition must be comprised of the same number of requirements.

To go from the abstract equations above to a decision tree, we must consider what a conjunctive condition represents. In general, a conjunctive condition defines a region in a space. Using Figure \ref{fig:tree-for-bicycle-consideration} as an example once more, consider the space formed by the variables $x_1 = \textrm{Rain}$ and $x_2 = \textrm{Travel Time}$. The conjunctive condition that leads to output node 2 is $x_1 == \textrm{False}$ AND $x_2 \leq 30$. This condition will define a rectangular region in the graph of $\left( x_1, x_2 \right)$ comprised of the area where $x_1$ is False, and the area where $x_2$ is less than 30. For more examples of regions formed by conjunctive conditions, see Figure \ref{fig:regions-of-tree} above. Now, when we have multiple conjunctive conditions, we have multiple regions in space. These regions will either be mutually exclusive, or they will overlap. It is crucial to note that any region defined by a set of overlapping conjunctive criteria can be expressed as a region defined by a set of mutually exclusive criteria. For instance, let $\left\lbrace p \right\rbrace = \left\lbrace p_1, p_2 \right\rbrace$ be a region defined by a set of overlapping conjunctive conditions, $p_1$ and $p_2$. This region can be re-expressed as a set of mutually exclusive conjunctive conditions, $\left\lbrace \tilde{p} \right\rbrace = \left\lbrace \tilde{p}_1, \tilde{p}_2 \right\rbrace$. One such re-expression is $\tilde{p}_1 = p_1$ and $\tilde{p}_2 = p_2 * p_{1}'$, where $\tilde{p}_2$ is read as ``$\tilde{p}_2$ equals $p_2$ AND NOT $p_1$.'' Observations meeting the condition $\tilde{p}_2$ will therefore satisfy all the requirements of $p_2$, but they will not satisfy all the requirements of $p_1$.

With the possibility of re-expression in mind, recall that using disjunctions-of-conjunctions means making statements of the form ``if at least one of some set of conjunctive conditions, $\left\lbrace p \right\rbrace$, is met, then $y$''. As just noted, this statement can be reformulated as, ``if at least one of some set of conjunctive conditions, $\left\lbrace \tilde{p} \right\rbrace$, is met, then $y$''. Given the mutually exclusive conjunctive conditions of $\left\lbrace \tilde{p} \right\rbrace$, our reformulation can be expressed as a decision tree where each conjunctive condition in $\tilde{p}$ becomes an ``if'' statement in the tree with a corresponding ``then $y$'' statement. Note we will also need a final condition such as ``if $\bigcap _{i=1} ^{D} \tilde{p}_{i}'$ then $y'$,'' where $y' \neq y$. Here, the final condition ensures that the decision tree is comprised of a set of conditions that are both mutually exclusive and exhaustive. The condition $\bigcap _{i=1} ^{D} \tilde{p}_{i}'$ is read as ``NOT $\tilde{p}_1$ and NOT $\tilde{p}_2$ and ... and NOT $\tilde{p}_D$.'' Finally, we use $y'$ as the outcome for the remaining conditions that are added to ensure exhaustiveness, e.g. $\bigcap _{i=1} ^{D} \tilde{p}_{i}'$, simply because we assume that if there was any other condition that would result in $y$, then that condition would have been part of the original set of conditions, $\left\lbrace p \right\rbrace$.

\section{A review of how non-compensatory protocols have been incorporated in discrete choice}
\label{sec:literature-review}
In Section \ref{sec:decision-tree-economics}, we described conjunctive rules, disjunctive rules, subset-conjunctive rules, and disjunctions-of-conjunctions. However, researchers have gone beyond mere descriptions. These decision protocols have been incorporated into choice models and used to quantitatively study the concordance of non-compensatory processes with observed choices. In this section, we will review the ways that conjunctive rules, disjunctive rules, and their generalizations have been previously incorporated into discrete choice models. Afterwards, we will highlight drawbacks of the previous work that our paper seeks to address. Having said this, we state upfront that our review mainly focuses on the way that non-compensatory protocols have been used to model choice set generation as opposed to modeling the actual choice being made. The reason for our focus is that conjunctive rules, disjunctive rules, and their generalizations are (in general) not sufficient to uniquely choose a particular alternative. Multiple alternatives may meet an individual's non-compensatory rules, but (in our context) a decision strategy must still be employed to generate a single discrete choice. As a result, conjunctive rules, disjunctive rules, and their generalizations have almost exclusively been used in the discrete choice literature to winnow a decision maker's choice set before another strategy is used (if necessary) to make the final choice. In Subsection \ref{sec:choice-set-generation-review}, we review this approach of choice set generation followed by compensatory choice amongst the considered alternatives, and we revisit this notion in Section \ref{sec:dtree-variants-preferences} when we describe the decision tree variant known as ``model trees.'' In Subsection \ref{sec:direct-modeling-review}, we will briefly review the few ways that observed choices have been directly\footnote{I.e., without estimating any rules or parameters that implicitly or explicitly determine one's choice set.} modeled with conjunctive rules, disjunctive rules, and their generalizations. 

\subsection{Choice-set generation via non-compensatory protocols}
\label{sec:choice-set-generation-review}
Across the literature, two main approaches have been used to incorporate conjunctive, disjunctive, and related protocols into discrete choice models. These two approaches differ primarily based on whether they explicitly model an individual's decision making using two-stages as prescribed by \citet{manski_1977_structure} or whether they use a single-stage model that implicitly performs choice-set generation. We will begin by first describing the single-stage models, also known as the ``reduced-form'' approach \citep{swait_2001_non}.

Pioneered by Swait (\citeyear{swait_2001_non}), single-stage models implement conjunctive and/or disjunctive rules by altering the systematic utility of an alternative. When representing strict non-compensatory behaviors, these models combine attribute values and attribute thresholds to set the systematic utility of an alternative to -/+ infinity, effectively removing an alternative from one's choice set or removing all other alternatives from one's choice set. Through the years, multiple single-stage models have been proposed, each with their own set of unique additions. \citet{swait_2001_non} allowed for non-strict non-compensatory behavior where violation of an attribute threshold was allowed but resulted in penalties to one's systematic utility. \citet{elrod_2004_new} estimated the attribute thresholds from choice data only, whereas \citet{swait_2001_non} required individuals to report their attribute thresholds. Moreover, \citet{elrod_2004_new} did not allow violation of one's attribute threshold and even penalized or rewarded the systematic utility when the value of an attribute approached that attribute's threshold, based on whether a conjunctive or disjunctive rule was being implemented. When allowing violation of one's attribute thresholds, \citet{martinez_2009_constrained} used non-linear penalty functions in contrast to the linear penalty functions of \citet{swait_2001_non}. Most recently, \citet{truong_2015_modeling} proposed a novel way to estimate the attribute thresholds in the context of Swait's original (\citeyear{swait_2001_non}) formulation. Common to all these implementations, however, is the fact that conjunctive or disjunctive behavior was operationalized through the systematic utility function.

The second approach used in the literature to represent conjunctive, disjunctive, and similar behaviors is the two-stage approach where one formally models the choice set generation process. To date, the vast majority of such two-stage models have relied on the Probabilistic Independent Availability Logit (PIAL) model \citep{swait_1984_probabilistic, swait_2009_choice}. Here, the two-stage models use non-compensatory decision rules to determine whether each alternative will be present in an individual's choice set. The randomness underlying the probability that an alternative is in one's choice set is explained as coming from analyst uncertainty over the attribute thresholds used by each individual to evaluate the non-compensatory rules. Moreover, the probability of an alternative being in one's choice set is considered to be independent of the probability that any other alternative is in one's choice set, hence the name PIAL. Despite this independence assumption, PIAL models still suffer from the curse of dimensionality since they typically require one to enumerate all possible subsets of one's universal choice set. As a result, important differences can be seen in the way that various authors have dealt with this computational hardship. Some authors have used simulation techniques to avoid full enumeration of the various consideration sets, other authors have made no attempts at avoiding computational difficulties in estimating PIAL models, and still other authors have tried to minimize the number of possible consideration sets by collecting explicit consideration set information from decision makers. Our review below will be structured around these modeling differences.

To the best of our knowledge, the first paper to incorporate conjunctive and disjunctive rules into a two-stage model was the \citeyear{gilbride_2004_choice} paper of Gilbride and Allenby. As described above, these authors parametrize the probability of an alternative being available as the probability of an alternative satisfying the conjunctive or disjunctive rules that are made up by the (unobserved) attribute thresholds for each attribute. To sidestep the computationally prohibitive step of enumerating each possible consideration set, Gilbride and Allenby use a bayesian estimation method. In particular, the authors use a MCMC sampling method to explore the space of possible thresholds, and each set of sampled thresholds induces a particular choice set that can be used in the second-stage choice process. While apparently successful in dealing with the curse of dimensionality, most models after \citet{gilbride_2004_choice} take a different (i.e. a frequentist) approach.

For an example of this frequentist approach, we can look at the second paper on this topic, by \citet{cantillo_2005_semi}. These authors estimate a frequentist version of the Gilbride and Allenby model, using standard maximum likelihood estimation as opposed to a simulation-based optimization method. As a result, these authors are forced to enumerate all possible consideration sets, thereby incurring all estimation difficulties from the curse of dimensionality. On a positive note, however, Cantillo and Ort{\'u}zar are able to parameterize the attribute thresholds as a function of socioeconomic variables and choice conditions (e.g. trip purpose, time restrictions, etc.). This allows them to give greater behavioral interpretation to the estimated thresholds. Shortly thereafter, \citet{jedidi_2005_probabilistic} use a PIAL model where they allow for subset-conjunctive rules and for individual heterogeneity through the use of latent classes. To accommodate uncertainty in the number of requirements that need to be satisfied, Jedidi and Kohli estimate this parameter as well. Their approach amounts to full enumeration of all possible choice sets under each possible set of criteria and each possible number of requirements. Later, \citet{swait_2009_choice} returns to the issue of choice set generation with a two-stage choice model called a k-Mix model. This model is a PIAL model at its core, albeit with a couple of important differences. First, favorable conjunctive or disjunctive rules can be used to not only allow for consideration of alternatives but to place them in a ``dominance'' state wherein alternatives are preferred to all other alternatives that are not in a dominant state. Secondly, unfavorable non-compensatory rules can be used to place alternatives in a ``rejection'' state where alternatives are completely disregarded unless all other alternatives are also placed into the ``rejection'' state.

Finally, some authors have tried to retain a frequentist modeling framework while avoiding the curse of dimensionality that often plagues PIAL models. The approach taken by these authors has been to elicit information from individual decision makers that allows the analyst to specify the decision maker's choice set exactly. The underlying assumption that is made by these authors is that all alternatives that meet the conjunctive or disjunctive criteria are deemed to be in an individual's consideration set. Given this assumption, the observation of the exact thresholds used by an individual permits one to specify an individual's consideration set with certainty. Prominent examples of models estimated in this vein include the series of papers by Kaplan et al. (\citeyear{kaplan_2009_two, kaplan_2012_closing, kaplan_2012_development}). In addition to making use of the observed thresholds, Kaplan et al. model the choice of threshold, thereby allowing the model to be used for prediction with observations for whom thresholds have not been elicited. Another model that is estimated according to this approach is the model of \citet{zolfaghari_2013_simplified}. Though similar to the Kaplan et al. models, Zolfaghari et al. allow for the possibility that individuals do not make use of all elicited attribute thresholds. As in the \citet{jedidi_2005_probabilistic} model, Zolfaghari et al. deal with the uncertainty over the number and composition of criteria being used by fully enumerating all possible combinations of number and sets of criteria. This leads to a formulation that is similar to that of a subset-conjunctive rule with uncertainty over the number of criteria that must to be met.

Across the aforementioned one-stage and two-stage models, there are two key issues that this paper seeks to address. The first issue is that the aforementioned models primarily represent only conjunctive or disjunctive rules. Only the model by \citet{jedidi_2005_probabilistic} allowed for subset-conjunctive rules, and none of the models allowed for disjunctions-of-conjunctions as described in Section \ref{sec:decision-tree-economics}. Secondly, the one-stage models described above suffer from theoretical issues due to their use of constraints to implement strict non-compensatory behavior. In particular, imagine that there are two attributes, $x_1$ and $x_2$, and that violating the threshold for attribute $x_1$ leads to a systematic utility of positive infinity while violating the threshold for attribute $x_2$ leads to negative infinity. Although none of the observations in one's original dataset may violate both of these estimated thresholds, there is no guarantee that these thresholds will not be simultaneously violated by one or more observations when making predictions. In a situation where both thresholds are simultaneously violated, it is not clear what value the systematic utility should be set to and how calculation of choice probabilities should proceed. The decision tree models described in Section \ref{sec:decision-tree-explanation} and \ref{sec:decision-tree-variants} avoid this issue by using sets of conjunctive conditions that are all mutually exclusive, thus ensuring that no observation is ever described by more than one condition.

\subsection{Direct choice modeling via non-compensatory protocols}
\label{sec:direct-modeling-review}
As mentioned in the beginning of this section, few models have directly used conjunctive rules, disjunctive rules, or their generalizations to predict the probability of a given choice without estimating any rules or parameters that explicitly or implicitly determine an individual's choice set. To the best of our knowledge, there have only been two such modeling approaches: the cognitive process model of \citet{zhu_2010_cognitive} and the decision tree models of \citet{arentze_2004_learning, arentze_2007_parametric}. These will briefly be described below.

The cognitive process model first creates a new set of discrete features comprised of the originally discrete features and discretizations of the originally continuous features. The continuous features are discretized using estimated thresholds. Then, each alternative's set of discrete features are weighted using estimated weights, and a systematic utility for each alternative is created by summing the weighted, discretized features. Next, the systematic utilities are compared to estimated thresholds to determine the ``state'' that an alternative is determined to be in. In \citet{zhu_2010_cognitive}, it is assumed that there is only a reject or accept state. Based on the estimated thresholds and estimated weights, conjunctive or disjunctive rules may be expressed, and some\footnote{Note, we use the qualifier ``some'' because it is not clear to us that all disjunctions-of-conjunctions can be expressed using some combination of weights and thresholds in the cognitive process model.} disjunctions-of-conjunctions can also be expressed. A drawback of this model is that it is not clear how it works when there are more than two alternatives. In particular, it is not clear what would happen if two or more alternatives are placed into the ``accept'' state, and it is not clear what process would be used to determine a particular choice from the multiple acceptable alternatives.

In contrast to the cognitive process model, which is quite different from the models described in this paper, the decision tree models of \citet{arentze_2004_learning, arentze_2007_parametric} are highly related to our work. Using either decision trees by themselves or in combination with standard discrete choice models such as the MNL model, \citeauthor{arentze_2004_learning} directly predict the probability of a given alternative. Though not heavily emphasized in the original works of \citet{arentze_2004_learning, arentze_2007_parametric}, these models do permit the same microeconomic interpretations that we are describing in this paper. However, the models in \citet{arentze_2007_parametric} were motivated mostly by an attempt to the estimate the effect of discrete variables on one's systematic utilities using a non-parametric function that is adept at detecting interactions. In particular, when a decision tree is combined with standard discrete choice models in \citet{arentze_2007_parametric}, the decision tree is estimated based only on the explanatory variables that are originally discrete, and then a dummy variable for each output node of the tree is added to the systematic utilities of the various alternatives. The coefficients of these dummy variables are then estimated along with the usual parameters of one's choice model. As we will explain in Section \ref{sec:decision-tree-variants}, the models of \citet{arentze_2004_learning, arentze_2007_parametric} are actually special cases of the more general decision tree variant known as ``model trees.'' Moreover, as we will further explain in Section \ref{sec:decision-tree-variants}, our paper is the first (as far as we know) to interpret model trees as operationalizing a type of non-compensatory, context-dependent preference heterogeneity.

\section{Decision Tree Variants and Economic Considerations}
\label{sec:decision-tree-variants}
In Section \ref{sec:literature-review}, we described the way that discrete choice models have incorporated conjunctive rules, disjunctive rules, and their generalizations, and in Section \ref{sec:decision-tree-economics} we showed that these non-compensatory protocols can be expressed as decision trees. In this section, we concentrate on economic considerations that are likely to arise when choice modelers consider using decision trees in their own modeling activities. In particular, we will use Subsection \ref{sec:dtree-major-considerations} to focus on the ways that decision trees can (1) make probabilistic predictions, (2) represent heterogeneity in a population's non-compensatory rules, (3) represent estimation uncertainty, (4) represent context-dependent preference heterogeneity, and (5) satisfy monotonicity constraints. After this, we use Subsection \ref{sec:dtree-combinatorics} to discuss the ways that certain combinations of these considerations have been jointly accounted for by existing decision tree variants. Additionally, since choice modelers will likely need to account for all of these considerations simultaneously, we will end this section by pointing out the remaining methodological gaps that prevent these considerations from being addressed concurrently.

\subsection{Major Considerations}
\label{sec:dtree-major-considerations}

\subsubsection{Probabilistic predictions}
Some readers may note that, thus far, all of our decision tree and disjunction-of-conjunction examples have involved deterministic outputs. However, people with the same values for their explanatory variables may nevertheless make different choices. As a result, models of individual decision making need to be capable of producing probabilistic predictions. Fortunately, decision trees can and often do make probabilistic predictions in their output nodes. Conditional on a particular output node, the probability of a given alternative is often predicted to be the fraction of observations in that output node who chose the alternative in question \citep{arentze_2004_learning, strobl_2009_introduction}.

To economically motivate the move from deterministic outputs to the more general case of probabilistic outputs, we make two observations. First, we note that individuals may explicitly have probabilistic outputs in mind when they are using disjunctions-of-conjunctions. For instance, individuals may well say ``if any of these conjunctive conditions are met, then it is highly likely that I will do $y$,'' where $y$ is some outcome. In this case, the estimated decision tree will be estimating what ``highly likely'' means for this population. Secondly, it has long been noted that people violate their stated thresholds and attribute cutoffs when using non-compensatory protocols such as conjunctive and disjunctive rules \citep{green_1988_completely, huber_1991_adapting, swait_2001_non}. One implication of such cutoff violations is that even if an individual consciously operates as if satisfaction of some set of conjunctive conditions will result in a deterministic outcome $y$, there is still some probability that an individual in may choose another alternative $y'$ because he or she is violating their own conditions. In either motivating case\footnote{We are aware that in random utility maximization models, probabilistic outputs are often motivated through the argument that an analyst is unable to observe all of the variables that lead to an individual's deterministic choice. We believe that a lack of analyst omniscience will also lead to probabilistic outputs for decision tree models, but this reasoning also begs the question of how decision tree models behave when important explanatory variables are omitted. Such an investigation is beyond the scope of this paper, so for ease of exposition, we assume analysts using decision tree techniques observe all relevant explanatory variables.}, a decision tree will estimate the probability that each alternative is chosen from a given set of options.

\subsubsection{Heterogenous non-compensatory rules}
\label{sec:dtree-variants-heterogeneity}
When describing human behavior, it is often unreasonable to expect that all individuals in a population will use exactly the same non-compensatory rules. For example, imagine that the decision tree shown earlier in Figure \ref{fig:tree-for-bicycle-consideration} is generally accurate for two individuals: one who is fit and the other who is not fit. In this case, perhaps the fit individual believes commuting by bicycle for more than 45 minutes is unacceptable whereas the unfit individual thinks bicycling longer than 20 minutes is unacceptable. Here, the two individuals differ in the value that \textit{Travel Time} is split on in the decision tree. We will refer to this heterogeneity in the split point for an explanatory variable as local heterogeneity. In contrast, we will use the term global heterogeneity to describe the situation where even the structure of the decision tree differs across individuals. For instance, perhaps the unfit individual does not consider bicycling if the topography is hilly, regardless of the travel time. This would be heterogeneity in the set of conjunctive conditions that must be met in order for the individuals to consider bicycling. Below, we will discuss how both local and global heterogeneity have been accounted for by existing decision tree variants.

To begin, we note that local heterogeneity is fully accounted for by ``soft decision trees'' \citep{quinlan_1990_probabilistic, villandre_2012_soft}, also known as decision trees with ``soft splits'' \citep{kindermann_1998_model} or ``fuzzy decision trees'' \citep{jang_1994_structure, olaru_2003_complete}. These decision trees place a probability distribution over the splitting point of each continuous explanatory variable. Continuing the bicycle consideration example, these probability distributions enable soft decision trees to account for more realistic scenarios where 30 minutes is unacceptable to some people, 29 minutes is unacceptable to some other people, and yet still other people find 31 minutes to be acceptable. In these scenarios, the basic structure of the tree is correct, but individuals differ on the exact point at which their requirements are met. In order to account for this situation, one can make predictions as if a split point is known, and then one can use the given distributions to marginalize over the possible split points. When using this process, one eventually ends up still using formulas such as Equation \ref{eq:consideration-likelihood}, but now the probability of being in a given region (i.e. a given output node) will be some value between 0\% and 100\% instead of being deterministic.

Turning now to considerations of global heterogeneity, we find that this concern is accommodated by decision tree ensembles \citep{rokach_2010_ensemble}. In particular, ensembles of decision trees such as random forests \citep{breiman_2001_random} or boosted trees \citep{buhlmann_2007_boosting} represent global heterogeneity in much the same way that ensembles of discrete choice models (i.e. latent class choice models) represent heterogeneity amongst the compensatory decision protocols being used by differing market segments in a population \citep{vij_2013_incorporating}. The basic feature of tree ensembles is that many trees are estimated, and then predictions are made by averaging the predictions of each tree in the ensemble. However, a second feature of ensembles that we highlight is the ensemble's asymptotic behavior. What happens as the number of observations being used to estimate the trees goes to infinity\footnote{Note, this discussion is closely related to the notions of model averaging versus model combination \citep{minka_2002_bayesian}. Asymptotically, ensembles that implement model averaging will reduce to the estimation of a single tree, while ensembles that implement model combination will still estimate multiple, distinct decision trees. Model averaging is therefore seen as way to reduce estimation uncertainty while model combination accounts for global heterogeneity.} \citep{minka_2002_bayesian}? Asymptotically, decision tree ensembles such as bayesian decision trees and ``bagging'' (a portmanteau of ``bootstrap aggregation'') lead to the estimation of a single tree. We interpret these ensemble methods as catering for estimation uncertainty, so these methods will be described in Section \ref{sec:dtree-variants-uncertainty}. In contrast, global heterogeneity is represented by the ensemble methods that estimate multiple decision trees, even as the number of observations grows without bound. Analogously, as the number of observations tends to infinity, a latent class model still returns estimates for the different market segments in a population---it does not collapse to a choice model with one class.

Despite the similarities between latent class models and decision tree ensemble methods, there are some salient implementation differences between the two types of techniques. One of the most obvious differences is that latent class models often estimate a relatively small number of classes \citep{allenby_1999_marketing}, but ensemble methods usually result in models with hundreds of decision trees. While perhaps initially disconcerting, we note that having many trees makes sense behaviorally. The disjunctions-of-conjunctions used by individuals can differ in many ways. Even the simple difference between how the fit and unfit cyclists processed topography information in our earlier example would lead to two separate decision trees. As a result, a population can be expected to have many different decision trees being used by different people.

\subsubsection{Estimation uncertainty}
\label{sec:dtree-variants-uncertainty}
In many statistical applications, quantifying one's inferential uncertainty is important. For models that depend on continuous parameters, uncertainty is often quantified by the sampling distribution of one's estimator. However, unlike traditional models that are indexed by continuous parameters, decision trees are made up of discrete parameters such as the depth of the decision tree, the variables that the tree is split on, the values of the variables that are being split on, etc. In such discrete settings, uncertainty is quantified by the probability of a given combination of parameters being the data-generating parameters.  In other words, we need the probability of any given tree being the ``correct tree.'' Unfortunately, as with estimation of the tree, one will have to make approximations since complete enumeration of the possible decision trees is typically prohibitive \citep[p. 960]{chipman_1998_bayesian}.

Here, as noted in Section \ref{sec:dtree-variants-heterogeneity}, ensembles methods such as bayesian decision trees and bagging can provide a measure of estimation uncertainty. That bayesian decision trees provide the desired uncertainty quantification is due to the fact that bayesian methods explicitly estimate posterior probabilities of particular parameter values being true. The link between uncertainty quantification and bootstrap aggregation (i.e bagging) comes from the fact that the bootstrap is equivalent to a traditional bayesian analysis using a particular prior \citep{rubin_1981_bayesian, newton_1994_approximate}. In both cases, one would take the fraction of times a particular decision tree appears in the ensemble as being an estimate of the probability that the given decision tree is the ``true'' tree. These methods provide an approximate measure of the estimation uncertainty because there is no guarantee that these ensembles will contain all possible decision trees \citep[p. 960]{chipman_1998_bayesian}.

\subsubsection{Context-dependent preference heterogeneity}
\label{sec:dtree-variants-preferences}
In the discrete choice literature, and in the broader literature concerning human decision-making, it has long been acknowledged that ``the context in which a decision is made is an important determinant of outcomes'' \citep{swait_2002_context}. In particular, one's choice context may affect one's preferences or sensitivities to a given set of explanatory variables, and we use the term ``context-dependent preference heterogeneity'' to refer to this phenomenon. As an example, consider an individual making a choice of travel mode for his/her commute. When the cost of a given travel mode is low, perhaps the individual is most sensitive to that mode's travel time. However, when the cost of the travel mode is high, perhaps the individual becomes more sensitive to changes in travel cost than to changes in travel time. For such a simple scenario, a piecewise linear function for one's systematic utility may be sufficient. However, for scenarios where preferences are dependent on arbitrarily complex conditions, potentially involving multiple variables, we do not know of any accommodating methods within the traditional discrete choice literature.

Looking instead to the literature on decision tree methods, we note that decision tree variants known as ``hybrid,'' ``model,'' or ``functional'' trees \citep{zeilis_2008_model, rusch_2013_gaining} are able to account for such notions of context-dependent preference heterogeneity. Model trees are decision trees where the output at a given output node is a statistical model \citep{chan_2004_lotus, landwehr_2005_logistic, zeilis_2008_model, yu_2016_logit}. To make predictions, the decision tree is used to determine the output node that corresponds to the given observation, and then that output node's statistical model is used to provide the final outcome probabilities for the observation. In the specific case where discrete choice models are used in the output nodes, preference heterogeneity is represented by differing systematic utility functions in the models used in different nodes. Returning to our example from the previous paragraph, imagine that we had a decision tree that was split on the \textit{Travel Cost} variable at a value that distinguished ``low'' versus ``high'' travel costs. The model at the low-travel-cost output node might have a systematic utility function that is linear-in-parameters with a coefficient $\beta _{\textrm{LowCost}}$ being multiplied by the travel-cost variable. Conversely, the model at the high-travel-cost output node might also have a linear-in-parameters systematic utility function, with a coefficient $\beta _{\textrm{HighCost}}$ being multiplied by the travel-cost variable, where $\beta _{\textrm{HighCost}} > \beta _{\textrm{LowCost}}$. Such a model tree would capture the notion that preferences (in this case, the travel-cost coefficients) are dependent on the context in which the choice is being made---a low travel cost context versus a high travel cost context.

Beyond the general description provided in the previous paragraph, we pause here to note that many decision tree methods and discrete choice methods can be seen as special cases of model trees. First, the standard decision tree described in Section \ref{sec:decision-tree-explanation} can be seen as a model tree where discrete choice models such as the MNL are used in each node, and each alternative's systematic utility is only comprised of an alternative specific constant (ASC). For decision trees with deterministic outputs, these constants are either infinity or negative infinity. For decision trees with probabilistic outputs, the relative values of these constants can be determined by constraining a reference alternative's ASC to zero, and determining what ASCs of the other alternatives will lead to the decision tree's estimated choice probabilities. Secondly, other proposed models estimate a decision tree and then place a dummy variable for each output node into one's systematic utility functions in a discrete choice model. This methodology includes models such as the parametric-action decision tree \citep{arentze_2007_parametric}, the hybrid CART-logit model \citep{steinberg_hybrid_1998}, the tree-augmented logistic model \citep{su_2007_tree}, and the two-stage MNL model\citep{kim_2009_two, kim_2011_two}. Such models can be seen as special cases of model trees that allow for context-dependent heterogeneity in the ASCs but enforce homogeneity on the remaining parameters in the choice models. Finally, the semi-compensatory models used in the discrete choice literature are also special cases of model trees. In these semi-compensatory models, described in Section \ref{sec:literature-review}, conjunctions, disjunctions, or disjunctions-of-conjunctions are used to screen alternatives and then a compensatory discrete choice model is used to select from any remaining alternatives. This can be seen as a model tree where the parameters of the systematic utility function for available alternatives are constrained to be equal across the various output nodes, and output nodes that result in a given alternative not being available simply set the systematic utility for that alternative to negative infinity.

\subsubsection{Monotonicity}
\label{sec:dtree-variants-monotonicity}
Lastly, we note that models of human decision making are often subject to constraints based on economic theory. For instance, all else equal, as the price of a normal good increases, the probability that this good is chosen should decrease or, at worst, stay the same. This is a monotonicity constraint. In discrete choice models that use linear-in-parameters systematic utility functions, such monotonicity constraints are operationalized through constraints on the sign of the model coefficients. These sign constraints allow one to quickly check if one's estimated parameters comply with economic theory about the relationship between an explanatory variable and an outcome of interest. And as noted in the introduction, discrete choice modelers are highly unlikely to use a model that does not demonstrate compliance with economic theory.

Fortunately, decision tree variants that can incorporate monotonicity constraints have been created \citep{potharst_2002_classification, velikova_2004_decision, hu_2012_rank, marsala_2015_rank, pei_2016_multivariate}. Such monotonic decision trees are constructed by altering the estimation process to ensure that the desired monotonicity constraints are not violated. By using monotonic decision trees, one can estimate the disjunctions-of-conjunctions that may be in use in one's population, while at the same time guaranteeing compliance with economic theory. The ability to ensure the monotonicity of key relationships should go a long way towards easing the concerns of choice modelers who are considering using decision trees in their analyses but want  to make sure that their estimated trees ``make sense.''

\subsection{Combining considerations}
\label{sec:dtree-combinatorics}
In Subsection \ref{sec:dtree-major-considerations}, we sequentially detailed how various types of decision trees allow researchers to (1) make probabilistic predictions, (2) represent heterogeneity in a population's non-compensatory rules, (3) represent estimation uncertainty, (4) represent context-dependent preference heterogeneity, and (5) satisfy monotonicity constraints. However, in real applications, analysts may wish to simultaneously account for all of the considerations described above. In this subsection, we will briefly detail the ways that such goals can and cannot yet be met. Our discussion will point out advanced decision tree variants as well as point to methodological gaps that must be filled in order to make decision trees maximally useful to discrete choice researchers.

To begin, we first point out that all decision tree variants allow for the use of probabilistic predictions. Accordingly, we will focus our discussion on considerations (2) - (5), listed above. Next, we will make the point upfront that there are no decision tree variants that currently account for all four of the remaining considerations. The best that can be done with available methods is to account for combinations of two or three of considerations (2) - (5). Moving swiftly through such combinations, the only three considerations that have been combined are the representation of local heterogeneity, the representation of estimation uncertainty and the representation of context-dependent preference heterogeneity. These three concerns are simultaneously accounted for in the decision tree variant known as a bayesian hierarchical mixture-of-experts model \citep{bishop_2003_bayesian}. Such a model makes use of model trees with soft-splits and uses bayesian estimation techniques to account for estimation uncertainty. Moving to combinations of two of the four considerations, only three of the six possible combinations have been accounted for in the literature. First, bayesian soft decision trees \citep{kindermann_1998_model} and bagged soft decision trees \citep{yildiz_2016_bagging} allow for estimation uncertainty and representations of local heterogeneity. Furthermore, soft tree ensembles such as a random forest of soft trees \citep{seyedhosseini_2015_disjunctive, kumar_2016_ensemble} allow for representations of both local and global heterogeneity. Secondly, soft model trees known as mixtures of experts or hierarchical mixtures of experts \citep{jordan_1994_hierarchical, yuksel_2012_twenty} allow for context-dependent preferences and local heterogeneity. Thirdly, global heterogeneity and monotonicity have been jointly represented by monotonic random forests \citep{gonzalez_2015_monotonic}.

To the best of our knowledge, no combination of considerations has been addressed beyond those detailed in the last paragraph. As a result, by developing decision tree models that account for the missing combinations of economic considerations, discrete choice researchers can help advance the fields of computer science and statistics while simultaneously catering for properties they wish to have in their own analyses. In Section \ref{sec:empirical-application}, we illustrate such development by formulating and estimating what we believe is the first bayesian model tree. This allows us to account for estimation uncertainty and context-dependent preference heterogeneity. While not simultaneously addressing all of considerations (2) - (5) mentioned above, our model nevertheless fills a missing rung in the methodological ladder of existing decision trees.

\section{Empirical Application}
\label{sec:empirical-application}
In the last section, we showed how common economic concerns can be addressed by existing variants of decision trees. Additionally, we pointed out gaps in existing decision tree methodologies that need to be filled in order to make decision trees most useful when modeling economic phenomena. In this section, we switch focus and review our paper's empirical application. Given the economic interpretation of decision trees representing disjunctions-of-conjunctions, we study whether such rules appear to be used by commuters in the San Francisco Bay Area. In particular, we model how disjunctions-of-conjunctions are used to choose whether or not bicycle would be considered as a travel mode and, if bicycle was considered, how the disjunctions-of-conjunctions affect the overall preference for bicycling when choosing between the considered travel modes. Moreover, we take pains to capture our uncertainty in the estimated disjunctions-of-conjunctions. As a result, our application contributes to the literature by creating the framework and estimation techniques for the first decision tree variant that accounts for both context-dependent preference heterogeneity and model uncertainty. 

In the following subsections, we first review the motivation for our proposed semi-compensatory model (i.e. the combination of a decision tree with a standard mode choice model). Next, Section \ref{sec:model-framework} reviews the details of how our proposed model works, and Section \ref{sec:estimation-methods} details the proposed and implemented estimation techniques for our new model. In Section \ref{sec:data-and-specification} we detail the model specification and data used in our application, and in Section \ref{sec:results-and-discussion} we present our results and discussion.

\subsection{Motivation}
\label{sec:model-motivation}
As previously noted, our application concerns the choice of travel mode in the San Francisco Bay Area. Specifically, we are interested in whether people choose to commute by bicycle. Of critical importance are two phenomena. First, individuals may (for a variety of reasons) exclude bicycling from consideration, thereby removing all possibility that they will use a bicycle to commute to work/school. If such differences in consideration are not accounted for, then one will make incorrect inferences regarding the amount by which any project can be expected to increase the expected number of cyclists. Secondly, individuals may find themselves in situations that lead them to be more or less amenable to the idea of commuting by bicycle. If an individual has a very low general preference for bicycling, then policies to increase bicycling rates may only have a minor impact on this individual's probability of bicycling. In other words, before judging the ability of an intervention to increase the probability that the individual actually bikes, one must be sure that an individual is considering bicycling as a commuting option, and one should attempt to judge an individual's general preference for bicycling.

In previous discrete choice research that allowed for heterogeneous consideration sets, mode choice models have been operationalized based on assumptions regarding: the existence of latent market segments that each have their own consideration sets and utility coefficients \citep{vij_2013_incorporating, vij_2014_preference}, the existence of individuals that have either complete choice sets or who irrationally only consider a single travel mode \citep{swait1987incorporating}, or whether alternatives are independently chosen for inclusion in one's consideration set \citep{swait1987empirical, swait_2001_choice, swait_2009_choice}. With these formulations, researchers have already found support for the hypothesis that, beyond deterministic differences in the travel modes which are available to a given person, individuals differ in whether they consider bicycling as a commuting option and in how much they generally prefer cycling \citep{swait_2009_choice, vij_2013_incorporating, vij_2014_preference, mahmoud_2016_myopic}.

In all the modeling efforts just described, the probability of an individual considering a particular mode was always based on a compensatory model. These models are curious in light of the fact that when asked about why they don't commute by bicycle, individuals do not state that the issues which make them avoid bicycling to work can be compensated for by other commonly used variables in mode choice models. Individuals commonly state that they live too far away to commute by bicycle, that roadway conditions are too dangerous for them to commute by bike, that cycling would require too much physical exertion, that they have to transport children to some place, and so on \citep{goldsmith_1992_reasons, cleland_2004_why}. It is not clear  \textit{a-priori} that these type of concerns can be incrementally compensated for by changes in sociodemographic variables or level-of-service variables for the various travel modes. As a result, it is reasonable to think that non-compensatory models of consideration set formation may be better able to emulate the actual decision making process of individuals. Our goal for this application was to develop a policy analysis tool for bicycling that could capture the effect of non-compensatory protocols on choice set formation and on the general preference for bicycling. We used disjunctions-of-conjunctions as our non-compensatory protocol in order to account for the ``if-then'' nature of people's stated reasons for not bicycling. Beyond using decision trees to model the consideration of the bicycle alternative, we wanted to be sure to account for the effect of the attributes of the non-bicycle alternatives. As a result, we follow the lead of the semi-compensatory models reviewed in Section \ref{sec:literature-review} by using a compensatory model to predict the final choice between any alternatives that are considered.

\subsection{Model Framework}
\label{sec:model-framework}

In the last subsection's discussion, we reviewed why we desire a semi-compensatory model that combines decision trees and discrete choice models. In this subsection, we will review our desired model in more detail so readers are clear about how it works and so that readers of Section \ref{sec:estimation-methods} have enough context to understand why we chose the estimation methods that we chose.

First, as described in Section \ref{sec:dtree-variants-preferences}, our proposed type of model is known in the decision tree literature as a model tree. Model trees are decision trees that use statistical models in their output nodes to predict the outcome of interest. Here, the statistical models in the output nodes typically differ from one another. In our application, the model tree will function as follows. There will be a decision tree with mode choice models in the output nodes. The tree will be used to winnow the bicycle from an individual's choice set, and across the different situations where bicycle is considered, the general preference for bicycling will be allowed to differ. This results in differing bicycle ASCs in the mode choice models of the different output nodes of the decision tree. For simplicity, we have constrained the other parameters in our choice model to remain constant across the various output nodes. In other words, accounting for context-dependent preference heterogeneity in the parameters other than the bicycle ASC is left for future research, as is accounting for global and local heterogeneity in the estimated disjunctions-of-conjunctions or accounting for a-priori monotonicity constraints.

Second, we go beyond the mere use of model trees as they have already been implemented. Instead, we contribute to the literature of decision tree methodologies by developing a bayesian model tree. By using bayesian estimation techniques, we can account for estimation uncertainty about which model tree is the ``true'' tree. These various candidate trees, denoted by $m$, represent different non-compensatory decision protocols, and we are using the bayesian estimation to compute the probabilities of these different protocols being the one used in our population. In addition, as is always done when estimating bayesian choice models, the bayesian estimation also accounts for the estimation uncertainty in the choice model parameters.

Now, because we are estimating a model tree, we can partition the model parameters into those that describe the tree and the parameters that describe the choice models at the output nodes of the tree. We will start with the tree parameters. Using the notation from Section \ref{sec:dtrees-and-disjunctions-of-conjunctions}, a decision tree is uniquely identified by three sets of parameters. The first parameter is how many conjunctive conditions (i.e. output nodes) are in the tree. We denote this as $D^m$. The second set of parameters is how many requirements are in each conjunctive condition. We denote these parameters as $\mid p_i ^m \mid$, where $i \in \left\lbrace 1, 2, ..., D^m \right\rbrace$. Lastly, the third set of parameters is the primitive boolean conditions that make up each requirement. We denote these parameters as $b_j ^{i, m}$ where $j \in \left\lbrace 1, 2, ..., \mid p_i ^m \mid \right\rbrace$.

Next, we will move onto the parameters of the choice models at the output nodes of the tree. We denote these parameters as $\gamma ^m$, and we note that in our application, we are only allowing the bicycle ASC to differ across output nodes. As a result, we can further partition the parameters that describe the choice models at the output nodes. Conditional on a tree ($m$), there will be one parameter per output node ($i$), and these parameters will determine the bicycle ASC for the given node. We will denote these node-varying parameters by $\eta _i ^m$. Additionally, there will be the remaining choice model parameters that do not change from one output node to the next. We will denote these parameters by $\beta$. All together, we have $\gamma ^m = \left( \eta _i ^m, \beta \right)$. Combining this paragraph with the last, the parameters to be estimated are $\theta = \left( D^m, \mid p_i ^m \mid, b_j ^{i, m}, \eta _i ^m, \beta \right)$ for all $i \in \left\lbrace 1, 2, ..., D^m \right\rbrace$ and for all $j \in \left\lbrace 1, 2, ..., \mid p_i ^m \mid \right\rbrace$.

\begin{figure}
\centering
\includegraphics[width=0.75\textwidth]{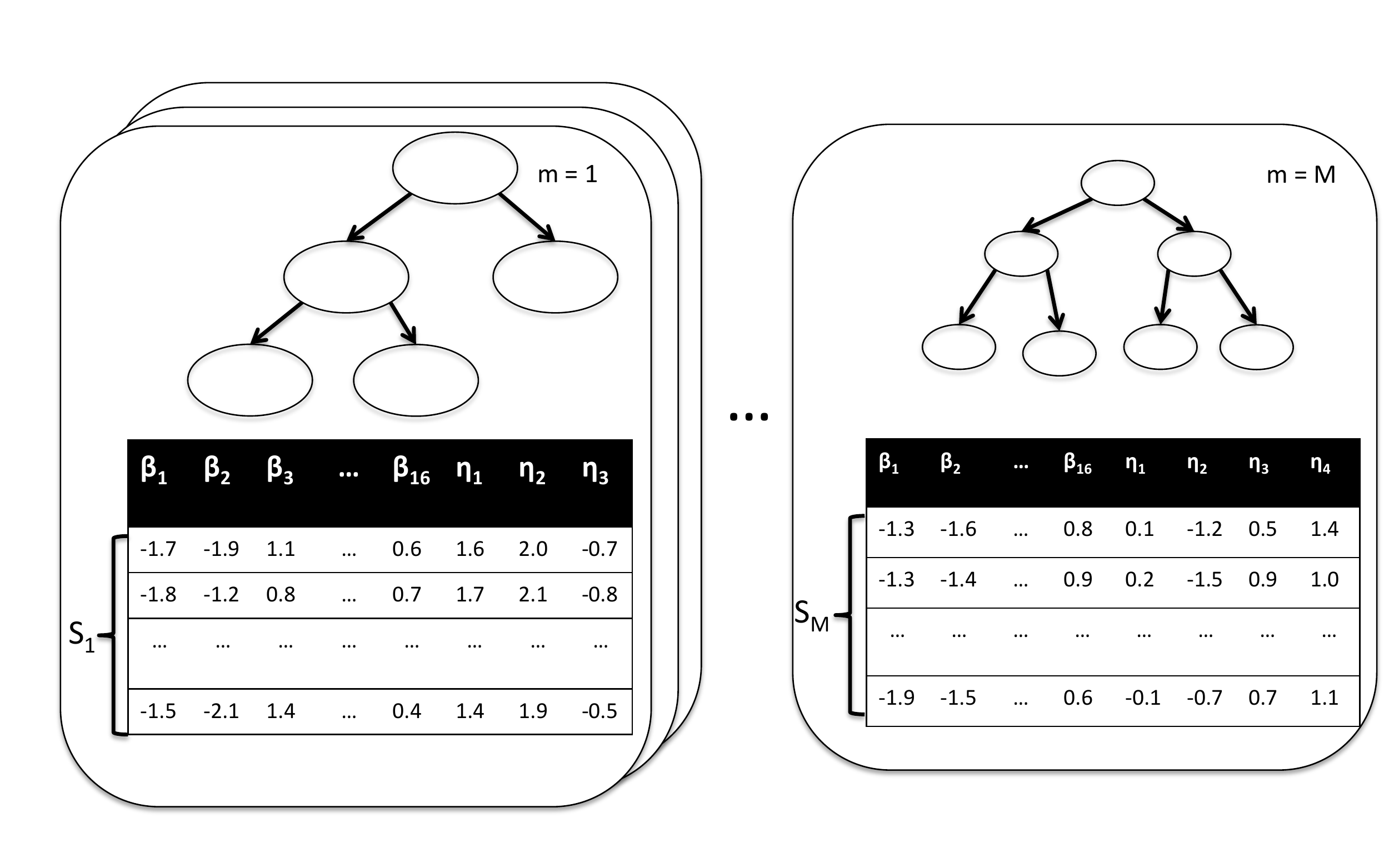}
\caption{Procedural diagram of bayesian model trees}
\label{fig:sampling-process}
\end{figure}

Due to the bayesian estimation techniques, our estimation results will now be a posterior distribution that reflects our uncertainty in the ``true tree'' and in the ``true'' parameters of the choice models in that tree's output nodes. Moreover, since we do not have a closed-form expression for this posterior distribution, it will be represented by a sample from this joint distribution of trees and choice model parameters. Each sampled element ($s$) will be a decision tree ($m$) and the parameters of the choice models at that tree's output nodes ($\gamma _{s} ^m$). We denote the number of sampled elements containing tree $m$ as $S_m$. Next, we can use the fraction of times that a specific tree appears in the posterior sample to estimate the posterior probability of a given tree ($P_{\textrm{Post}} \left( m \right) = \frac{S_m}{\sum _{\ell} S_{\ell}}$). Finally, in a bayesian model tree setting, we calculate the predicted probability of outcome $Y$ given explanatory variables $X$ using the following formula:
\begin{equation}
\label{eq:model-tree-prediction}
\begin{aligned}
\hat{P} \left( Y \mid X \right) &= \sum _{m = 1} ^M P_{\textrm{Post}} \left( Y \mid X, m \right) P_{\textrm{Post}} \left( m \right)\\
&= \sum _{m = 1} ^M \left[ \frac{1}{S_m} \sum _{s=1} ^{S_m} P \left( Y \mid X, \gamma_{s} ^m, m \right) \right] P_{\textrm{Post}} \left( m \right)\\
\textrm{where } M &= \textrm{The total number of unique trees in one's sample.}\\
P \left( Y \mid X, \gamma_{s} ^m, m \right) &= \textrm{The choice model probability of $Y$ given $X$, $\gamma _{s} ^m$, and tree $m$. }
\end{aligned}
\end{equation}
For a graphical depiction of this process, see the diagram in Figure \ref{fig:sampling-process}.

\subsection{Estimation Methods}
\label{sec:estimation-methods}
The previous subsection reviewed the overall framework, mechanics, and parameters of our proposed bayesian model tree. In this subsection, we detail our estimation techniques. These details are discussed at length because we found estimation of this new model to be a nontrivial challenge, and we want other researchers to be able to replicate and build off our work. Readers who would like to immediately get to the results and `big-picture' discussion may feel free to skip ahead to Section \ref{sec:results-and-discussion}.

Subsection \ref{sec:model-framework}'s formulation of $\theta$ shows that the total number of parameters being estimated depends on the decision tree. In particular, as we change from tree to tree, the number of conjunctive conditions ($D^m$) will vary, and as a result, the dimensionality of the parameter vector will vary. Unfortunately, such changing dimensionality necessitates the use of specialized estimation techniques \citep[p.5]{das_2017_transdimensional}. Of these, the reversible-jump algorithm \citep{green_1995_reversible} is the most common bayesian estimation technique for problems of varying dimensionality, both overall \citep{sisson_2005_transdimensional} and for decision trees in particular \citep{denison_1998_bayesian, wu_2007_bayesian, mohammadi_2016_comment}.

As noted by \citet[p.72-73]{fan_2011_reversible}, efficient reversible-jump algorithms require a way for one to propose parameter values of high posterior probability while switching between parameter spaces of varying dimensions, and creating such proposal mechanisms is not straightforward. Our initial attempts at using a reversible-jump algorithm to estimate our model tree failed because we were unable to devise a good way to propose new choice model parameters when switching from one tree to another. How should we propose new bicycle ASCs when the groups of individuals in each output node are completely different? The creation of an efficient, reversible-jump proposal mechanism for bayesian model trees remains an open problem, and it is one that we would be happy to collaborate with others on.

Given our difficulties with the reversible-jump algorithm, we instead sought an alternative estimation strategy. The approach we settled on was to split the problem into two sub-problems, each of which was more easily solved than the original problem. Specifically, as we noted in Section \ref{sec:how-to-estimate-dtrees}, there are existing methods for performing a bayesian estimation of decision trees. Additionally, one can estimate the parameters of a given choice model using almost all existing bayesian estimation techniques for fixed-dimensional problems. In light of these two facts, we sought to break our model tree estimation into a first step where we estimate the decision trees by themselves and a second step where, conditional on a given decision tree, we estimate the choice models that belong in each output node of the tree. Finally, some procedure would be needed to tie these two estimation tasks together.

To implement this divide-and-conquer approach, our original (and idealized) plan was as follows. First, we would use the techniques of \citet{letham_2015_interpretable} to perform a bayesian estimation of the decision trees. Then, conditional on each tree, we would use the techniques of \citet{braun_2016_scalable} to estimate our mode choice model with varying bicycle ASCs. And lastly, we would use importance sampling to adjust the original posterior distribution of decision trees in light of the information provided by the choice models at the output nodes at each tree. Below, we briefly justify each of these choices.

Beginning with the estimation of the decision trees, we chose to use the techniques of \citet{letham_2015_interpretable} for two reasons. First, their methods were implemented in freely available python scripts, so we would not have to re-invent their techniques. Secondly, their approach requires researchers to specify the possible requirements that can be used in the conjunctive conditions that comprise the decision tree. This specification gives researchers the ability to check for sensible relationships between the explanatory variables and the outcomes of interest. For example, by specifying the regions of parameter space that the travel distance is split into, the researcher can empirically check whether the fraction of individuals bicycling decreases as one moves from the region where travel distance is between 2 and 3 miles to the region where travel distance is between 3 and 4 miles.

Moving to the choice model estimation, we (again) had two reasons for choosing the techniques of \citet{braun_2016_scalable}. First, unlike typical MCMC procedures that only generate dependent samples from the posterior distribution of one's choice model parameters, the techniques of \citet{braun_2016_scalable} generate independent samples, resulting in higher effective sample sizes per unit of computational time. Secondly, the methods of \citet{braun_2016_scalable} automatically provide accurate estimates of the total probability of the data given one's decision tree (i.e. after marginalizing over the parameters in the choice model). This probability is needed for our last step: importance sampling.

After the initial estimation of the decision trees and the mode choice models, conditional on the decision trees, we need to link these two estimation procedures. In particular, we want a sample from the joint distribution of decision trees and their accompanying choice models. However, our original sample of decision trees was produced without using any information from the choice models at the output nodes. As a result, our original sample of decision trees is (in general) drawn from an incorrect distribution. We use importance sampling \citep{gelman_1992_iterative, hesterberg_1995_weighted} to weight our original sample of decision trees such that the weighted sample comes from our desired distribution. Since our original sample was drawn from a distribution $P \left( \textrm{tree without choice models} \mid \textrm{data} \right)$ instead of $P \left( \textrm{tree with choice models} \mid \textrm{data} \right)$, we will weight each tree by the ratio $\tfrac{P \left( \textrm{tree with choice models} \mid \textrm{data} \right)}{P \left( \textrm{tree without choice models} \mid \textrm{data} \right)}$. The probabilities in the numerator and denominator are computed up to a constant of proportionality using Bayes rule, and then the importance weights are normalized such that they sum to one across all the trees in our sample. At this point, estimation is complete and the weighted sample is then available for prediction or further inference tasks.

As just described, this three step procedure is our current, ideal method for estimating bayesian model trees. However, this procedure is computationally expensive. For example, our initial sample of trees contained more than 5,000 unique decision trees. On average, for a single decision tree, it took approximately 2 hours to perform the bayesian estimation of the choice models at the output nodes. The total estimation time would have taken more than a week for our dataset and choice model specification (described in Section \ref{sec:data-and-specification}). Given our current computing resources (a single laptop), we deemed this estimation time unreasonable, so we made further approximations to speed up the estimation process. In particular, we selected a subset of 10 decision trees from the total set of unique trees so that the total estimation time would be less than a day. Then, we then estimated the choice models at the output nodes of these trees, and we proceeded as if these ten trees were the complete set of possible trees for our data. As far as we know, it is impossible to account for the existence of the other trees without performing the estimation of those trees' choice models, which is exactly what we wished to avoid. While numerous ways of choosing the ten trees are possible, we tried to follow the intuition of \citet{breiman_2001_random} who noted that the accuracy of a set of trees ``depends on the strength of the individual tree classifiers and a measure of the dependence between them.'' Specifically, we chose the ten trees as follows. We chose top three trees in terms of their (approximate) log-posterior from step 1, and we also chose the top three trees in terms of their (approximate) log-likelihood from step 1\footnote{Note, we use the term `approximate' because the log-posterior values and log-likelihood values of step 1 do not take into account the choice models at the output nodes of the tree.}. We chose the final four trees by first selecting the trees that had approximately the posterior mean number of output nodes ($D^m$) and then, from the selected trees, choosing the 4 trees with the highest log-posterior. This procedure closely follows the recommendation of \citet{letham_2015_interpretable} for selecting a single decision tree to be a point estimate for the posterior distribution of trees. In the end, our selected trees were all ``strong'' in some way, whether that be high log-likelihoods or high log-posterior values, and across the three selection criteria, the trees were quite different from one another. We will refer to the procedure described in this paragraph as our ``actual'' estimation methodology, whereas the procedures described in the paragraphs above are our ``ideal'' estimation methodology. As we will see in Section \ref{sec:results-and-discussion}, despite our radical simplifications, our actual estimation methodology still produces a model that provides quantitatively more accurate and qualitatively more reasonable inferences than the traditional MNL model.

\subsection{Data and model specification}
\label{sec:data-and-specification}
In the previous subsections, we described our model framework and estimation methods. In this section we describe the data used in our application and the precise specification (i.e. model priors and choice model specification) of our bayesian model trees.

\subsubsection{Data}
\label{subsec:data}
Starting with the data, we are using 1,015 observations from the California Household Travel Survey (\citeyear{california_department_of_transportation_2010-2012_2013}). Each individual in our sample lives in Oakland, Berkeley, or San Francisco, CA, and the observations represent home to work or school commute tours. For level-of-service variables (such as travel time, cost, and distance) we use estimates provided by the San Francisco Metropolitan Transportation Commission (MTC) (\citeyear{san_francisco_metropolitan_transportation_commission_travel_2012}). Basing our set of possible alternatives on the alternatives used by MTC, we classify observations as having traveled via one of eight travel modes. There were three driving modes, each differentiated by the number of passengers: drive-alone, shared-ride with two passengers, and shared-ride with three or more passengers. There were also three transit modes, each differentiated by their access and egress modes: walk-transit-walk (where walking is used for access and egress), drive-transit-walk, and walk-transit-drive. Finally, there were two non-motorized modes: walking and bicycling. For each tour, the travel mode that was used for the longest distance was used as the ``chosen travel mode'' for that tour.

Importantly, one of our uses for non-compensatory rules is to determine whether or not an individual considers bicycling as a travel mode. Accordingly, our decision trees are based on spatial variables and socio-demographics that have been mentioned in reasons why individuals did not consider bicycling. In particular, the trees are based on spatial variables such as distance, roadway slopes, elevation, on-street bicycle infrastructure, speed limits, and socio-demographics such as the number of children. Post-processing of the raw spatial data was done using a novel concept called the zone of likely travel. The main idea is, for each individual, to form a buffer around the shortest path between one's home and work or school. This buffer is constrained to follow the roadway network instead of merely being laid atop of a map, and the buffer is constructed so its perimeter is based on each user's likely, maximum deviation from the shortest path. In other words, the zone should contain the roadways over which one is likely to travel. All spatial variables are then calculated over the roadways in one's zone of likely travel. In general, the details of this post-processing procedure are not related to the main purpose of this paper, so we will not review them any further. However, we instead encourage interested readers to review the details of this processing in \citet{brathwaite_2018_holy}.

\subsubsection{Model Specification}
Traditionally, when discrete choice modelers talk about model specification, they mean the specification of one's utility functions. However, in a bayesian paradigm, one also needs to specify his/her model priors. These priors are probability distributions that encapsulate the modeler's prior beliefs about the true value of the model parameters. Together with the utility specifications and likelihood function, these specification choices allow for model estimation. Below, we will note each our specifications in turn, starting with the choice model.

Specifically, we specify the systematic utility functions in our choice model as follows:

\begin{equation}
\label{eq:utility-specifications}
\begin{aligned}
V_{DA} &= \beta_{\textrm{travel-time-auto}} \textrm{TravelTime}_{\textrm{DA}} + \beta_{\textrm{autos-per-driver}} \textrm{AutosPerDriver}\\
V_{SR2} &= \textrm{ASC}_{\textrm{shared-ride-2}} + \beta_{\textrm{travel-time-auto}} \textrm{TravelTime}_{\textrm{SR2}} + \beta_{\textrm{autos-per-driver}} \textrm{AutosPerDriver} \\
& \quad + \beta _{\textrm{cross-bay}} \textrm{CrossBay} + \beta _{\textrm{num-kids}} \textrm{NumberKids} + \beta_{\textrm{household-size}} \textrm{HouseholdSize}\\
V_{SR3} &= \textrm{ASC}_{\textrm{shared-ride-3}} + \beta_{\textrm{travel-time-auto}} \textrm{TravelTime}_{\textrm{SR3}} + \beta_{\textrm{autos-per-driver}} \textrm{AutosPerDriver} \\
& \quad + \beta _{\textrm{cross-bay}} \textrm{CrossBay} + \beta _{\textrm{num-kids}} \textrm{NumberKids} + \beta_{\textrm{household-size}} \textrm{HouseholdSize}\\
V_{WTW} &= \textrm{ASC}_{\textrm{walk-transit-walk}} + \beta_{\textrm{travel-time-transit}} \textrm{TravelTime}_{\textrm{WTW}} + \beta_{\textrm{travel-cost-transit}} \textrm{TravelCost}_{\textrm{WTW}}\\
V_{WTD} &= \textrm{ASC}_{\textrm{walk-transit-drive}} + \beta_{\textrm{travel-time-transit}} \textrm{TravelTime}_{\textrm{WTD}} + \beta_{\textrm{travel-cost-transit}} \textrm{TravelCost}_{\textrm{WTD}}\\
V_{DTW} &= \textrm{ASC}_{\textrm{drive-transit-walk}} + \beta_{\textrm{travel-time-transit}} \textrm{TravelTime}_{\textrm{DTW}} + \beta_{\textrm{travel-cost-transit}} \textrm{TravelCost}_{\textrm{DTW}}\\
V_{walk} &= \textrm{ASC}_{\textrm{walk}} + \beta_{\textrm{distance-walk}} \textrm{TravelDistance} _{\textrm{walk}}\\
V_{bike} &= \textrm{ASC}_{\textrm{bike}} + \beta_{\textrm{distance-walk}} \textrm{TravelDistance} _{\textrm{bike}}\\
\end{aligned}
\end{equation}
In the systematic utility equations above, DA means ``drive alone,'' SR2 means ``shared-ride with two passengers,'' SR3 means ``shared-ride with three or more passengers,'' WTW means ``walk-transit-walk,'' WTD means ``walk-transit-drive,'' and DTW means ``drive-transit-walk.'' Though not indicated using subscripts on the variables, all of these systematic utility equations are specific to a given individual.

Next, we note that our specifications above were not made arbitrarily. Travel cost was excluded from the driving alternatives because it was too collinear with the travel time variable to permit estimates that had the correct sign. This is to be expected since MTC calculates both its travel cost and travel time estimates for driving modes as a function of travel distance. Secondly, income and gender are not present in our specifications because it was missing for numerous individuals in our dataset.

Finally, the systematic utility specifications shown in Equation \ref{eq:utility-specifications} are common across both the MNL model and the bayesian model trees used in this paper. There are only two differences between the systematic utility specification of the MNL and the bayesian model trees. First, as mentioned above, the $ASC _{\textrm{bike}}$ is allowed to differ from output node to output node. In other words, the bayesian model trees replace $ASC _{\textrm{bike}}$ with $\sum _{i=1} ^{D^m} \delta _i ASC _{\textrm{bike}, i}$ where $i$ denotes a particular output node of decision tree $m$ and $\delta_i$ is a dummy variable that indicates whether or not an individual is in output node $i$. Briefly, we note that we do not directly estimate the parameters $ASC _{\textrm{bike}, i} \ \forall i \in \left\lbrace 1, 2, ..., D^m \right\rbrace$. This would correspond to using a \textit{no-pooling} estimator that treats the output nodes as being completely different from one another. Instead, we would rather estimate how different the nodes are from one another. To do this, we use a hierarchical logit estimator (i.e. a \textit{partial-pooling} estimator) \citep{bafumi_2006_fitting, gelman_2006_multilevel, gelman_2014_bayesian} that combines (i.e. pools) information about the overall bicycle preference across output nodes. The $ASC _{\textrm{bike}, i}$ parameters are conceptualized as instances from an overall, normal distribution of bicycle ASCs with mean $ASC _{\textrm{bike}}$ and variance $\sigma _{\textrm{bike}} ^2$. Here, the mean and variance parameters are estimated along with the individual $ASC _{\textrm{bike}, i}$ parameters\footnote{To tie this paragraph back to Section \ref{sec:model-framework} where we first discussed our model parameters,  we define $ASC _{\textrm{bike}, i} = ASC _{\textrm{bike}} + \eta _i$. In our application we actually estimate $\eta _i$ and $ASC _{\textrm{bike}}$ instead of $ASC _{\textrm{bike}, i}$ and $ASC _{\textrm{bike}}$. In the statistical literature, this choice is referred to as the use of a non-centered parametrization \citep{papaspiliopoulos_2007_general}. We used the non-centered parametrization instead of the traditional approach of directly estimating $ASC _{\textrm{bike}, i}$ because this method led to faster estimation times.}. As $\sigma _{\textrm{bike}} ^2 \rightarrow \infty$, we are increasingly certain that the output nodes are completely different from one another, and as $\sigma _{\textrm{bike}} ^2 \rightarrow 0$ we are increasingly confident that the general preference for bicycling is actually the same across output nodes.

The second difference is, as noted in Section \ref{subsec:data}, that we use spatial variables in the construction of the decision trees. In order to fairly compare the MNL and the bayesian model tree, we include the spatial variables in the MNL model by placing these variables in the bicycle systematic utility. In particular, the bicycle utility of the MNL model is expanded to include the following variables and their coefficients: shortest path length, median slope, average speed limit, proportion of roadway miles on the shortest path with speed limits of 25 mile per hour or less, proportion of roadway miles with bicycle lanes, and the proportion of roadway miles with ``share the road'' markings (also known as ``sharrows''). These variables are excluded from the bicycle utility of the choice models in the bayesian model tree as they are already used when constructing the decision trees.

Next we state our model priors. In a bayesian setting, priors must be specified for all parameters that are being estimated. We start with the choice model parameters. For all choice model parameters, excluding $ASC _{\textrm{bike}, i} \ \forall i  \in \left\lbrace 1, 2, ..., D^m \right\rbrace$ and $\sigma _{\textrm{bike}} ^2$, we assumed independent priors of $\mathcal{N} \left( 0, 4 \right)$ where 4 is the variance of the normal distribution. This prior was chosen to reflect the fact that we think it is highly unlikely for a 1-unit change of any of our variables to cause a change of 4 in our systematic utility functions. Such changes would greatly increase or decrease the probability of choosing a given alternative, and we don't expect a 1 minute change in travel time, a 1 dollar change in travel cost, a change in 1 mile of travel distance, etc. to cause drastic changes in the probability of a given mode. For, the $ASC _{\textrm{bike}, i}$ parameters, we use the prior distribution mentioned above. That is, the prior distribution of $ASC _{\textrm{bike}, i}$ is $\mathcal{N} \left( \textrm{ASC}_{\textrm{bike}}, \sigma _{\textrm{bike}} ^2 \right)$. Here, we again use a $\mathcal{N} \left( 0, 4 \right)$ for the hyperprior on $ASC _{\textrm{bike}}$. The hyperprior for the variance is specified as $\ln \left[ \mathcal{N} \left( 0, 4 \right) \right]$, i.e. log-normal with a location parameter of zero and a scale parameter of 2 ($\sqrt{4} = 2$).

Moving to our priors for the parameters of the model trees, we need a prior distribution for $\left( D^m, \mid p_i ^m \mid, b_j ^{i, m} \right)$ for all $i \in \left\lbrace 1, 2, ..., D^m \right\rbrace$ and for all $j \in \left\lbrace 1, 2, ..., \mid p_i ^m \mid \right\rbrace$. To construct our prior, we precisely follow the methodology described in Section 2 of \citet{letham_2015_interpretable}. Unfortunately, this methodology took \citeauthor{letham_2015_interpretable} nearly four pages and much mathematical notation to describe. Additional pages would be needed to relate their original description to the characterization of decision trees that we have given in Sections \ref{sec:decision-tree-explanation} and \ref{sec:decision-tree-economics}. Since reviewing the techniques of \citet{letham_2015_interpretable} is not a primary focus of our article, we state upfront that the following description of our prior distribution of decision trees will be necessarily brief and will likely require a reader to consult \citet{letham_2015_interpretable} for full understanding. For readers who prefer reading code to reading verbal descriptions of our procedures, all scripts used in this application are available upon request.

Now, we begin with $D^m$, the number of output nodes (or conjunctive conditions) in our decision tree. Given that one of the arguments for non-compensatory rules is that humans are boundedly rational and only spend but so much mental effort making decisions, we do not think individuals are using overly complex rules. Our prior for $D^m$ was therefore specified as a truncated Poisson distribution with a rate parameter of 5, reflecting our prior belief that the expected number of output nodes in one's decision tree is approximately five. A truncated (as opposed to standard) Poisson distribution was used because the support of the standard Poisson distribution extends to positive infinity whereas the number of possible conjunctive conditions for the trees is limited by the finite number of possible requirements from which the conjunctive rules can be composed. See \citet[p. 1355]{letham_2015_interpretable} for the specific form of the truncated Poisson distribution and for more details on this prior specification.

Continuing to the next parameter, we have to specify a prior for $\mid p_i ^m \mid$: the number of requirements in each conjunctive condition. We will not delve into the details here, but the methods of \citet{letham_2015_interpretable} use a slightly different representation of decision trees than have been described in this paper. In their formulation, output nodes are evaluated sequentially, and $\mid p_i ^m \mid$ represents the number of requirements in output node $i$, conditional on the requirements of the previous nodes not being met. Given this set up, and given the assumption that people are using relatively simple rules to make their decisions, we specify our prior for $\mid p_i ^m \mid$ as a truncated Poisson distribution with a rate parameter of 2. In other words, besides the requirement of not meeting the conditions specified by the previous output nodes, we expect that a given output node will be described by approximately two requirements.

Next, we need prior distributions for the requirements ($b_j ^{i, m}$) that make up each conjunctive condition. We pause here to note that such prior distributions implicitly define a prior on the conjunctive conditions ($p_i$) that correspond to each output node. Alternatively, placing a prior directly on the conjunctive conditions ($p_i$) will implicitly define a prior on the requirements ($b_j ^{i, m}$) that make up these conditions. Following the procedures in \citet{letham_2015_interpretable}, we use a three-stage procedure to place a prior directly on the conjunctive conditions ($p_i$). First, we specify the possible requirements that a conjunctive condition can be composed of. These requirements are formed by discretizing the explanatory variables into various ranges (e.g. minimum distance greater than 4 miles). Second, we specify which of the possible combinations of requirements will be allowed as possible conjunctive conditions. And finally, we specify a prior distribution over the possible conjunctive conditions. We will discuss each of these three steps below.

To specify the possible requirements from which a conjunctive condition could be composed, our strategy\footnote{Note, we are aware that other strategies could have been used to discretize our variables in order to create requirements for use in the decision tree. Future researchers are free to use any such discretization strategies they prefer and to make a case for such strategies. We chose to follow the procedures of \citet{letham_2015_interpretable} who manually discretized their variables according to their a-priori beliefs.} was to subdivide the explanatory variables used to construct the decision tree into as many equal sized groups as possible. The only constraint was that the partition had to maintain the expected relationships between the groups and the outcome of bicycling or not. For example, the variable denoting the number of children was split into three groups: $\left[ 0, 1 \right]$, $\left( 1, 2 \right]$, and $\left( 2, \infty \right)$. For these categories of the number of children, the percentages of individuals in each category that owned a bicycle and actually bicycle to work or to school were approximately 16\%, 13\% and 0\%. Such trends follow our a-priori expectations that the probability of bicycling decreases as one has more children. Sub-dividing the number of children variable into 4 or more categories led to relationships that were deemed to be spurious since they did not match our a-priori beliefs about the relationship between number of children and the probability of bicycling commuting. All together, the possible requirements used to construct the decision tree were as follows (with numbers rounded to two decimal places, or more when necessary):
\begin{itemize}
\item Number of Kids: $\left[ 0, 1 \right]$, $\left[ 2 \right]$, and $\left[ 3, \infty \right)$

\item Minimum distance (miles): $\left[ 0, 1.17 \right]$, $\left( 1.17, 1.92 \right]$, $\left( 1.92, 3.00 \right]$, $\left( 3.00, 4.37 \right]$, $\left( 4.37, \infty \right)$

\item Average Speed Limit (miles per hour): $\left[ 23.01, 25.15 \right]$, $\left( 25.15, 25.78 \right]$, $\left( 25.78, \infty \right)$

\item Median Slope (meters per foot): $ \left[0, 0.01 \right]$, $\left( 0.01, 0.02 \right]$, $\left( 0.02, 0.03 \right]$, $\left( 0.03, 0.04 \right]$, $\left(0.04, \infty \right)$

\item Proportion of roadway miles along one's shortest path with speed limits $<$ 25 miles per hour: \newline $\left[ 0, 0.66 \right]$, $\left( 0.66, 0.83 \right]$, $\left( 0.83, 0.95 \right]$, $\left( 0.95, 0.9984 \right]$, $\left( 0.9984, 0.9986 \right]$, $\left( 0.9986, \infty \right)$

\item Proportion of roadway miles with bicycle lanes: $\left[ 0, 0.04 \right]$, $\left(0.04, 0.11 \right]$, $\left( 0.11, \infty \right)$

\item Proportion of roadway miles with ``share the road'' markings: $\left[ 0, 0.08 \right]$, $\left(0.08, 0.14 \right]$, $\left( 0.14, \infty \right)$

\end{itemize}
These requirements are all binary boolean conditions that are to be read as ``\textit{variable} in \textit{range}.'' For instance, ``minimum distance in [0, 1.17].'' 

Given the possible requirements specified above, the next task is to specify the combinations of these requirements that will be allowed as possible conjunctive conditions in our decision trees. Going along with the notion of non-compensatory rules are at least partially motivated by a desire to minimize cognitive effort, we hypothesize that no individual conjunctive condition will be made up of a large number of requirements. As a result, we specify the maximum number of requirements in a conjunctive condition to be 2. Moreover, since we are trying to estimate a decision tree that (by assumption) is used by our entire population, we limit the possible conjunctive conditions to those conjunctions that apply to a large percentage of the population. In particular, we only consider those conjunctive conditions that apply to (1) 10\% or more of those who bicycle or (2) 10\% or more of those who did not bicycle. As is done in \citet{letham_2015_interpretable}, we use the FP-growth algorithm implemented by \citet{borgelt_2005_implementation} to enumerate these conjunctive conditions.

Now, given the possible conjunctive conditions, our remaining task is to assign a prior probability to each of these conjunctive rule. Because we do not have any prior information about whether one conjunctive condition would be used more than any other, we use a uniform distribution as our prior. In particular, we follow Equation 2.2 of \citet{letham_2015_interpretable} and place a uniform distribution prior over all conjunctive conditions that (1) have $\mid p_i ^m \mid$ requirements and (2) have not already been used in the decision tree.

Lastly, in order to use the methods of \citet{letham_2015_interpretable}, we initially use the decision trees to predict the choice of bicycling or not, for those individuals who own a bicycle. This bicycle focused prediction is performed for two reasons. First, we are being agnostic (initially) about the presence of a more general choice model at the output nodes of the decision tree, and unlike our mode choice models, our trees are not constructed with the relevant variables for predicting all travel modes. Secondly, we focus on the choice of bicycling because the tree is will ultimately be used specifically to determine whether bicycle is considered or not and to what extent bicycling is generally preferred when it is considered. Either way, to make predictions about whether or not an individual bicycles to work or to school, the methods of \citet{letham_2015_interpretable} require us to specify a prior for the probability that an individual chooses to bicycle. For a fully unknown person, we chose our prior to express maximal ignorance about his/her probability of bicycling. Our prior for the probability that an individual commutes by bicycle is $\textrm{Beta} \left( 1, 1 \right)$: a uniform distribution over the range $\left( 0, 1 \right)$.

For further clarification of how we initially sampled the decision trees, see \citet{letham_2015_interpretable}.
\subsection{Results and Discussion}
\label{sec:results-and-discussion}

In this subsection, we discuss the results of our empirical application. Specifically, we compare the MNL model with our proposed bayesian model trees in four ways. We first quantitatively compare these two models in terms of in-sample fit. Note that we do not compare the two models in terms of out-of-sample predictive ability simply because our long estimation times and small sample size made both cross-validation and the use of a holdout sample unappealing. Moreover, it is well known that while frequentist estimation techniques such as maximum likelihood are prone to over-fitting, bayesian estimation techniques are much less likely to overfit, and bayesian model selection techniques automatically penalize model complexity (\citealp{dawid_2002_comment}; \citealp[Section 3.2]{wagenmakers_2008_bayesian}). Secondly, we qualitatively compare the two models in terms of their forecasted relationship between public investments in bicycle lanes and expected bicycle mode shares. Thirdly, in an attempt to uncover the model differences that lead to the divergent forecasts, we compare the estimation results of those coefficients that are common to the two models. Finally, we conclude this subsection by discussing the behavioral differences that lead to greater plausibility of our bayesian model tree forecasts as compared to the forecasts of the MNL model.

To begin, we start with the in-sample results. In a frequentist setting, models are commonly compared using log-likelihood ratios. For non-nested models, one might use the Vuong test (a generalization of the likelihood ratio test) to test which of two models is closer in terms of Kullback-Leibler divergence to the true data generating process \citep{vuong_1989_likelihood}. In a bayesian setting, the same interpretation can be given to the posterior probability of a model. Simply, the posterior probability of a model is the probability that, out of one's set of models, a given model is closest in terms of Kullback-Leibler divergence to the true data generating process \citep{walker_2013_bayesian}. Because we use the scalable rejection sampling algorithm of \citet{braun_2016_scalable}, we automatically get an estimate of $P \left( Y \mid X, m \right)$ for each of our ten model trees, and because we have the prior of each tree $m$, we can calculate $P \left( Y \mid X \right)$ given our bayesian model tree. Likewise, the scalable rejection sampling algorithm provides an estimate of $P \left( Y \mid X \right)$ for our MNL model. Combining these probabilities with prior probabilities of $\tfrac{1}{2}$ for each model (to reflect maximal uncertainty about which model is closest to the data generating process), we find that the posterior probability of our bayesian model trees is 99.91\% compared to the posterior probability of 0.09\% for the MNL model. In other words, based on our data, the bayesian model tree is overwhelmingly more likely to be closer to the true data generating process than the MNL model.

\begin{figure}
\centering
\includegraphics[width=0.7\textwidth]{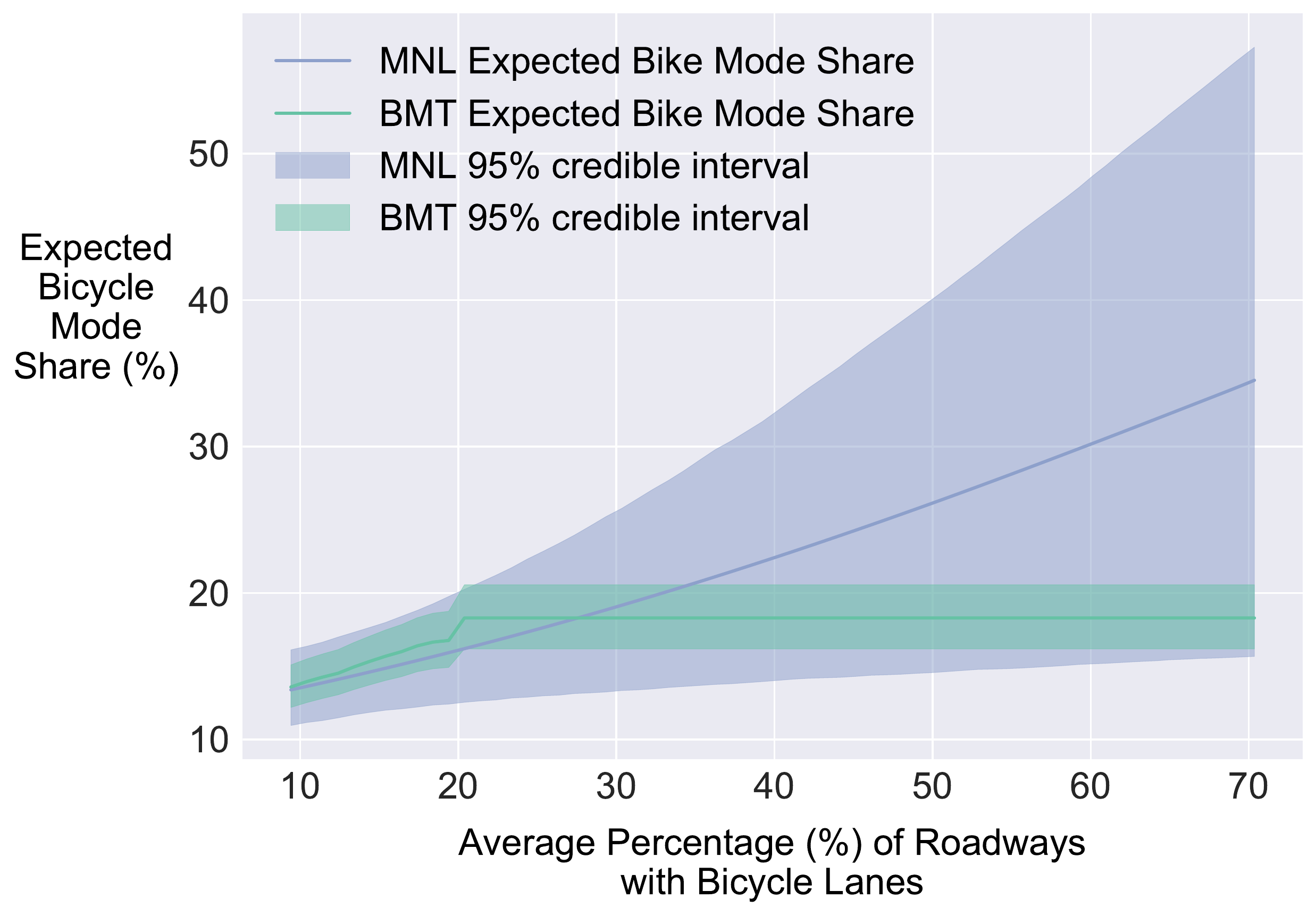}
\caption{Expected Bicycle Mode Share versus Mean Percentage of Bicycle Lanes}
\label{fig:mode-share-plot}
\end{figure}

Given that the bayesian model trees are likely to be a better representation of the true data generating process, we now turn to the question of whether this model leads to different policy implications as compared to the MNL model. For our policy application, we considered the effect of increasing the proportion of bicycle lanes for the individuals in our sample. In particular, we raised the proportion of bicycle lanes for each individual in our sample, one percentage point at a time, until each individual's proportion of bicycle lanes was approximately 70\% (the maximum value observed in our estimation dataset). After each incremental increase in the proportion of roadways with bicycle lanes, we predicted the average bicycle mode share across the dataset. In Figure \ref{fig:mode-share-plot} we plot the expected bicycle mode shares, as predicted by the bayesian model trees and the MNL model, along with their associated credible intervals\footnote{Note that credible intervals are the bayesian analog of confidence intervals}. Note that in this plot, we use the acronym ``BMT'' to refer to the bayesian model trees. Naming aside, Figure \ref{fig:mode-share-plot} shows that the two models lead to very different forecasts. In particular, as one begins to install bicycle lanes, both models show an increase in the expected bicycle mode share, but eventually, the bayesian model trees predicts that the expected bicycle mode share will flatline. In contrast, the MNL model predicts that the expected bicycle mode share will increase continually. A-priori, the predictions of the bayesian model trees appear more plausible than the predictions of the MNL model. In particular, we expect diminishing returns from increasing the proportion of roadways with bicycle lanes since individuals will eventually come to feel safe on public roadways but will still reject the bicycling alternative due to other factors such as time pressures due to childcare obligations, concerns about sweating, etc.

To corroborate our a-priori expectations, we note that in the United States (U.S.) and internationally, solely having many bicycle lanes does not lead to the huge bicycle mode shares predicted by the MNL model. For instance, take the case of Davis, California. Davis has lead the U.S. in the installation of on-street bicycle infrastructure. The first on-street bicycle lanes, the first bicycle traffic signal, and the first `protected intersection' for bicyclists were all installed in Davis \citep{caltrans_2017_toward}. Accordingly, out of all cities in the U.S. with populations of greater than 20,000 individuals, Davis has the highest bicycle commuting mode share. Depending on one's source, Davis' bicycle commuting mode share is 17\% - 19\% \citep{mckenzie_2014_modes, mcleod_2016_where}, a value that precisely matches the predictions of our bayesian model trees. Looking internationally, we can observe countries such as the Netherlands that lead the world in on-street bicycle infrastructure investments. Here, we are quick to note that bicycle infrastructure in the Netherlands is often of much higher quality than in the U.S. Bike lanes in the Netherlands are often `protected' in the sense that they are physically-separated from motor vehicles \citep{pucher_2008_making}. Additionally, the entire travel context in the Netherlands is more supportive of bicycling: fuel and automobile-ownership costs are much higher than in the U.S., more downtown areas are designated as automobile-free, local roads are often `traffic-calmed,' and travel by bicycle is often more direct than by automobile \citep{pucher_2008_making}. Even with all of these advantages, only about 36\% percent of all trips are taken by bicycle in the Netherlands \citep{directorate_2015_quality}. We are immediately sceptical of any model, such as our MNL model, that predicts a similar level of bicycling based solely on the installation of `unprotected' bicycle lanes (the only type of bicycle lane present in our study area at the time the data was collected).

Now, to investigate the causes of the differing forecasts shown in Figure \ref{fig:mode-share-plot}, we start with the estimated choice model coefficients ($\beta$) of the MNL model and our bayesian model trees. A summary of the posterior distribution of the choice model parameters for both the MNL and bayesian model trees is given in Table \ref{table:posterior-summaries}. Briefly, Table \ref{table:posterior-summaries} shows the posterior mean, the 2.5th percentile, and the 97.5th percentile of the posterior samples for each choice model parameter. To calculate the posterior summary for the bayesian model trees, we calculated a weighted posterior mean and weighted percentiles where the posterior samples of each choice model parameter, for each decision tree, were weighted by the posterior probability of that tree. To make the display manageable, Table \ref{table:posterior-summaries} only displays the parameters estimated in the MNL model. The parameters that are specific to the model trees, such as the bicycle ASCs that are specific to each output-node ($\textrm{ASC}_{\textrm{bike}, i}$), are not shown. Now, the main finding from Table \ref{table:posterior-summaries} is that the estimation results of parameters that are common to both models are very similar. The only parameter whose posterior mean shows large differences between the two models is $\textrm{ASC}_{\textrm{bike}}$, and this is because the $\textrm{ASC}_{\textrm{bike}}$ in the bayesian model tree plays a different role than it does in the MNL model. Recall that in the bayesian model tree, $\textrm{ASC}_{\textrm{bike}}$ is just the group mean that the output-node specific $\textrm{ASC}_{\textrm{bike}, i}$ are centered around. Details aside, knowing that the estimated choice models are the largely the same means that we should look towards the decision trees themselves to find out why the two models have such differing forecasts. This line of investigation is pursued below.

Behaviorally, we believe that four qualities of our bayesian model trees lead to its differing forecasts from the MNL model. The first quality we note is that the bicycle lane variable is almost never included in the conjunctive condition that splits the root node. In other words, the bicycle lane variable is almost never in the conditions at the top of our decision trees. In nine of our ten decision trees, there were nodes that filtered out individuals before they could get to an output node that depended on bicycle lanes. Furthermore, the one tree that had a bicycle lane requirement for the first output node actually had a low probability (0.2\%) of being the tree that is closest to representing the true data generating process. The behavioral interpretation of this finding is that bicycle lanes are not the most important variable in an individual's decision making process about whether or not to commute by bike. Variables that appear to take precedence over bicycle lanes when deciding whether or not to commute by bike include topography, the number of children an individual has, and the minimum distance between an individual's home and work/school. As a result, the impact of installing bicycle lanes will be moderated by these other variables.

The second quality that we note about our bayesian model trees is that the bicycle lane variable never appeared by itself. In particular, when the proportion of roadways containing bicycle lanes was present in a conjunctive condition, it always appeared alongside another variable. For instance, in seven out of ten decision trees, the bicycle lane variable appeared in the following conjunctive condition: `proportion of roadways with bicycle lanes is greater than 0.11 and the number of children is 0 or 1.' The posterior probability that the true data generating process was most closely represented by one of these seven trees was over 99\%. Such a finding emphasizes that the proportion of roadways containing bicycle lanes is not always important. In particular, if a person has 2 or more children, then bicycle lanes are unlikely to affect whether the individual commutes by bicycle. Presumably, childcare pressures will be a bigger determining factor of those individuals' choice of travel mode.

\begin{landscape}
\begin{table}
\centering

\caption{Posterior Summaries of Choice Model Parameters}
\label{table:posterior-summaries}

\begin{tabular}{K{0.45\linewidth}rrrrrr}
\toprule
{} & \multicolumn{3}{c}{MNL} & \multicolumn{3}{c}{Bayesian Model Trees}\\
Variables & \multicolumn{1}{c}{Mean} & \multicolumn{1}{c}{2.5\%} & \multicolumn{1}{c}{97.5\%} & \multicolumn{1}{c}{Mean} & \multicolumn{1}{c}{2.5\%} & \multicolumn{1}{c}{97.5\%} \tabularnewline
\midrule

\multicolumn{7}{l}{Alternative Specific Constants}\\
\quad Shared Ride: 2 & -1.746*\hphantom{*} & -2.282 & -1.231 & -1.787*\hphantom{*} & -2.322 & -1.259\\
\quad Shared Ride: 3+ & -1.865*\hphantom{*} & -2.378 & -1.345 & -1.909*\hphantom{*} & -2.453 & -1.378\\
\quad Walk-Transit-Walk & 1.070*\hphantom{*} & 0.495 & 1.623 & 1.078*\hphantom{*} & 0.497 & 1.666\\
\quad Drive-Transit-Walk & -2.183*\hphantom{*} & -2.969 & -1.431 & -2.223*\hphantom{*} & -3.060 & -1.418\\
\quad Walk-Transit-Drive & -2.677*\hphantom{*} & -3.537 & -1.886 & -2.734*\hphantom{*} & -3.602 & -1.910\\
\quad Walk & 2.428*\hphantom{*} & 1.826 & 3.042 & 2.504*\hphantom{*} & 1.885 & 3.149\\
\quad Bike & 1.087\hphantom{*}\hphantom{*} & -2.573 & 4.854 & 0.171\hphantom{*}\hphantom{*} & -3.419 & 3.738\\

\multicolumn{7}{l}{Travel Time, units:0.1min}\\
\quad All Auto Modes & -1.113*\hphantom{*} & -1.358 & -0.875 & -1.129*\hphantom{*} & -1.383 & -0.894\\
\quad All Transit Modes & -0.374*\hphantom{*} & -0.479 & -0.273 & -0.378*\hphantom{*} & -0.486 & -0.276\\

\multicolumn{7}{l}{Travel Cost, units:\$}\\
\quad All Transit Modes & -0.173*\hphantom{*} & -0.300 & -0.053 & -0.173*\hphantom{*} & -0.301 & -0.048\\

\multicolumn{7}{l}{Travel Distance, units:mi}\\
\quad Walk & -1.125*\hphantom{*} & -1.299 & -0.960 & -1.151*\hphantom{*} & -1.337 & -0.979\\
\quad Bike & -0.242\hphantom{*}\hphantom{*} & -0.503 & 0.052 & -0.356*\hphantom{*} & -0.469 & -0.245\\

\multicolumn{7}{l}{Systematic Heterogeneity}\\
\quad Autos per licensed drivers (All Auto Modes) & 1.181*\hphantom{*} & 0.784 & 1.582 & 1.221*\hphantom{*} & 0.822 & 1.621\\
\quad Cross-Bay Tour (Shared Ride 2 \& 3+) & -0.517\hphantom{*}\hphantom{*} & -1.264 & 0.167 & -0.518\hphantom{*}\hphantom{*} & -1.286 & 0.201\\
\quad Household Size (Shared Ride 2 \& 3+) & 0.108\hphantom{*}\hphantom{*} & -0.097 & 0.311 & 0.127\hphantom{*}\hphantom{*} & -0.079 & 0.330\\
\quad Number of Kids in Household (Shared Ride 2 \& 3+) & 0.662*\hphantom{*} & 0.414 & 0.903 & 0.634*\hphantom{*} & 0.394 & 0.892\\

\multicolumn{7}{l}{Spatial Variables}\\
\quad Minimum Distance units:mi (Bike) & -0.232\hphantom{*}\hphantom{*} & -0.822 & 0.294 & \_ & \_ & \_\\
\quad Median Slope units:m/ft (Bike) & -1.433\hphantom{*}\hphantom{*} & -5.046 & 2.304 & \_ & \_ & \_\\
\quad Mean Speed Limit units:mph (Bike) & -0.064\hphantom{*}\hphantom{*} & -0.206 & 0.072 & \_ & \_ & \_\\
\quad Proportion of Shortest Path Roads slower than 25mph (Bike) & 0.484\hphantom{*}\hphantom{*} & -0.701 & 1.791 & \_ & \_ & \_\\
\quad Proportion of Roadways with Bike Lanes (Bike) & 2.218*\hphantom{*} & 0.363 & 4.048 & \_ & \_ & \_\\
\quad Proportion of Roadways with Bicycle Chevrons (Bike) & -0.765\hphantom{*}\hphantom{*} & -2.647 & 1.108 & \_ & \_ & \_\\

\bottomrule
\multicolumn{7}{l}{Note: * means the equal-tailed 95\% credible interval excludes zero.}\tabularnewline
\multicolumn{7}{l}{\qquad \hphantom{*} Additionally, bicycle chevrons are another name for ``share the road'' arrows.}
\end{tabular}

\end{table}
\end{landscape}

Third, we point out that decision trees, by their very nature, incorporate a notion of threshold effects. These threshold effects can be seen in our application by the fact that all of our decision trees with posterior probabilities of greater than 0.2\% all feature the requirement that the `proportion of roadways with bicycle lanes is greater than 0.11.' Undoubtedly, the presence of this sharp discontinuity at 0.11 is partly an artifact of our discretization methods. However, as described in Section \ref{sec:dtree-variants-heterogeneity}, even when using soft decision trees that don't implement such ``hard'' thresholds, the interpretation is that individuals do use hard thresholds, but we are merely uncertain about what those hard thresholds are. Either way, the presence of these threshold effects leads to two features of our forecasts. First, the threshold effects lead to the discrete jump in the expected bicycle mode share when the average percentage of all roadways containing bicycle lanes is about 20\%. At this point, almost everyone's proportion of roadways with bicycle lanes rises above 0.11, so all individuals now belong to output nodes with the highest chance of bicycling. Secondly, the threshold effects also cause the flatline in expected bicycle mode share. Because further increases in the proportion of roadways with bicycle lanes do not cause any more changes in the output node's of an individual, there are no more changes in the probability that an individual chooses to bike.

Finally, the last major difference between the forecasts of the bayesian model trees and the MNL model is that as the average percentage of roadways with bicycle lanes increases, the variance in the expected bicycle mode share increases for the MNL model but not for the bayesian model trees. This finding is perhaps best explained mathematically. Whether operating in a frequentist or bayesian setting, the parameters of one's choice model will have an associated probability distribution. In a frequentist setting, this will be the sampling distribution of $\hat{\beta}$ and in a bayesian setting, this will be the posterior distribution of $\beta$. For ease of exposition, we will continue our discussion from a bayesian perspective, but our explanation is equally valid from a frequentist perspective. Since the MNL model multiplies the proportion of roadways with bicycle lanes ($X_{\textrm{bike-lanes}}$) by $\beta_{\textrm{bike-lanes}}$, we can calculate the variance of the product as $\textrm{Var}  \left[ X_{\textrm{bike-lanes}} \beta_{\textrm{bike-lanes}} \right] = X_{\textrm{bike-lanes}} ^2 \textrm{Var}  \left[ \beta_{\textrm{bike-lanes}} \right]$. This means that as $X_{\textrm{bike-lanes}}$ increases, the variance of $X_{\textrm{bike-lanes}} \beta_{\textrm{bike-lanes}}$ increases. Because the MNL's probability that an individual commutes by bicycle is dependent on $X_{\textrm{bike-lanes}} \beta_{\textrm{bike-lanes}}$, the increase in the variance of $X_{\textrm{bike-lanes}} \beta_{\textrm{bike-lanes}}$ leads to an increase in the variance of the probability that an individual commutes by bicycle. Aggregated over all individuals, the increases in the variances of the bicycle probabilities lead to an increase in the variance of the expected bicycle mode share.

In contrast to the process described above, changing the value of $X_{\textrm{bike-lanes}}$ when forecasting with the bayesian model trees only changes what output node one falls into for a given decision tree. The structure of the trees remains unchanged, the posterior probabilities across the trees remains unchanged, and the variance of the $\textrm{ASC}_{\textrm{bike}, i}$ remain mostly constant across output nodes that have bicycle available as an alternative. As a result, the variance of the expected bicycle mode share remains mostly constant as $X_{\textrm{bike-lanes}}$ is increased for the various individuals in our dataset. Behaviorally, these differences in forecast uncertainty can be attributed to the fact that when using a compensatory model, one is uncertain about the extent to which the proportion of roadways with bicycle lanes compensates for the other variables that affect one's probability of bicycling. In other words, one is uncertain about the value of $\hat{\beta}_{\textrm{bike-lanes}}$ or $\beta_{\textrm{bike-lanes}}$. Since $X_{\textrm{bike-lanes}}$, only appears in the non-compensatory portion of our bayesian model trees (i.e. in the decision trees as opposed to the choice model), we are always certain that the bike lane proportion does not compensate for other variables. I.e. our bayesian model trees are based on the assumption that $\beta_{\textrm{bike-lanes}} = 0$. This constant level of uncertainty with respect to the compensatory nature of $X_{\textrm{bike-lanes}}$ leads directly to the constant level of forecast uncertainty for the bayesian model trees in Figure \ref{fig:mode-share-plot}.

\section{Conclusion}
\label{sec:conclusion}
In this paper, we have made three contributions to the literature. First, we have provided a micro-economic framework for interpreting a class of machine learning models known as decision trees. In particular, we reviewed how decision trees are used and estimated (Section \ref{sec:decision-tree-explanation}), we showed how decision trees represent a non-compensatory decision protocol known as disjunctions-of-conjunctions (Section \ref{sec:decision-tree-economics}), and we discussed how existing decision tree variants can account for economic considerations that discrete choice modelers are likely to have (Section \ref{sec:decision-tree-variants}).

Secondly, we contributed to both the discrete choice and decision tree literatures by formulating and estimating the first bayesian model tree: a semi-compensatory, two-stage model of human decision making. Our model uses a non-compensatory, disjunctions-of-conjunctions protocol to determine one's choice set, and conditional on a given choice set, it uses a compensatory discrete choice model (e.g. an MNL model) to make a final selection if more than one alternative is available. Beyond one's choice set, our bayesian model tree allows the non-compensatory rules to influence one's preferences, as embodied in the choice model parameters, and our model allows for quantification of one's uncertainty over which set of disjunctions-of-conjunctions are actually being used in a population. To the best of our knowledge, this is the first time a bayesian model tree has ever been proposed and estimated.

Finally, beyond the mere proposition of the bayesian model tree, our paper carried out an empirical application of this model. We made three major findings. First, our proposed bayesian model tree is more than 1,000-times more likely ($\frac{99.01}{0.09} \approx 1,100$) to be closer to our application's true data-generating process than the MNL model. Second, our bayesian model trees provide forecasts that are consistent with observed bicycle mode shares in areas with abundant bike lanes such as Davis, CA and the Netherlands. In comparison, the forecasts of the MNL model were overly optimistic. Third, our bayesian model trees provide insights that are qualitatively different than the MNL model. Specifically, our bayesian model trees suggest that (1) investments in on-street bicycle lanes will eventually suffer from diminishing returns and (2) that factors such as travel distance, child-related pressures, and topography may all prevent individuals from bicycling even if there are many bicycle lanes. These insights are missing from the more traditional (and compensatory) MNL model.

Moving forward, we note that in the decades after McFadden revealed the economic implications of the conditional logit model, discrete choice modelers moved swiftly to create needed extensions. As a result, we can now avoid many of the troubling assumptions and properties of the conditional logit model, leading to more accurate analyses and more sensible policy implications. Analogously, by linking decision trees to economics, our paper brings decision trees to a similar, infantile stage. As noted in Section \ref{sec:dtree-combinatorics}, there remain a number of economic concerns (or more specifically, combinations of concerns) that must be confronted before decision trees will be maximally useful in policy settings. By detailing the microeconomic implications of decision trees, we aim to draw the attention of choice modelers. Hopefully, our paper will encourage the use and extension of current decision tree methodologies, thereby increasing the accuracy and usefulness of such models for policy analyses.

\section{Acknowledgements}
We thank Paul Waddell, the MacArthur Foundation, and UCCONNECT's Dissertation Grant for funding this research. We also thank Feras El Zarwi, Sreeta Gorripaty, and Madeleine Sheehan for their helpful discussions in the beginning stages of this research endeavor.

\newpage
\section*{\refname}
\bibliography{timothyb_dtree_v2}

\begin{thebibliography}{149}
\expandafter\ifx\csname natexlab\endcsname\relax\def\natexlab#1{#1}\fi
\providecommand{\url}[1]{\texttt{#1}}
\providecommand{\href}[2]{#2}
\providecommand{\path}[1]{#1}
\providecommand{\DOIprefix}{doi:}
\providecommand{\ArXivprefix}{arXiv:}
\providecommand{\URLprefix}{URL: }
\providecommand{\Pubmedprefix}{pmid:}
\providecommand{\doi}[1]{\href{http://dx.doi.org/#1}{\path{#1}}}
\providecommand{\Pubmed}[1]{\href{pmid:#1}{\path{#1}}}
\providecommand{\bibinfo}[2]{#2}
\ifx\xfnm\relax \def\xfnm[#1]{\unskip,\space#1}\fi
\bibitem[{Allenby and Rossi(1999)}]{allenby_1999_marketing}
\bibinfo{author}{Allenby, G.M.}, \bibinfo{author}{Rossi, P.E.},
  \bibinfo{year}{1999}.
\newblock \bibinfo{title}{Marketing models of consumer heterogeneity}.
\newblock \bibinfo{journal}{Journal of Econometrics} \bibinfo{volume}{89},
  \bibinfo{pages}{57--78}.
\bibitem[{Angelino et~al.(2017)Angelino, Larus-Stone, Alabi, Seltzer and
  Rudin}]{angelino_2017_learning}
\bibinfo{author}{Angelino, E.}, \bibinfo{author}{Larus-Stone, N.},
  \bibinfo{author}{Alabi, D.}, \bibinfo{author}{Seltzer, M.},
  \bibinfo{author}{Rudin, C.}, \bibinfo{year}{2017}.
\newblock \bibinfo{title}{Learning certifiably optimal rule lists}, in:
  \bibinfo{booktitle}{Proceedings of the 23rd ACM SIGKDD International
  Conference on Knowledge Discovery and Data Mining},
  \bibinfo{organization}{ACM}. pp. \bibinfo{pages}{35--44}.
\bibitem[{Arentze and Timmermans(2004)}]{arentze_2004_learning}
\bibinfo{author}{Arentze, T.A.}, \bibinfo{author}{Timmermans, H.J.},
  \bibinfo{year}{2004}.
\newblock \bibinfo{title}{A learning-based transportation oriented simulation
  system}.
\newblock \bibinfo{journal}{Transportation Research Part B: Methodological}
  \bibinfo{volume}{38}, \bibinfo{pages}{613--633}.
\bibitem[{Arentze and Timmermans(2007)}]{arentze_2007_parametric}
\bibinfo{author}{Arentze, T.A.}, \bibinfo{author}{Timmermans, H.J.},
  \bibinfo{year}{2007}.
\newblock \bibinfo{title}{Parametric action decision trees: Incorporating
  continuous attribute variables into rule-based models of discrete choice}.
\newblock \bibinfo{journal}{Transportation Research Part B: Methodological}
  \bibinfo{volume}{41}, \bibinfo{pages}{772--783}.
\bibitem[{Bafumi and Gelman(2006)}]{bafumi_2006_fitting}
\bibinfo{author}{Bafumi, J.}, \bibinfo{author}{Gelman, A.},
  \bibinfo{year}{2006}.
\newblock \bibinfo{title}{Fitting multilevel models when predictors and group
  effects correlate}.
\newblock \bibinfo{type}{Technical Report}.
\newblock \URLprefix \url{http://dx.doi.org/10.2139/ssrn.1010095}.
\bibitem[{Bajari et~al.(2015a)Bajari, Nekipelov, Ryan and
  Yang}]{bajari_2015_demand}
\bibinfo{author}{Bajari, P.}, \bibinfo{author}{Nekipelov, D.},
  \bibinfo{author}{Ryan, S.P.}, \bibinfo{author}{Yang, M.},
  \bibinfo{year}{2015}a.
\newblock \bibinfo{title}{Demand estimation with machine learning and model
  combination}.
\newblock \bibinfo{type}{Technical Report}. National Bureau of Economic
  Research.
\bibitem[{Bajari et~al.(2015b)Bajari, Nekipelov, Ryan and
  Yang}]{bajari_2015_machine}
\bibinfo{author}{Bajari, P.}, \bibinfo{author}{Nekipelov, D.},
  \bibinfo{author}{Ryan, S.P.}, \bibinfo{author}{Yang, M.},
  \bibinfo{year}{2015}b.
\newblock \bibinfo{title}{Machine learning methods for demand estimation}.
\newblock \bibinfo{journal}{The American Economic Review}
  \bibinfo{volume}{105}, \bibinfo{pages}{481--485}.
\bibitem[{Barros et~al.(2012)Barros, Basgalupp, De~Carvalho and
  Freitas}]{barros_2012_survey}
\bibinfo{author}{Barros, R.C.}, \bibinfo{author}{Basgalupp, M.P.},
  \bibinfo{author}{De~Carvalho, A.C.}, \bibinfo{author}{Freitas, A.A.},
  \bibinfo{year}{2012}.
\newblock \bibinfo{title}{A survey of evolutionary algorithms for decision-tree
  induction}.
\newblock \bibinfo{journal}{IEEE Transactions on Systems, Man, and Cybernetics,
  Part C (Applications and Reviews)} \bibinfo{volume}{42},
  \bibinfo{pages}{291--312}.
\bibitem[{Ben-Akiva(1973)}]{ben_1973_structure}
\bibinfo{author}{Ben-Akiva, M.E.}, \bibinfo{year}{1973}.
\newblock \bibinfo{title}{Structure of passenger travel demand models}.
\newblock \bibinfo{publisher}{Ph.D. Thesis, Dept. of Civil Engineering,
  Massachusetts Institute of Technology}, \bibinfo{address}{Cambridge, MA}.
\bibitem[{Bhat(1998)}]{bhat_1998_accommodating}
\bibinfo{author}{Bhat, C.R.}, \bibinfo{year}{1998}.
\newblock \bibinfo{title}{Accommodating variations in responsiveness to
  level-of-service measures in travel mode choice modeling}.
\newblock \bibinfo{journal}{Transportation Research Part A: Policy and
  Practice} \bibinfo{volume}{32}, \bibinfo{pages}{495--507}.
\bibitem[{Bhat(2015)}]{bhat2015comprehensive}
\bibinfo{author}{Bhat, C.R.}, \bibinfo{year}{2015}.
\newblock \bibinfo{title}{A comprehensive dwelling unit choice model
  accommodating psychological constructs within a search strategy for
  consideration set formation}.
\newblock \bibinfo{journal}{Transportation Research Part B: Methodological}
  \bibinfo{volume}{79}, \bibinfo{pages}{161--188}.
\bibitem[{Bishop(2006)}]{bishop_2006_pattern}
\bibinfo{author}{Bishop, C.M.}, \bibinfo{year}{2006}.
\newblock \bibinfo{title}{Pattern recognition and machine learning}.
\newblock \bibinfo{publisher}{Springer}.
\bibitem[{Bishop and Svens{\'e}n(2003)}]{bishop_2003_bayesian}
\bibinfo{author}{Bishop, C.M.}, \bibinfo{author}{Svens{\'e}n, M.},
  \bibinfo{year}{2003}.
\newblock \bibinfo{title}{{Bayesian Hierarchical Mixtures of Experts}}, in:
  \bibinfo{booktitle}{Proceedings of the Nineteenth conference on Uncertainty
  in Artificial Intelligence}, \bibinfo{organization}{Morgan Kaufmann
  Publishers Inc.}. pp. \bibinfo{pages}{57--64}.
\bibitem[{Borgelt(2005)}]{borgelt_2005_implementation}
\bibinfo{author}{Borgelt, C.}, \bibinfo{year}{2005}.
\newblock \bibinfo{title}{{An implementation of the FP-growth algorithm}}, in:
  \bibinfo{booktitle}{{OSDM'05 Proceedings of the 1st International Workshop on
  Open Source Data Mining: Frequent Pattern Mining Implementations}},
  \bibinfo{publisher}{ACM, New York}. pp. \bibinfo{pages}{1--5}.
\bibitem[{Brathwaite(2018)}]{brathwaite_2018_holy}
\bibinfo{author}{Brathwaite, T.}, \bibinfo{year}{2018}.
\newblock \bibinfo{title}{{The Holy Trinity: Blending Statistics, Machine
  Learning and Discrete Choice with Applications to Strategic Bicycle
  Planning}}.
\newblock \bibinfo{publisher}{Ph.D. Thesis, Dept. of Civil Engineering,
  University of California, Berkeley}, \bibinfo{address}{Berkeley, CA}.
\bibitem[{Braun and Damien(2016)}]{braun_2016_scalable}
\bibinfo{author}{Braun, M.}, \bibinfo{author}{Damien, P.},
  \bibinfo{year}{2016}.
\newblock \bibinfo{title}{Scalable rejection sampling for bayesian hierarchical
  models}.
\newblock \bibinfo{journal}{Marketing Science} \bibinfo{volume}{35},
  \bibinfo{pages}{427--444}.
\bibitem[{Breiman(2001)}]{breiman_2001_random}
\bibinfo{author}{Breiman, L.}, \bibinfo{year}{2001}.
\newblock \bibinfo{title}{Random forests}.
\newblock \bibinfo{journal}{Machine Learning} \bibinfo{volume}{45},
  \bibinfo{pages}{5--32}.
\bibitem[{Bronner(1982)}]{bronner_1982_decision}
\bibinfo{author}{Bronner, A.}, \bibinfo{year}{1982}.
\newblock \bibinfo{title}{Decision styles in transport mode choice}.
\newblock \bibinfo{journal}{Journal of Economic Psychology}
  \bibinfo{volume}{2}, \bibinfo{pages}{81--101}.
\bibitem[{B{\"u}hlmann and Hothorn(2007)}]{buhlmann_2007_boosting}
\bibinfo{author}{B{\"u}hlmann, P.}, \bibinfo{author}{Hothorn, T.},
  \bibinfo{year}{2007}.
\newblock \bibinfo{title}{{Boosting Algorithms: Regularization, Prediction and
  Model Fitting} (with discussion)}.
\newblock \bibinfo{journal}{Statistical Science} \bibinfo{volume}{22},
  \bibinfo{pages}{477--505}.
\bibitem[{{California Department of
  Transportation}(2013)}]{california_department_of_transportation_2010-2012_2013}
\bibinfo{author}{{California Department of Transportation}},
  \bibinfo{year}{2013}.
\newblock \bibinfo{title}{2010-2012 {California} {Household} {Travel} {Survey}
  {Final} {Report}}.
\newblock \bibinfo{type}{Technical Report}.
\newblock \URLprefix \url{http://www.dot.ca.gov/hq/tsip/FinalReport.pdf}.
\bibitem[{Caltrans(2017)}]{caltrans_2017_toward}
\bibinfo{author}{Caltrans}, \bibinfo{year}{2017}.
\newblock \bibinfo{title}{Toward An Active California: State Bicycle +
  Pedestrian Plan}.
\newblock \bibinfo{type}{Technical Report}. California State Department of
  Transportation.
\newblock \URLprefix
  \url{http://www.dot.ca.gov/activecalifornia/documents/Hi-Res_Final_ActiveCA.pdf}.
\bibitem[{Cantillo and de~Dios~Ort{\'u}zar(2005)}]{cantillo_2005_semi}
\bibinfo{author}{Cantillo, V.}, \bibinfo{author}{de~Dios~Ort{\'u}zar, J.},
  \bibinfo{year}{2005}.
\newblock \bibinfo{title}{A semi-compensatory discrete choice model with
  explicit attribute thresholds of perception}.
\newblock \bibinfo{journal}{Transportation Research Part B: Methodological}
  \bibinfo{volume}{39}, \bibinfo{pages}{641--657}.
\bibitem[{Cantillo et~al.(2006)Cantillo, Heydecker and
  de~Dios~Ort{\'u}zar}]{cantillo2006discrete}
\bibinfo{author}{Cantillo, V.}, \bibinfo{author}{Heydecker, B.},
  \bibinfo{author}{de~Dios~Ort{\'u}zar, J.}, \bibinfo{year}{2006}.
\newblock \bibinfo{title}{A discrete choice model incorporating thresholds for
  perception in attribute values}.
\newblock \bibinfo{journal}{Transportation Research Part B: Methodological}
  \bibinfo{volume}{40}, \bibinfo{pages}{807--825}.
\bibitem[{Cascetta and Papola(2009)}]{cascetta2009dominance}
\bibinfo{author}{Cascetta, E.}, \bibinfo{author}{Papola, A.},
  \bibinfo{year}{2009}.
\newblock \bibinfo{title}{Dominance among alternatives in random utility
  models}.
\newblock \bibinfo{journal}{Transportation Research Part A: Policy and
  Practice} \bibinfo{volume}{43}, \bibinfo{pages}{170--179}.
\bibitem[{Chan and Loh(2004)}]{chan_2004_lotus}
\bibinfo{author}{Chan, K.}, \bibinfo{author}{Loh, W.}, \bibinfo{year}{2004}.
\newblock \bibinfo{title}{{LOTUS: An Algorithm for Building Accurate and
  Comprehensible Logistic Regression Trees}}.
\newblock \bibinfo{journal}{Journal of Computational and Graphical Statistics}
  \bibinfo{volume}{13}, \bibinfo{pages}{826--852}.
\bibitem[{Chipman et~al.(1998)Chipman, George and
  McCulloch}]{chipman_1998_bayesian}
\bibinfo{author}{Chipman, H.A.}, \bibinfo{author}{George, E.I.},
  \bibinfo{author}{McCulloch, R.E.}, \bibinfo{year}{1998}.
\newblock \bibinfo{title}{{Bayesian CART model search (with discussion)}}.
\newblock \bibinfo{journal}{Journal of the American Statistical Association}
  \bibinfo{volume}{93}, \bibinfo{pages}{935--960}.
\bibitem[{Chorus(2012)}]{chorus_2012_random}
\bibinfo{author}{Chorus, C.}, \bibinfo{year}{2012}.
\newblock \bibinfo{title}{Random regret minimization: an overview of model
  properties and empirical evidence}.
\newblock \bibinfo{journal}{Transport Reviews} \bibinfo{volume}{32},
  \bibinfo{pages}{75--92}.
\bibitem[{Cleland and Walton(2004)}]{cleland_2004_why}
\bibinfo{author}{Cleland, B.S.}, \bibinfo{author}{Walton, D.},
  \bibinfo{year}{2004}.
\newblock \bibinfo{title}{Why don{\textquoteright}t people walk and cycle}.
\newblock \bibinfo{type}{Technical Report} \bibinfo{number}{528007}. Central
  Laboratories. \bibinfo{address}{New Zealand}.
\newblock \URLprefix
  \url{http://can.org.nz/system/files/Why%20dont%20people%20walk%20and%20cycle.pdf}.
\bibitem[{Commission and
  Brinckerhoff(2012)}]{san_francisco_metropolitan_transportation_commission_travel_2012}
\bibinfo{author}{Commission, S.F.M.T.}, \bibinfo{author}{Brinckerhoff, P.},
  \bibinfo{year}{2012}.
\newblock \bibinfo{title}{Travel {Model} {Development}: {Calibration} and
  {Validation}}.
\newblock \bibinfo{type}{Technical Report}. San Francisco Metropolitan
  Transportaton Commission.
\newblock \URLprefix
  \url{http://mtcgis.mtc.ca.gov/foswiki/pub/Main/Documents/2012_05_18_RELEASE_DRAFT_Calibration_and_Validation.pdf}.
\bibitem[{Conlisk(1996)}]{conlisk_1996_why}
\bibinfo{author}{Conlisk, J.}, \bibinfo{year}{1996}.
\newblock \bibinfo{title}{Why bounded rationality?}
\newblock \bibinfo{journal}{Journal of Economic Literature}
  \bibinfo{volume}{34}, \bibinfo{pages}{669—--700}.
\bibitem[{Coombs(1951)}]{coombs_1951_mathematical}
\bibinfo{author}{Coombs, C.H.}, \bibinfo{year}{1951}.
\newblock \bibinfo{title}{Mathematical models in psychological scaling}.
\newblock \bibinfo{journal}{Journal of the American Statistical Association}
  \bibinfo{volume}{46}, \bibinfo{pages}{480--489}.
\bibitem[{Cox(1966)}]{cox_1966_some}
\bibinfo{author}{Cox, D.R.}, \bibinfo{year}{1966}.
\newblock \bibinfo{title}{Some procedures connected with the logistic
  qualitative response curve}, in: \bibinfo{editor}{David, F.N.} (Ed.),
  \bibinfo{booktitle}{Research Papers in Statistics}, \bibinfo{publisher}{John
  Wiley \& Sons, New York}. pp. \bibinfo{pages}{55--71}.
\bibitem[{Das and Bhattacharya(2017)}]{das_2017_transdimensional}
\bibinfo{author}{Das, M.}, \bibinfo{author}{Bhattacharya, S.},
  \bibinfo{year}{2017}.
\newblock \bibinfo{title}{{Transdimensional Transformation based Markov Chain
  Monte Carlo}}.
\newblock \bibinfo{journal}{arXiv preprint arXiv:1403.5207v5} .
\bibitem[{Dawes(1964)}]{dawes_1964_social}
\bibinfo{author}{Dawes, R.M.}, \bibinfo{year}{1964}.
\newblock \bibinfo{title}{Social selection based on multidimensional criteria.}
\newblock \bibinfo{journal}{The Journal of Abnormal and Social Psychology}
  \bibinfo{volume}{68}, \bibinfo{pages}{104}.
\bibitem[{Dawid(2002)}]{dawid_2002_comment}
\bibinfo{author}{Dawid, A.P.}, \bibinfo{year}{2002}.
\newblock \bibinfo{title}{Comment on bayesian measures of complexity and fit}.
\newblock \bibinfo{journal}{{Journal of the Royal Statistics Society: Series B
  (Methodological)}} \bibinfo{volume}{64}, \bibinfo{pages}{583--639}.
\bibitem[{Denison et~al.(1998)Denison, Mallick and
  Smith}]{denison_1998_bayesian}
\bibinfo{author}{Denison, D.G.}, \bibinfo{author}{Mallick, B.K.},
  \bibinfo{author}{Smith, A.F.}, \bibinfo{year}{1998}.
\newblock \bibinfo{title}{A bayesian cart algorithm}.
\newblock \bibinfo{journal}{Biometrika} \bibinfo{volume}{85},
  \bibinfo{pages}{363--377}.
\bibitem[{Einav and Levin(2014)}]{einav_2014_data}
\bibinfo{author}{Einav, L.}, \bibinfo{author}{Levin, J.}, \bibinfo{year}{2014}.
\newblock \bibinfo{title}{The data revolution and economic analysis}.
\newblock \bibinfo{journal}{Innovation Policy and the Economy}
  \bibinfo{volume}{14}, \bibinfo{pages}{1--24}.
\bibitem[{Elrod et~al.(2004)Elrod, Johnson and White}]{elrod_2004_new}
\bibinfo{author}{Elrod, T.}, \bibinfo{author}{Johnson, R.D.},
  \bibinfo{author}{White, J.}, \bibinfo{year}{2004}.
\newblock \bibinfo{title}{A new integrated model of noncompensatory and
  compensatory decision strategies}.
\newblock \bibinfo{journal}{Organizational Behavior and Human Decision
  Processes} \bibinfo{volume}{95}, \bibinfo{pages}{1--19}.
\bibitem[{Esposito et~al.(1997)Esposito, Malerba, Semeraro and
  Kay}]{esposito_1997_comparative}
\bibinfo{author}{Esposito, F.}, \bibinfo{author}{Malerba, D.},
  \bibinfo{author}{Semeraro, G.}, \bibinfo{author}{Kay, J.},
  \bibinfo{year}{1997}.
\newblock \bibinfo{title}{A comparative analysis of methods for pruning
  decision trees}.
\newblock \bibinfo{journal}{IEEE transactions on pattern analysis and machine
  intelligence} \bibinfo{volume}{19}, \bibinfo{pages}{476--491}.
\bibitem[{Fan and Sisson(2011)}]{fan_2011_reversible}
\bibinfo{author}{Fan, Y.}, \bibinfo{author}{Sisson, S.A.},
  \bibinfo{year}{2011}.
\newblock \bibinfo{title}{{Reversible Jump MCMC}}, in: \bibinfo{editor}{Brooks,
  S.}, \bibinfo{editor}{Gelman, A.}, \bibinfo{editor}{Jones, G.L.},
  \bibinfo{editor}{Meng, X.L.} (Eds.), \bibinfo{booktitle}{{Handbook of Markov
  Chain Monte Carlo}}. \bibinfo{publisher}{Chapman \& Hall/CRC},
  \bibinfo{address}{New York}, pp. \bibinfo{pages}{67--91}.
\bibitem[{Fern{\'a}ndez-Delgado et~al.(2014)Fern{\'a}ndez-Delgado, Cernadas,
  Barro and Amorim}]{fernandez_2014_do}
\bibinfo{author}{Fern{\'a}ndez-Delgado, M.}, \bibinfo{author}{Cernadas, E.},
  \bibinfo{author}{Barro, S.}, \bibinfo{author}{Amorim, D.},
  \bibinfo{year}{2014}.
\newblock \bibinfo{title}{Do we need hundreds of classifiers to solve real
  world classification problems?}
\newblock \bibinfo{journal}{Journal of Machine Learning Research}
  \bibinfo{volume}{15}, \bibinfo{pages}{3133--3181}.
\bibitem[{Foerster(1979)}]{foerster1979mode}
\bibinfo{author}{Foerster, J.F.}, \bibinfo{year}{1979}.
\newblock \bibinfo{title}{Mode choice decision process models: a comparison of
  compensatory and non-compensatory structures}.
\newblock \bibinfo{journal}{Transportation Research Part A: General}
  \bibinfo{volume}{13}, \bibinfo{pages}{17--28}.
\bibitem[{Friedman et~al.(2008)Friedman, Hastie and
  Tibshirani}]{friedman_2008_elements}
\bibinfo{author}{Friedman, J.}, \bibinfo{author}{Hastie, T.},
  \bibinfo{author}{Tibshirani, R.}, \bibinfo{year}{2008}.
\newblock \bibinfo{title}{{The Elements of Statistical Learning: Data Mining,
  Inference, and Prediction}}.
\newblock \bibinfo{edition}{2} ed., \bibinfo{publisher}{{Springer Series in
  Statistics}}.
\bibitem[{Gelman(1992)}]{gelman_1992_iterative}
\bibinfo{author}{Gelman, A.}, \bibinfo{year}{1992}.
\newblock \bibinfo{title}{Iterative and non-iterative simulation algorithms}.
\newblock \bibinfo{journal}{{Computing Science and Statistics}}
  \bibinfo{volume}{24}, \bibinfo{pages}{433--438}.
\bibitem[{Gelman(2006)}]{gelman_2006_multilevel}
\bibinfo{author}{Gelman, A.}, \bibinfo{year}{2006}.
\newblock \bibinfo{title}{Multilevel (hierarchical) modeling: what it can and
  cannot do}.
\newblock \bibinfo{journal}{Technometrics} \bibinfo{volume}{48},
  \bibinfo{pages}{432--435}.
\bibitem[{Gelman et~al.(2014)Gelman, Carlin, Stern, Dunson, Vehtari and
  Rubin}]{gelman_2014_bayesian}
\bibinfo{author}{Gelman, A.}, \bibinfo{author}{Carlin, J.B.},
  \bibinfo{author}{Stern, H.S.}, \bibinfo{author}{Dunson, D.B.},
  \bibinfo{author}{Vehtari, A.}, \bibinfo{author}{Rubin, D.B.},
  \bibinfo{year}{2014}.
\newblock \bibinfo{title}{Bayesian data analysis}. volume~\bibinfo{volume}{3}.
\newblock \bibinfo{publisher}{CRC press Boca Raton, FL}.
\bibitem[{Gigerenzer and Goldstein(1996)}]{gigerenzer_1996_reasoning}
\bibinfo{author}{Gigerenzer, G.}, \bibinfo{author}{Goldstein, D.G.},
  \bibinfo{year}{1996}.
\newblock \bibinfo{title}{Reasoning the fast and frugal way: models of bounded
  rationality.}
\newblock \bibinfo{journal}{Psychological review} \bibinfo{volume}{103},
  \bibinfo{pages}{650}.
\bibitem[{Gilbride and Allenby(2004)}]{gilbride_2004_choice}
\bibinfo{author}{Gilbride, T.J.}, \bibinfo{author}{Allenby, G.M.},
  \bibinfo{year}{2004}.
\newblock \bibinfo{title}{A choice model with conjunctive, disjunctive, and
  compensatory screening rules}.
\newblock \bibinfo{journal}{Marketing Science} \bibinfo{volume}{23},
  \bibinfo{pages}{391--406}.
\bibitem[{Goldsmith(1992)}]{goldsmith_1992_reasons}
\bibinfo{author}{Goldsmith, S.A.}, \bibinfo{year}{1992}.
\newblock \bibinfo{title}{Reasons why bicycling and walking are and are not
  being used more extensively as travel modes}.
\newblock \bibinfo{number}{1}, \bibinfo{publisher}{Federal Highway
  Administration}.
\newblock \URLprefix \url{http://safety.fhwa.dot.gov/ped_bike/docs/case1.pdf}.
\bibitem[{Gonz{\'a}lez et~al.(2015)Gonz{\'a}lez, Herrera and
  Garc{\'i}a}]{gonzalez_2015_monotonic}
\bibinfo{author}{Gonz{\'a}lez, S.}, \bibinfo{author}{Herrera, F.},
  \bibinfo{author}{Garc{\'i}a, S.}, \bibinfo{year}{2015}.
\newblock \bibinfo{title}{{Monotonic Random Forest with an Ensemble Pruning
  Mechanism based on the Degree of Monotonicity}}.
\newblock \bibinfo{journal}{New Generation Computing} \bibinfo{volume}{33},
  \bibinfo{pages}{367--388}.
\bibitem[{Gordon and Olshen(1980)}]{gordon_1980_consistent}
\bibinfo{author}{Gordon, L.}, \bibinfo{author}{Olshen, R.A.},
  \bibinfo{year}{1980}.
\newblock \bibinfo{title}{Consistent nonparametric regression from recursive
  partitioning schemes}.
\newblock \bibinfo{journal}{Journal of Multivariate Analysis}
  \bibinfo{volume}{10}, \bibinfo{pages}{611--627}.
\bibitem[{Gordon and Olshen(1984)}]{gordon_1984_almost}
\bibinfo{author}{Gordon, L.}, \bibinfo{author}{Olshen, R.A.},
  \bibinfo{year}{1984}.
\newblock \bibinfo{title}{Almost surely consistent nonparametric regression
  from recursive partitioning schemes}.
\newblock \bibinfo{journal}{Journal of Multivariate Analysis}
  \bibinfo{volume}{15}, \bibinfo{pages}{147--163}.
\bibitem[{Green et~al.(1988)Green, Krieger and Bansal}]{green_1988_completely}
\bibinfo{author}{Green, P.E.}, \bibinfo{author}{Krieger, A.M.},
  \bibinfo{author}{Bansal, P.}, \bibinfo{year}{1988}.
\newblock \bibinfo{title}{Completely unacceptable levels in conjoint analysis:
  A cautionary note}.
\newblock \bibinfo{journal}{Journal of Marketing Research} ,
  \bibinfo{pages}{293--300}.
\bibitem[{Green(1995)}]{green_1995_reversible}
\bibinfo{author}{Green, P.J.}, \bibinfo{year}{1995}.
\newblock \bibinfo{title}{{Reversible jump Markov chain Monte Carlo computation
  and Bayesian model determination}}.
\newblock \bibinfo{journal}{Biometrika} \bibinfo{volume}{82},
  \bibinfo{pages}{711--732}.
\bibitem[{Hansen(1982)}]{hansen_1982_large}
\bibinfo{author}{Hansen, L.P.}, \bibinfo{year}{1982}.
\newblock \bibinfo{title}{Large sample properties of generalized method of
  moments estimators}.
\newblock \bibinfo{journal}{Econometrica} \bibinfo{volume}{50},
  \bibinfo{pages}{1029--1054}.
\bibitem[{Hauser et~al.(2010)Hauser, Toubia, Evgeniou, Befurt and
  Dzyabura}]{hauser_2010_disjunctions}
\bibinfo{author}{Hauser, J.R.}, \bibinfo{author}{Toubia, O.},
  \bibinfo{author}{Evgeniou, T.}, \bibinfo{author}{Befurt, R.},
  \bibinfo{author}{Dzyabura, D.}, \bibinfo{year}{2010}.
\newblock \bibinfo{title}{Disjunctions of conjunctions, cognitive simplicity,
  and consideration sets}.
\newblock \bibinfo{journal}{Journal of Marketing Research}
  \bibinfo{volume}{47}, \bibinfo{pages}{485--496}.
\bibitem[{Hess et~al.(2012)Hess, Stathopoulos and Daly}]{hess2012allowing}
\bibinfo{author}{Hess, S.}, \bibinfo{author}{Stathopoulos, A.},
  \bibinfo{author}{Daly, A.}, \bibinfo{year}{2012}.
\newblock \bibinfo{title}{Allowing for heterogeneous decision rules in discrete
  choice models: an approach and four case studies}.
\newblock \bibinfo{journal}{Transportation} \bibinfo{volume}{39},
  \bibinfo{pages}{565--591}.
\bibitem[{Hesterberg(1995)}]{hesterberg_1995_weighted}
\bibinfo{author}{Hesterberg, T.}, \bibinfo{year}{1995}.
\newblock \bibinfo{title}{Weighted average importance sampling and defensive
  mixture distributions}.
\newblock \bibinfo{journal}{{Technometrics}} \bibinfo{volume}{37},
  \bibinfo{pages}{185--194}.
\bibitem[{Hu et~al.(2012)Hu, Che, Zhang, Zhang, Guo and Yu}]{hu_2012_rank}
\bibinfo{author}{Hu, Q.}, \bibinfo{author}{Che, X.}, \bibinfo{author}{Zhang,
  L.}, \bibinfo{author}{Zhang, D.}, \bibinfo{author}{Guo, M.},
  \bibinfo{author}{Yu, D.}, \bibinfo{year}{2012}.
\newblock \bibinfo{title}{{Rank Entropy Based Decision Trees for Monotonic
  Classification}}.
\newblock \bibinfo{journal}{IEEE Transactions on Knowledge and Data
  Engineering} \bibinfo{volume}{24}, \bibinfo{pages}{2052--2064}.
\bibitem[{Huber and Klein(1991)}]{huber_1991_adapting}
\bibinfo{author}{Huber, J.}, \bibinfo{author}{Klein, N.M.},
  \bibinfo{year}{1991}.
\newblock \bibinfo{title}{Adapting cutoffs to the choice environment: the
  effects of attribute correlation and reliability}.
\newblock \bibinfo{journal}{Journal of Consumer Research} \bibinfo{volume}{18},
  \bibinfo{pages}{346--357}.
\bibitem[{Ittner and Schlosser(1996)}]{ittner_1996_non}
\bibinfo{author}{Ittner, A.}, \bibinfo{author}{Schlosser, M.},
  \bibinfo{year}{1996}.
\newblock \bibinfo{title}{Non-linear decision trees-ndt}, in:
  \bibinfo{booktitle}{ICML}, \bibinfo{organization}{Citeseer}. pp.
  \bibinfo{pages}{252--257}.
\bibitem[{Jang(1994)}]{jang_1994_structure}
\bibinfo{author}{Jang, J.S.}, \bibinfo{year}{1994}.
\newblock \bibinfo{title}{Structure determination in fuzzy modeling: a fuzzy
  cart approach}, in: \bibinfo{booktitle}{Fuzzy Systems, 1994. IEEE World
  Congress on Computational Intelligence., Proceedings of the Third IEEE
  Conference on}, \bibinfo{organization}{IEEE}. pp. \bibinfo{pages}{480--485}.
\bibitem[{Jedidi and Kohli(2005)}]{jedidi_2005_probabilistic}
\bibinfo{author}{Jedidi, K.}, \bibinfo{author}{Kohli, R.},
  \bibinfo{year}{2005}.
\newblock \bibinfo{title}{Probabilistic subset-conjunctive models for
  heterogeneous consumers}.
\newblock \bibinfo{journal}{Journal of Marketing Research}
  \bibinfo{volume}{42}, \bibinfo{pages}{483--494}.
\bibitem[{Jordan and Jacobs(1994)}]{jordan_1994_hierarchical}
\bibinfo{author}{Jordan, M.I.}, \bibinfo{author}{Jacobs, R.A.},
  \bibinfo{year}{1994}.
\newblock \bibinfo{title}{{Hierarchical Mixtures of Experts}}.
\newblock \bibinfo{journal}{Neural Computation} \bibinfo{volume}{6},
  \bibinfo{pages}{181--214}.
\bibitem[{Kamakura et~al.(1996)Kamakura, Kim and Lee}]{kamakura_1996_modeling}
\bibinfo{author}{Kamakura, W.A.}, \bibinfo{author}{Kim, B.D.},
  \bibinfo{author}{Lee, J.}, \bibinfo{year}{1996}.
\newblock \bibinfo{title}{Modeling preference and structural heterogeneity in
  consumer choice}.
\newblock \bibinfo{journal}{Marketing Science} \bibinfo{volume}{15},
  \bibinfo{pages}{152--172}.
\bibitem[{Kaplan et~al.(2009)Kaplan, Bekhor and Shiftan}]{kaplan_2009_two}
\bibinfo{author}{Kaplan, S.}, \bibinfo{author}{Bekhor, S.},
  \bibinfo{author}{Shiftan, Y.}, \bibinfo{year}{2009}.
\newblock \bibinfo{title}{Two-stage model for jointly revealing determinants of
  noncompensatory conjunctive choice set formation and compensatory choice}.
\newblock \bibinfo{journal}{Transportation Research Record: Journal of the
  Transportation Research Board} , \bibinfo{pages}{153--163}.
\bibitem[{Kaplan and Prato(2012)}]{kaplan_2012_closing}
\bibinfo{author}{Kaplan, S.}, \bibinfo{author}{Prato, C.G.},
  \bibinfo{year}{2012}.
\newblock \bibinfo{title}{Closing the gap between behavior and models in route
  choice: The role of spatiotemporal constraints and latent traits in choice
  set formation}.
\newblock \bibinfo{journal}{Transportation Research Part F: traffic psychology
  and behaviour} \bibinfo{volume}{15}, \bibinfo{pages}{9--24}.
\bibitem[{Kaplan et~al.(2012)Kaplan, Shiftan and
  Bekhor}]{kaplan_2012_development}
\bibinfo{author}{Kaplan, S.}, \bibinfo{author}{Shiftan, Y.},
  \bibinfo{author}{Bekhor, S.}, \bibinfo{year}{2012}.
\newblock \bibinfo{title}{Development and estimation of a semi-compensatory
  model with a flexible error structure}.
\newblock \bibinfo{journal}{Transportation Research Part B: Methodological}
  \bibinfo{volume}{46}, \bibinfo{pages}{291--304}.
\bibitem[{Kim and Kim(2011)}]{kim_2011_two}
\bibinfo{author}{Kim, J.H.}, \bibinfo{author}{Kim, M.}, \bibinfo{year}{2011}.
\newblock \bibinfo{title}{{Two-Stage Multinomial Logit Model}}.
\newblock \bibinfo{journal}{Expert Systems with Applications}
  \bibinfo{volume}{38}, \bibinfo{pages}{6439--6446}.
\bibitem[{Kim(2009)}]{kim_2009_two}
\bibinfo{author}{Kim, M.}, \bibinfo{year}{2009}.
\newblock \bibinfo{title}{{Two-Stage Logistic Regression Model}}.
\newblock \bibinfo{journal}{Expert Systems with Applications}
  \bibinfo{volume}{36}, \bibinfo{pages}{6727--6734}.
\bibitem[{Kindermann and Paass(1998)}]{kindermann_1998_model}
\bibinfo{author}{Kindermann, J.}, \bibinfo{author}{Paass, G.},
  \bibinfo{year}{1998}.
\newblock \bibinfo{title}{Model switching for bayesian classification trees
  with soft splits}.
\newblock \bibinfo{journal}{Principles of Data Mining and Knowledge Discovery}
  , \bibinfo{pages}{148--157}.
\bibitem[{Kohli and Jedidi(2007)}]{kohli2007representation}
\bibinfo{author}{Kohli, R.}, \bibinfo{author}{Jedidi, K.},
  \bibinfo{year}{2007}.
\newblock \bibinfo{title}{Representation and inference of lexicographic
  preference models and their variants}.
\newblock \bibinfo{journal}{Marketing Science} \bibinfo{volume}{26},
  \bibinfo{pages}{380--399}.
\bibitem[{Kumar et~al.(2016)Kumar, Viswanath and Rao}]{kumar_2016_ensemble}
\bibinfo{author}{Kumar, G.K.}, \bibinfo{author}{Viswanath, P.},
  \bibinfo{author}{Rao, A.A.}, \bibinfo{year}{2016}.
\newblock \bibinfo{title}{{Ensemble of randomized soft decision trees for
  robust classification}}.
\newblock \bibinfo{journal}{Sadhana} \bibinfo{volume}{41},
  \bibinfo{pages}{273--282}.
\bibitem[{Landwehr et~al.(2005)Landwehr, Hall and
  Frank}]{landwehr_2005_logistic}
\bibinfo{author}{Landwehr, N.}, \bibinfo{author}{Hall, M.},
  \bibinfo{author}{Frank, E.}, \bibinfo{year}{2005}.
\newblock \bibinfo{title}{{Logistic Model Trees}}.
\newblock \bibinfo{journal}{Machine Learning} \bibinfo{volume}{59},
  \bibinfo{pages}{161--205}.
\bibitem[{Lemon et~al.(2003)Lemon, Roy, Clark, Friedmann and
  Rakowski}]{lemon_2003_classification}
\bibinfo{author}{Lemon, S.C.}, \bibinfo{author}{Roy, J.},
  \bibinfo{author}{Clark, M.A.}, \bibinfo{author}{Friedmann, P.D.},
  \bibinfo{author}{Rakowski, W.}, \bibinfo{year}{2003}.
\newblock \bibinfo{title}{Classification and regression tree analysis in public
  health: methodological review and comparison with logistic regression}.
\newblock \bibinfo{journal}{Annals of behavioral medicine}
  \bibinfo{volume}{26}, \bibinfo{pages}{172--181}.
\bibitem[{Leong and Hensher(2012)}]{leong_2012_embedding}
\bibinfo{author}{Leong, W.}, \bibinfo{author}{Hensher, D.A.},
  \bibinfo{year}{2012}.
\newblock \bibinfo{title}{Embedding decision heuristics in discrete choice
  models: A review}.
\newblock \bibinfo{journal}{Transport Reviews} \bibinfo{volume}{32},
  \bibinfo{pages}{313--331}.
\bibitem[{Letham et~al.(2015)Letham, Rudin, McCormick and
  Madigan}]{letham_2015_interpretable}
\bibinfo{author}{Letham, B.}, \bibinfo{author}{Rudin, C.},
  \bibinfo{author}{McCormick, T.H.}, \bibinfo{author}{Madigan, D.},
  \bibinfo{year}{2015}.
\newblock \bibinfo{title}{Interpretable classifiers using rules and bayesian
  analysis: Building a better stroke prediction model}.
\newblock \bibinfo{journal}{The Annals of Applied Statistics}
  \bibinfo{volume}{9}, \bibinfo{pages}{1350--1371}.
\bibitem[{Loh(2011)}]{loh_2011_classification}
\bibinfo{author}{Loh, W.Y.}, \bibinfo{year}{2011}.
\newblock \bibinfo{title}{Classification and regression trees}.
\newblock \bibinfo{journal}{Wiley Interdisciplinary Reviews: Data Mining and
  Knowledge Discovery} \bibinfo{volume}{1}, \bibinfo{pages}{14--23}.
\bibitem[{Loh(2014)}]{loh_2014_fifty}
\bibinfo{author}{Loh, W.Y.}, \bibinfo{year}{2014}.
\newblock \bibinfo{title}{Fifty years of classification and regression trees}.
\newblock \bibinfo{journal}{International Statistical Review}
  \bibinfo{volume}{82}, \bibinfo{pages}{329—--348}.
\bibitem[{Lomax and Vadera(2013)}]{lomax_2013_survey}
\bibinfo{author}{Lomax, S.}, \bibinfo{author}{Vadera, S.},
  \bibinfo{year}{2013}.
\newblock \bibinfo{title}{A survey of cost-sensitive decision tree induction
  algorithms}.
\newblock \bibinfo{journal}{ACM Computing Surveys (CSUR)} \bibinfo{volume}{45},
  \bibinfo{pages}{16}.
\bibitem[{Mahmoud et~al.(2016)Mahmoud, Weiss and Habib}]{mahmoud_2016_myopic}
\bibinfo{author}{Mahmoud, M.S.}, \bibinfo{author}{Weiss, A.},
  \bibinfo{author}{Habib, K.N.}, \bibinfo{year}{2016}.
\newblock \bibinfo{title}{{Myopic choice or rational decision making? An
  investigation into mode choice preference structures in competitive modal
  arrangements in a multimodal urban area, the City of Toronto}}.
\newblock \bibinfo{journal}{Canadian Journal of Civil Engineering}
  \bibinfo{volume}{43}, \bibinfo{pages}{420--428}.
\bibitem[{Manski(1977)}]{manski_1977_structure}
\bibinfo{author}{Manski, C.F.}, \bibinfo{year}{1977}.
\newblock \bibinfo{title}{The structure of random utility models}.
\newblock \bibinfo{journal}{Theory and decision} \bibinfo{volume}{8},
  \bibinfo{pages}{229--254}.
\bibitem[{Manski(2001)}]{manski_2001_daniel}
\bibinfo{author}{Manski, C.F.}, \bibinfo{year}{2001}.
\newblock \bibinfo{title}{Daniel mcfadden and the econometric analysis of
  discrete choice}.
\newblock \bibinfo{journal}{The Scandinavian Journal of Economics}
  \bibinfo{volume}{103}, \bibinfo{pages}{217--229}.
\bibitem[{Marsala and Petturiti(2015)}]{marsala_2015_rank}
\bibinfo{author}{Marsala, C.}, \bibinfo{author}{Petturiti, D.},
  \bibinfo{year}{2015}.
\newblock \bibinfo{title}{{Rank discrimination measures for enforcing
  monotonicity in decision tree induction}}.
\newblock \bibinfo{journal}{Information Sciences} \bibinfo{volume}{291},
  \bibinfo{pages}{143--171}.
\bibitem[{Marschak(1960)}]{marschak_1960_binary}
\bibinfo{author}{Marschak, J.}, \bibinfo{year}{1960}.
\newblock \bibinfo{title}{Binary-choice constraints and random utility
  indicators}, in: \bibinfo{booktitle}{Stanford Symposium on Mathematical
  Methods in the Social Sciences}, \bibinfo{publisher}{Stanford University
  Press}, \bibinfo{address}{Stanford, California}.
\bibitem[{Mart{\'\i}nez et~al.(2009)Mart{\'\i}nez, Aguila and
  Hurtubia}]{martinez_2009_constrained}
\bibinfo{author}{Mart{\'\i}nez, F.}, \bibinfo{author}{Aguila, F.},
  \bibinfo{author}{Hurtubia, R.}, \bibinfo{year}{2009}.
\newblock \bibinfo{title}{The constrained multinomial logit: A
  semi-compensatory choice model}.
\newblock \bibinfo{journal}{Transportation Research Part B: Methodological}
  \bibinfo{volume}{43}, \bibinfo{pages}{365--377}.
\bibitem[{McFadden(2001)}]{mcfadden_2001_economic}
\bibinfo{author}{McFadden, D.}, \bibinfo{year}{2001}.
\newblock \bibinfo{title}{Economic choices}.
\newblock \bibinfo{journal}{The American Economic Review} \bibinfo{volume}{91},
  \bibinfo{pages}{351--378}.
\bibitem[{McKenzie(2014)}]{mckenzie_2014_modes}
\bibinfo{author}{McKenzie, B.}, \bibinfo{year}{2014}.
\newblock \bibinfo{title}{{Modes less traveled---Bicycling and walking to work
  in the United States: 2008-2012}}.
\newblock \bibinfo{type}{Technical Report}. United States Census Bureau.
  \bibinfo{address}{Suitland, MD}.
\bibitem[{McLeod(2016)}]{mcleod_2016_where}
\bibinfo{author}{McLeod, K.}, \bibinfo{year}{2016}.
\newblock \bibinfo{title}{{Where We Ride: Analysis of bicycle commuting in
  American Cities}}.
\newblock \bibinfo{type}{Technical Report}. League of American Bicyclists.
\bibitem[{Meila and Jordan(2000)}]{meila_2000_learning}
\bibinfo{author}{Meila, M.}, \bibinfo{author}{Jordan, M.I.},
  \bibinfo{year}{2000}.
\newblock \bibinfo{title}{Learning with mixtures of trees}.
\newblock \bibinfo{journal}{Journal of Machine Learning Research}
  \bibinfo{volume}{1}, \bibinfo{pages}{1--48}.
\bibitem[{Mingers(1989)}]{mingers_1989_empirical}
\bibinfo{author}{Mingers, J.}, \bibinfo{year}{1989}.
\newblock \bibinfo{title}{An empirical comparison of pruning methods for
  decision tree induction}.
\newblock \bibinfo{journal}{Machine Learning} \bibinfo{volume}{4},
  \bibinfo{pages}{227--243}.
\bibitem[{Minka(2002)}]{minka_2002_bayesian}
\bibinfo{author}{Minka, T.P.}, \bibinfo{year}{2002}.
\newblock \bibinfo{title}{Bayesian model averaging is not model combination}.
\newblock \URLprefix
  \url{https://pdfs.semanticscholar.org/e30a/1d14dd097608583d6c200b43fada35dac444.pdf}.
\bibitem[{Mohammadi and Kaptein(2016)}]{mohammadi_2016_comment}
\bibinfo{author}{Mohammadi, A.}, \bibinfo{author}{Kaptein, M.},
  \bibinfo{year}{2016}.
\newblock \bibinfo{title}{{Comment on ``Efficient Metropolis--Hastings proposal
  mechanisms for Bayesian regression tree models''}}.
\newblock \bibinfo{journal}{Bayesian analysis} \bibinfo{volume}{11},
  \bibinfo{pages}{938--940}.
\bibitem[{Murphy(2012)}]{murphy_2012_machine}
\bibinfo{author}{Murphy, K.P.}, \bibinfo{year}{2012}.
\newblock \bibinfo{title}{Machine learning: a probabilistic perspective}.
\newblock \bibinfo{publisher}{MIT press}.
\bibitem[{Murthy(1998)}]{murthy_1998_automatic}
\bibinfo{author}{Murthy, S.K.}, \bibinfo{year}{1998}.
\newblock \bibinfo{title}{Automatic construction of decision trees from data: A
  multi-disciplinary survey}.
\newblock \bibinfo{journal}{Data mining and knowledge discovery}
  \bibinfo{volume}{2}, \bibinfo{pages}{345--389}.
\bibitem[{Murthy et~al.(1994)Murthy, Kasif and Salzberg}]{murthy_1994_system}
\bibinfo{author}{Murthy, S.K.}, \bibinfo{author}{Kasif, S.},
  \bibinfo{author}{Salzberg, S.}, \bibinfo{year}{1994}.
\newblock \bibinfo{title}{A system for induction of oblique decision trees}.
\newblock \bibinfo{journal}{Journal of Artificial Intelligence Research}
  \bibinfo{volume}{2}, \bibinfo{pages}{1--32}.
\bibitem[{Newton and Raftery(1994)}]{newton_1994_approximate}
\bibinfo{author}{Newton, M.A.}, \bibinfo{author}{Raftery, A.E.},
  \bibinfo{year}{1994}.
\newblock \bibinfo{title}{Approximate bayesian inference with the weighted
  likelihood bootstrap}.
\newblock \bibinfo{journal}{Journal of the Royal Statistical Society. Series B
  (Methodological)} , \bibinfo{pages}{3--48}.
\bibitem[{Olaru and Wehenkel(2003)}]{olaru_2003_complete}
\bibinfo{author}{Olaru, C.}, \bibinfo{author}{Wehenkel, L.},
  \bibinfo{year}{2003}.
\newblock \bibinfo{title}{A complete fuzzy decision tree technique}.
\newblock \bibinfo{journal}{Fuzzy sets and systems} \bibinfo{volume}{138},
  \bibinfo{pages}{221--254}.
\bibitem[{Papaspiliopoulos et~al.(2007)Papaspiliopoulos, Robers and
  Sk{\"o}ld}]{papaspiliopoulos_2007_general}
\bibinfo{author}{Papaspiliopoulos, O.}, \bibinfo{author}{Robers, G.O.},
  \bibinfo{author}{Sk{\"o}ld, M.}, \bibinfo{year}{2007}.
\newblock \bibinfo{title}{{A General Framework for the Parametrization of
  Hierarchical Models}}.
\newblock \bibinfo{journal}{Statistical Science} \bibinfo{volume}{22},
  \bibinfo{pages}{59--73}.
\bibitem[{Pei et~al.(2016)Pei, Hu and Chen}]{pei_2016_multivariate}
\bibinfo{author}{Pei, S.}, \bibinfo{author}{Hu, Q.}, \bibinfo{author}{Chen,
  C.}, \bibinfo{year}{2016}.
\newblock \bibinfo{title}{{Multivariate decision trees with monotonicity
  constraints}}.
\newblock \bibinfo{journal}{Knowledge-Based Systems} \bibinfo{volume}{112},
  \bibinfo{pages}{14--25}.
\bibitem[{Potharst and Feelders(2002)}]{potharst_2002_classification}
\bibinfo{author}{Potharst, R.}, \bibinfo{author}{Feelders, A.},
  \bibinfo{year}{2002}.
\newblock \bibinfo{title}{{Classification trees for problems with monotonicity
  constraints}}.
\newblock \bibinfo{journal}{SIGKDD Explorations Newsletter}
  \bibinfo{volume}{4}, \bibinfo{pages}{1--10}.
\bibitem[{Pratola(2016)}]{pratola_2016_efficient}
\bibinfo{author}{Pratola, M.T.}, \bibinfo{year}{2016}.
\newblock \bibinfo{title}{Efficient metropolis--hastings proposal mechanisms
  for bayesian regression tree models}.
\newblock \bibinfo{journal}{Bayesian Analysis} \bibinfo{volume}{11},
  \bibinfo{pages}{885--911}.
\bibitem[{Pucher and Buehler(2008)}]{pucher_2008_making}
\bibinfo{author}{Pucher, J.}, \bibinfo{author}{Buehler, R.},
  \bibinfo{year}{2008}.
\newblock \bibinfo{title}{Making cycling irresistible: lessons from the
  netherlands, denmark and germany}.
\newblock \bibinfo{journal}{Transport reviews} \bibinfo{volume}{28},
  \bibinfo{pages}{495--528}.
\bibitem[{Quinlan(1990)}]{quinlan_1990_probabilistic}
\bibinfo{author}{Quinlan, J.R.}, \bibinfo{year}{1990}.
\newblock \bibinfo{title}{Probabilistic decision trees}.
\newblock \bibinfo{journal}{{Machine Learning: an artificial intelligence
  approach}} \bibinfo{volume}{3}, \bibinfo{pages}{140--152}.
\bibitem[{Rivest(1987)}]{rivest_1987_learning}
\bibinfo{author}{Rivest, R.L.}, \bibinfo{year}{1987}.
\newblock \bibinfo{title}{Learning decision lists}.
\newblock \bibinfo{journal}{Machine Learning} \bibinfo{volume}{2},
  \bibinfo{pages}{229--246}.
\bibitem[{Rokach(2010)}]{rokach_2010_ensemble}
\bibinfo{author}{Rokach, L.}, \bibinfo{year}{2010}.
\newblock \bibinfo{title}{Ensemble-based classifiers}.
\newblock \bibinfo{journal}{Artificial Intelligence Review}
  \bibinfo{volume}{33}, \bibinfo{pages}{1--39}.
\bibitem[{Rokach and Maimon(2005)}]{rokach_2005_top}
\bibinfo{author}{Rokach, L.}, \bibinfo{author}{Maimon, O.},
  \bibinfo{year}{2005}.
\newblock \bibinfo{title}{Top-down induction of decision trees classifiers-a
  survey}.
\newblock \bibinfo{journal}{IEEE Transactions on Systems, Man, and Cybernetics,
  Part C (Applications and Reviews)} \bibinfo{volume}{35},
  \bibinfo{pages}{476--487}.
\bibitem[{Rokach and Maimon(2014)}]{rokach_2014_data}
\bibinfo{author}{Rokach, L.}, \bibinfo{author}{Maimon, O.},
  \bibinfo{year}{2014}.
\newblock \bibinfo{title}{Data mining with decision trees: theory and
  applications}.
\newblock \bibinfo{publisher}{World Scientific}.
\bibitem[{Rubin(1981)}]{rubin_1981_bayesian}
\bibinfo{author}{Rubin, D.B.}, \bibinfo{year}{1981}.
\newblock \bibinfo{title}{The bayesian bootstrap}.
\newblock \bibinfo{journal}{The Annals of Statistics} \bibinfo{volume}{9},
  \bibinfo{pages}{130--134}.
\bibitem[{Ruggieri(2017)}]{ruggieri_2017_enumerating}
\bibinfo{author}{Ruggieri, S.}, \bibinfo{year}{2017}.
\newblock \bibinfo{title}{Enumerating distinct decision trees}, in:
  \bibinfo{editor}{Precup, D.}, \bibinfo{editor}{Teh, Y.W.} (Eds.),
  \bibinfo{booktitle}{Proceedings of the 34th International Conference on
  Machine Learning}, \bibinfo{publisher}{PMLR}, \bibinfo{address}{International
  Convention Centre, Sydney, Australia}. pp. \bibinfo{pages}{2960--2968}.
\newblock \URLprefix \url{http://proceedings.mlr.press/v70/ruggieri17a.html}.
\bibitem[{Rusch and Zeilis(2013)}]{rusch_2013_gaining}
\bibinfo{author}{Rusch, T.}, \bibinfo{author}{Zeilis, A.},
  \bibinfo{year}{2013}.
\newblock \bibinfo{title}{{Gaining insight with recursive partitioning of
  generalized linear models}}.
\newblock \bibinfo{journal}{Journal of Statistical Computation and Simulation}
  \bibinfo{volume}{83}, \bibinfo{pages}{1301--1315}.
\bibitem[{Seyedhosseini and Tasdizen(2015)}]{seyedhosseini_2015_disjunctive}
\bibinfo{author}{Seyedhosseini, M.}, \bibinfo{author}{Tasdizen, T.},
  \bibinfo{year}{2015}.
\newblock \bibinfo{title}{{Disjunctive normal random forests}}.
\newblock \bibinfo{journal}{Pattern Recognition} \bibinfo{volume}{48},
  \bibinfo{pages}{976--983}.
\bibitem[{Simon(1955)}]{simon_1955_behavioral}
\bibinfo{author}{Simon, H.A.}, \bibinfo{year}{1955}.
\newblock \bibinfo{title}{A behavioral model of rational choice}.
\newblock \bibinfo{journal}{The Quarterly Journal of Economics}
  \bibinfo{volume}{69}, \bibinfo{pages}{99—--118}.
\bibitem[{Sisson(2005)}]{sisson_2005_transdimensional}
\bibinfo{author}{Sisson, S.A.}, \bibinfo{year}{2005}.
\newblock \bibinfo{title}{Transdimensional markov chains}.
\newblock \bibinfo{journal}{Journal of the American Statistical Association}
  \bibinfo{volume}{100}, \bibinfo{pages}{1077--1089}.
\bibitem[{Steinberg and Cardell(1998)}]{steinberg_hybrid_1998}
\bibinfo{author}{Steinberg, D.}, \bibinfo{author}{Cardell, N.S.},
  \bibinfo{year}{1998}.
\newblock \bibinfo{title}{The hybrid {CART}-{Logit} model in classification and
  data mining}.
\newblock \bibinfo{journal}{Salford Systems White Paper} \URLprefix
  \url{http://media.salford-systems.com/pdf/the-hybrid-cart-logit-model-in-classification-and-data%20mining-1998.pdf}.
\bibitem[{Strobl et~al.(2009)Strobl, Malley and
  Tutz}]{strobl_2009_introduction}
\bibinfo{author}{Strobl, C.}, \bibinfo{author}{Malley, J.},
  \bibinfo{author}{Tutz, G.}, \bibinfo{year}{2009}.
\newblock \bibinfo{title}{An introduction to recursive partitioning: rationale,
  application, and characteristics of classification and regression trees,
  bagging, and random forests.}
\newblock \bibinfo{journal}{Psychological methods} \bibinfo{volume}{14},
  \bibinfo{pages}{323}.
\bibitem[{St{\"u}ttgen et~al.(2012)St{\"u}ttgen, Boatwright and
  Monroe}]{stuttgen2012satisficing}
\bibinfo{author}{St{\"u}ttgen, P.}, \bibinfo{author}{Boatwright, P.},
  \bibinfo{author}{Monroe, R.T.}, \bibinfo{year}{2012}.
\newblock \bibinfo{title}{A satisficing choice model}.
\newblock \bibinfo{journal}{Marketing Science} \bibinfo{volume}{31},
  \bibinfo{pages}{878--899}.
\bibitem[{Su(2007)}]{su_2007_tree}
\bibinfo{author}{Su, X.}, \bibinfo{year}{2007}.
\newblock \bibinfo{title}{Tree-based model checking for logistic regression}.
\newblock \bibinfo{journal}{Statistics in medicine} \bibinfo{volume}{26},
  \bibinfo{pages}{2154--2169}.
\bibitem[{Swait(1984)}]{swait_1984_probabilistic}
\bibinfo{author}{Swait, J.}, \bibinfo{year}{1984}.
\newblock \bibinfo{title}{Probabilistic choice set formation in transportation
  demand models}.
\newblock \bibinfo{publisher}{Unpublished Ph.D. Thesis, Dept. of Civil
  Engineering, Massachusetts Institute of Technology},
  \bibinfo{address}{Cambridge, MA}.
\bibitem[{Swait(2001a)}]{swait_2001_choice}
\bibinfo{author}{Swait, J.}, \bibinfo{year}{2001}a.
\newblock \bibinfo{title}{Choice set generation within the generalized extreme
  value family of discrete choice models}.
\newblock \bibinfo{journal}{Transportation Research Part B: Methodological}
  \bibinfo{volume}{35}, \bibinfo{pages}{643--666}.
\newblock \URLprefix
  \url{http://www.sciencedirect.com/science/article/pii/S0191261500000291}.
\bibitem[{Swait(2001b)}]{swait_2001_non}
\bibinfo{author}{Swait, J.}, \bibinfo{year}{2001}b.
\newblock \bibinfo{title}{A non-compensatory choice model incorporating
  attribute cutoffs}.
\newblock \bibinfo{journal}{Transportation Research Part B: Methodological}
  \bibinfo{volume}{35}, \bibinfo{pages}{903--928}.
\bibitem[{Swait(2009)}]{swait_2009_choice}
\bibinfo{author}{Swait, J.}, \bibinfo{year}{2009}.
\newblock \bibinfo{title}{Choice models based on mixed discrete/continuous
  pdfs}.
\newblock \bibinfo{journal}{Transportation Research Part B: Methodological}
  \bibinfo{volume}{43}, \bibinfo{pages}{766--783}.
\bibitem[{Swait et~al.(2002)Swait, Adamowicz, Hanemann, Diederich, Krosnick,
  Layton, Provencher, Schkade and Tourangeau}]{swait_2002_context}
\bibinfo{author}{Swait, J.}, \bibinfo{author}{Adamowicz, W.},
  \bibinfo{author}{Hanemann, M.}, \bibinfo{author}{Diederich, A.},
  \bibinfo{author}{Krosnick, J.}, \bibinfo{author}{Layton, D.},
  \bibinfo{author}{Provencher, W.}, \bibinfo{author}{Schkade, D.},
  \bibinfo{author}{Tourangeau, R.}, \bibinfo{year}{2002}.
\newblock \bibinfo{title}{{Context Dependence and Aggregation in Disaggregate
  Choice Analysis}}.
\newblock \bibinfo{journal}{Marketing Letters} \bibinfo{volume}{13},
  \bibinfo{pages}{195--205}.
\bibitem[{Swait and Ben-Akiva(1987a)}]{swait1987empirical}
\bibinfo{author}{Swait, J.}, \bibinfo{author}{Ben-Akiva, M.},
  \bibinfo{year}{1987}a.
\newblock \bibinfo{title}{Empirical test of a constrained choice discrete
  model: mode choice in sao paulo, brazil}.
\newblock \bibinfo{journal}{Transportation Research Part B: Methodological}
  \bibinfo{volume}{21}, \bibinfo{pages}{103--115}.
\bibitem[{Swait and Ben-Akiva(1987b)}]{swait1987incorporating}
\bibinfo{author}{Swait, J.}, \bibinfo{author}{Ben-Akiva, M.},
  \bibinfo{year}{1987}b.
\newblock \bibinfo{title}{Incorporating random constraints in discrete models
  of choice set generation}.
\newblock \bibinfo{journal}{Transportation Research Part B: Methodological}
  \bibinfo{volume}{21}, \bibinfo{pages}{91--102}.
\bibitem[{Tibshirani and Knight(1999)}]{tibshirani_1999_model}
\bibinfo{author}{Tibshirani, R.}, \bibinfo{author}{Knight, K.},
  \bibinfo{year}{1999}.
\newblock \bibinfo{title}{Model search by bootstrap “bumping”}.
\newblock \bibinfo{journal}{Journal of Computational and Graphical Statistics}
  \bibinfo{volume}{8}, \bibinfo{pages}{671--686}.
\bibitem[{{TNS Opinion \& Social}(2015)}]{directorate_2015_quality}
\bibinfo{author}{{TNS Opinion \& Social}}, \bibinfo{year}{2015}.
\newblock \bibinfo{title}{Quality of transport}.
\newblock \bibinfo{type}{Technical Report}. {Directorate-General for Mobility
  and Transport (European Commission)}.
\newblock \URLprefix
  \url{https://publications.europa.eu/en/publication-detail/-/publication/fb18e0ed-52d0-41cd-be6c-61aad310fb53/language-en}.
\bibitem[{Toth and Eltinge(2011)}]{toth_2011_building}
\bibinfo{author}{Toth, D.}, \bibinfo{author}{Eltinge, J.L.},
  \bibinfo{year}{2011}.
\newblock \bibinfo{title}{Building consistent regression trees from complex
  sample data}.
\newblock \bibinfo{journal}{Journal of the American Statistical Association}
  \bibinfo{volume}{106}, \bibinfo{pages}{1626--1636}.
\bibitem[{Train(2009)}]{train_2009_discrete}
\bibinfo{author}{Train, K.}, \bibinfo{year}{2009}.
\newblock \bibinfo{title}{Discrete {Choice} {Methods} {With} {Simulation}}.
\newblock \bibinfo{edition}{2} ed., \bibinfo{publisher}{Cambridge University
  Press}, \bibinfo{address}{New York, NY, USA}.
\bibitem[{Truong et~al.(2015)Truong, Wiktor and Boxall}]{truong_2015_modeling}
\bibinfo{author}{Truong, T.D.}, \bibinfo{author}{Wiktor, L.},
  \bibinfo{author}{Boxall, P.C.}, \bibinfo{year}{2015}.
\newblock \bibinfo{title}{Modeling non-compensatory preferences in
  environmental valuation}.
\newblock \bibinfo{journal}{Resource and Energy Economics}
  \bibinfo{volume}{39}, \bibinfo{pages}{89--107}.
\bibitem[{Tversky(1972)}]{tversky_1972_elimination}
\bibinfo{author}{Tversky, A.}, \bibinfo{year}{1972}.
\newblock \bibinfo{title}{Elimination by aspects: A theory of choice.}
\newblock \bibinfo{journal}{Psychological review} \bibinfo{volume}{79},
  \bibinfo{pages}{281—--299}.
\bibitem[{Tversky and Kahneman(1986)}]{tversky_1986_rational}
\bibinfo{author}{Tversky, A.}, \bibinfo{author}{Kahneman, D.},
  \bibinfo{year}{1986}.
\newblock \bibinfo{title}{Rational choice and the framing of decisions}.
\newblock \bibinfo{journal}{The Journal of Business} \bibinfo{volume}{59},
  \bibinfo{pages}{S251—S278}.
\bibitem[{{University of California at
  Berkeley}(2000)}]{university_of_california_2000}
\bibinfo{author}{{University of California at Berkeley}}, \bibinfo{year}{2000}.
\newblock \bibinfo{title}{10.11.00 - daniel l. mcfadden wins nobel prize in
  economics}.
\newblock \URLprefix \url{http://www.berkeley.edu/news/features/2000/nobel/}.
\bibitem[{Velikova and Daniels(2004)}]{velikova_2004_decision}
\bibinfo{author}{Velikova, M.}, \bibinfo{author}{Daniels, H.},
  \bibinfo{year}{2004}.
\newblock \bibinfo{title}{{Decision trees for monotone price models}}.
\newblock \bibinfo{journal}{Computational Management Science}
  \bibinfo{volume}{1}, \bibinfo{pages}{231--244}.
\bibitem[{Vij et~al.(2013)Vij, Carrel and Walker}]{vij_2013_incorporating}
\bibinfo{author}{Vij, A.}, \bibinfo{author}{Carrel, A.},
  \bibinfo{author}{Walker, J.L.}, \bibinfo{year}{2013}.
\newblock \bibinfo{title}{Incorporating the influence of latent modal
  preferences on travel mode choice behavior}.
\newblock \bibinfo{journal}{Transportation Research Part A: General}
  \bibinfo{volume}{54}, \bibinfo{pages}{164--178}.
\bibitem[{Vij and Walker(2014)}]{vij_2014_preference}
\bibinfo{author}{Vij, A.}, \bibinfo{author}{Walker, J.L.},
  \bibinfo{year}{2014}.
\newblock \bibinfo{title}{Preference endogeneity in discrete choice models}.
\newblock \bibinfo{journal}{Transportation Research Part B: Methodological}
  \bibinfo{volume}{64}, \bibinfo{pages}{90--105}.
\newblock \URLprefix
  \url{http://www.sciencedirect.com/science/article/pii/S0191261514000344},
  \DOIprefix\doi{10.1016/j.trb.2014.02.008}.
\bibitem[{Villandr{\'e} et~al.(2012)Villandr{\'e}, Rich and
  Ciampi}]{villandre_2012_soft}
\bibinfo{author}{Villandr{\'e}, L.}, \bibinfo{author}{Rich, B.},
  \bibinfo{author}{Ciampi, A.}, \bibinfo{year}{2012}.
\newblock \bibinfo{title}{Soft classification trees}.
\newblock \bibinfo{journal}{Communications in Statistics-Theory and Methods}
  \bibinfo{volume}{41}, \bibinfo{pages}{3244--3258}.
\bibitem[{Vuong(1989)}]{vuong_1989_likelihood}
\bibinfo{author}{Vuong, Q.H.}, \bibinfo{year}{1989}.
\newblock \bibinfo{title}{Likelihood ratio tests for model selection and
  non-nested hypotheses}.
\newblock \bibinfo{journal}{Econometrica: Journal of the Econometric Society}
  \bibinfo{volume}{57}, \bibinfo{pages}{307--333}.
\bibitem[{Wagenmakers et~al.(2008)Wagenmakers, Lee, Lodewyckx and
  Iverson}]{wagenmakers_2008_bayesian}
\bibinfo{author}{Wagenmakers, E.J.}, \bibinfo{author}{Lee, M.},
  \bibinfo{author}{Lodewyckx, T.}, \bibinfo{author}{Iverson, G.J.},
  \bibinfo{year}{2008}.
\newblock \bibinfo{title}{Bayesian versus frequentist inference}, in:
  \bibinfo{editor}{Hoijtink, H.}, \bibinfo{editor}{Klugkist, I.},
  \bibinfo{editor}{Boelen, P.} (Eds.), \bibinfo{booktitle}{{Bayesian Evaluation
  of Informative Hypotheses}}. \bibinfo{publisher}{Springer}, pp.
  \bibinfo{pages}{181--207}.
\bibitem[{Wainer(2016)}]{wainer_2016_comparison}
\bibinfo{author}{Wainer, J.}, \bibinfo{year}{2016}.
\newblock \bibinfo{title}{Comparison of 14 different families of classification
  algorithms on 115 binary datasets}.
\newblock \bibinfo{journal}{arXiv preprint arXiv:1606.00930} .
\bibitem[{Walker(2013)}]{walker_2013_bayesian}
\bibinfo{author}{Walker, S.G.}, \bibinfo{year}{2013}.
\newblock \bibinfo{title}{Bayesian inference with misspecified models}.
\newblock \bibinfo{journal}{Journal of Statistical Planning and Inference}
  \bibinfo{volume}{143}, \bibinfo{pages}{1621--1633}.
\bibitem[{Wu et~al.(2007)Wu, Tjelmeland and West}]{wu_2007_bayesian}
\bibinfo{author}{Wu, Y.}, \bibinfo{author}{Tjelmeland, H.},
  \bibinfo{author}{West, M.}, \bibinfo{year}{2007}.
\newblock \bibinfo{title}{{Bayesian CART: Prior Specification and Posterior
  Simulation}}.
\newblock \bibinfo{journal}{Journal of Computational and Graphical Statistics}
  \bibinfo{volume}{16}, \bibinfo{pages}{44--66}.
\bibitem[{Yildiz et~al.(2016)Yildiz, Irsoy and Alpaydin}]{yildiz_2016_bagging}
\bibinfo{author}{Yildiz, O.T.}, \bibinfo{author}{Irsoy, O.},
  \bibinfo{author}{Alpaydin, E.}, \bibinfo{year}{2016}.
\newblock \bibinfo{title}{{Bagging Soft Decision Trees}}, in:
  \bibinfo{editor}{Holzinger, A.} (Ed.), \bibinfo{booktitle}{Machine Learning
  for Health Informatics}. \bibinfo{publisher}{Springer}, pp.
  \bibinfo{pages}{25--36}.
\bibitem[{Young(1984)}]{young_1984_non}
\bibinfo{author}{Young, W.}, \bibinfo{year}{1984}.
\newblock \bibinfo{title}{A non-tradeoff decision making model of residential
  location choice}.
\newblock \bibinfo{journal}{Transportation Research Part A: General}
  \bibinfo{volume}{18}, \bibinfo{pages}{1--11}.
\bibitem[{Yu et~al.(2016)Yu, Lee, Cheung, Lau, Mok and Hui}]{yu_2016_logit}
\bibinfo{author}{Yu, P.L.}, \bibinfo{author}{Lee, P.H.},
  \bibinfo{author}{Cheung, S.}, \bibinfo{author}{Lau, E.Y.},
  \bibinfo{author}{Mok, D.S.}, \bibinfo{author}{Hui, Harry, C.},
  \bibinfo{year}{2016}.
\newblock \bibinfo{title}{{Logit tree models for discrete choice data with
  application to advice-seeking preferences among Chinese Christians}}.
\newblock \bibinfo{journal}{Computational Statistics} \bibinfo{volume}{31},
  \bibinfo{pages}{799--827}.
\bibitem[{Yuksel et~al.(2012)Yuksel, Wilson and Gader}]{yuksel_2012_twenty}
\bibinfo{author}{Yuksel, S.E.}, \bibinfo{author}{Wilson, J.N.},
  \bibinfo{author}{Gader, P.D.}, \bibinfo{year}{2012}.
\newblock \bibinfo{title}{{Twenty Years of Mixture of Experts}}.
\newblock \bibinfo{journal}{IEEE Transactions on Neural Networks and Learning
  Systems} \bibinfo{volume}{23}, \bibinfo{pages}{1177--1193}.
\bibitem[{Zeilis et~al.(2008)Zeilis, Hothorn and Hornik}]{zeilis_2008_model}
\bibinfo{author}{Zeilis, A.}, \bibinfo{author}{Hothorn, T.},
  \bibinfo{author}{Hornik, K.}, \bibinfo{year}{2008}.
\newblock \bibinfo{title}{{Model-Based Recursive Partitioning}}.
\newblock \bibinfo{journal}{Journal of Computational and Graphical Statistics}
  \bibinfo{volume}{17}, \bibinfo{pages}{492--514}.
\bibitem[{Zhu and Timmermans(2010)}]{zhu_2010_cognitive}
\bibinfo{author}{Zhu, W.}, \bibinfo{author}{Timmermans, H.},
  \bibinfo{year}{2010}.
\newblock \bibinfo{title}{Cognitive process model of individual choice
  behaviour incorporating principles of bounded rationality and heterogeneous
  decision heuristics}.
\newblock \bibinfo{journal}{Environment and Planning B: Planning and Design}
  \bibinfo{volume}{37}, \bibinfo{pages}{59--74}.
\bibitem[{Zolfaghari et~al.(2013)Zolfaghari, Sivakumar and
  Polak}]{zolfaghari_2013_simplified}
\bibinfo{author}{Zolfaghari, A.}, \bibinfo{author}{Sivakumar, A.},
  \bibinfo{author}{Polak, J.}, \bibinfo{year}{2013}.
\newblock \bibinfo{title}{Simplified probabilistic choice set formation models
  in a residential location choice context}.
\newblock \bibinfo{journal}{Journal of choice modelling} \bibinfo{volume}{9},
  \bibinfo{pages}{3--13}.

\end{thebibliography}
\end{document}